\documentclass[a4paper,11pt]{article}
\pdfoutput=1 % if your are submitting a pdflatex (i.e. if you have
\usepackage{jheppub} % for details on the use of the package, please
\usepackage[T1]{fontenc} % if needed

\usepackage{braket}
\usepackage{todonotes}
\usepackage{braket}
\usepackage{appendix}
\usepackage{float}
\usepackage{array}
\usepackage{comment}
\usepackage{xcolor}
\usepackage{booktabs}
\usepackage{caption}
\usepackage{subcaption}
\usepackage{hhline}
\usepackage{enumerate}

\title{\boldmath Constraints from Entanglement Wedge Nesting for Holography at a Finite Cutoff}

\author[a]{Krishan Saraswat}

\affiliation[a]{Physics Department, Broida Hall, University of California, Santa Barbara, CA 93106, USA}

\emailAdd{ksaraswa@ualberta.ca}

\abstract{We explore constraints that arise from associating an entanglement wedge (EW) to subregions of a cutoff boundary at a finite distance in AdS/CFT, using a subcritical end-of-the-world (ETW) brane acting as a cutoff. 

In particular, we consider the case of two intervals in the holographic dual to a BCFT, with one interval $A$ located at the asymptotic boundary and the second interval $B$ located on the ETW brane. We discuss in detail subtleties that arise near the RT end-points when defining the EW for this configuration, particularly in the connected phase. Entanglement wedge nesting (EWN) requires that $\mathcal{W}_E(A) \cup \mathcal{W}_E(B) \subseteq \mathcal{W}_E(A\cup B)$. We demonstrate that already in the simplest example of an AdS$_3$ bulk geometry, EWN can be violated even if $A$ and $B$ are spacelike separated through the bulk and instead we must require the stronger condition that $\mathcal{W}_E(A)$ be spacelike separated from $\mathcal{W}_E(B)$, which highlights the non-local nature of the cutoff theory. Our prescription to associate EWs to subregions on the ETW brane is different from the restricted maximin procedure \cite{Grado-White:2020wlb} but will agree within the subset of parameter space where EWN is respected.  

Additionally, we study EWN in a two sided BTZ black hole geometry with an ETW brane in one of the exteriors. In the BTZ black hole example we find that our condition for EWN disallows configurations where the RT surface goes from the brane to the black hole singularity.}

\begin{document} 
\maketitle

\section{Introduction}
\subsection{Motivation and Background}
\label{BackgroundMotiSec}
While the AdS/CFT correspondence \cite{Maldacena:1997re,Witten:1998qj,Gubser:1999vj,Nastase:2007kj,Polchinski:2010hw,Ammon:2015wua,VanRaamsdonk:2016exw} is our best model of holography to date, it is -- in its best established form -- limited to describing a gravitational theory in an asymptotically AdS spacetime with prescribed boundary conditions at infinity \cite{Witten:1998qj,Gubser:1998bc,deHaro:2000vlm}. This presents a significant obstacle for studying quantum gravity in more realistic flat or de Sitter backgrounds, as well as localized subregions.

One attempt at generalizing the duality between quantum gravity in a bulk AdS spacetime and a quantum field theory on the asymptotic boundary is to consider spacetimes with boundaries at a finite proper distance. Understanding holography in such a setting would enable the study of quantum gravity in finite local subregions, and also potentially embed cosmologies in holography \cite{Gubser:1999vj,Cooper:2018cmb}. More generally, such ideas can be understood in the framework of the surface/state correspondence \cite{Miyaji:2015yva, Miyaji:2015fia}, which assigns a quantum state $\ket{\phi(\Sigma)}$ to convex codimension two submanifolds $\Sigma$ in a gravitational spacetime.

Concrete realizations for the surface/state correspondence can be found, apart from tensor networks which motivated the correspondence, for example in the study of $T \bar T$-de\-for\-ma\-tions of holographic CFTs \cite{Zamolodchikov:2004ce,Jiang2019LecturesOS,Bonelli:2018kik,Dubovsky:2017cnj,Grieninger:2019zts,Gross:2019ach,Cavaglia:2016oda,Pant:2024eno}. It has been proposed that a $T\bar T$ deformation of a holographic CFT should be understood as ``moving the boundary into the bulk'' \cite{McGough:2016lol,Guica:2019nzm,Lewkowycz:2019xse}. Furthermore, the appearance of cutoff surfaces have also played a key role in double holography \cite{Takayanagi:2011zk,Almheiri:2019hni,Geng:2020fxl,Geng:2020qvw,Chen:2020uac,Chen:2020hmv,Hernandez:2020nem,Grimaldi:2022suv,Geng:2021wcq,Geng:2021iyq,Neuenfeld2021DoubleHA,Neuenfeld:2021wbl,Omiya:2021olc,Geng:2023iqd,Geng:2023qwm}, where a convex cutoff surface in the form of an end-of-the-world (ETW) brane appears. Doubly-holographic models with AdS branes start by introducing the dual of a holographic $d$-dimensional boundary BCFT
in terms of a $d+1$-dimensional bulk AdS spacetime.\footnote{BCFTs are conformal field theories living on a space with boundary at which conformal boundary conditions are imposed. However, at least for a BCFT$_2$ one needs to introduce boundary degrees of freedom which break conformal invariance in order to obtain interesting models, e.g. by coupling an SYK model to the BCFT boundary. We will thus use the term BCFT more loosely to also account for situations like this.} The boundary of the BCFT is continued into the bulk by an ETW brane \cite{Takayanagi:2011zk} which removes part of the naive bulk spacetime dual to a CFT without a boundary.

Naturally, it is tempting to extend lessons learned from AdS/CFT to holography at a cutoff. Of particular interest are holographic computations of entanglement entropies. In AdS/CFT the holographic entanglement entropy of some subregion $U$ of the asymptotic boundary is given, at leading order in $N$, by the Hubeny-Rangamani-Ryu-Takayanagi formula\footnote{In fact, in certain situations such as black hole evaporation, the quantum corrected RT formula \cite{Engelhardt:2014gca} can yield results that differ at leading order from the RT formula. We will not be concerned with such situations in this paper.}  \cite{Ryu:2006bv,Hubeny:2007xt}
\begin{equation}
    \label{HRRT}
    S_\text{EE}(U)=\frac{\text{Area}(\chi(U))}{4G_N},
\end{equation}
where $\chi(U)$ is the minimal area extremal surface homologous to the boundary subregion $U$. The surface $\chi(U)$ is known as the HRRT surface or RT surface for short.

The surface/state correspondence proposes that in the case of a finite cutoff the von Neumann entropy of a reduced density matrix obtained from $\ket{\phi(\Sigma)}$ can be computed from co-dimension two extremal surfaces that are anchored at the boundary of subregions of (a slice of) the spacetime boundary $\Sigma$. A basic consistency check of this proposal with holographic subregion entropy computations has e.g.\ been performed in the context of $T \bar T$ in \cite{Donnelly:2018bef}. Also, in double holography, there is evidence that subregion entropies on the brane can be computed using the RT formula \cite{Chen:2020uac}. However, it is still unknown under which conditions the proposal of the surface/state correspondence is fully consistent. While it is not even clear what ``consistency'' entails in detail, strong subadditivity (SSA) of holographic entanglement entropy \cite{Headrick:2007km,Wall:2012uf} should certainly be a necessary condition if we want the holographic entanglement entropy to behave like an entropy in the quantum information theoretic sense.

In addition, in standard AdS/CFT we have a notion of subregion/subregion duality \cite{Headrick:2014cta,Akers:2016ugt,Akers:2017ttv,Chen:2021lnq,Leutheusser:2022bgi}. More precisely, a boundary subregion $U$ is dual to the \emph{entanglement wedge} $\mathcal W_E(U)$ which is defined as the domain of dependence of the partial bulk Cauchy slice bounded by $U$ and the corresponding RT surface $\chi(U)$. Moreover, given a subregion $U' \subset U$, $\mathcal{W}_E(U')$ is the bulk subregion dual to $U'$, and must satisfy $\mathcal{W}_E(U')\subset\mathcal{W}_E(U)$. This follows from the fact that tracing out parts of $U$ to $U'$ acts as a quantum channel on the reduced density matrix $\rho_U$. This property of entanglement wedges is called \emph{entanglement wedge nesting} (EWN).

The purpose of this paper is to first carefully describe the subtleties in explicitly defining and constructing entanglement wedges for cutoff subregions $B$, and second, to derive non-trivial conditions on the relative location of $B$ with respect to another subregion $A$ (disjoint from $B$ and living on a conformal boundary) imposed by EWN. For concreteness, we will focus on the simple geometries of AdS$_3$ and Planar BTZ cut off by a hyperbolic ETW brane where computations can be done easily and the constraints we derive can be interpreted in a straightforward manner. 

We will \emph{assume} that the RT formula in its simplest formulation, Eq.~\eqref{HRRT}, holds.\footnote{The authors of \cite{Bousso:2022hlz} proposed the notion of generalized entanglement wedges for gravitational theories. Their proposal appears to only work for subregions of a time-reflection symmetric Cauchy slice and thus is not applicable to our configurations. 
} Given this starting point, the RT surface which computes $S(A \cup B)$ for two disconnected, spacelike subregions $A$ and $B$ consists of two disconnected pieces which can be in one of two phases. Either, each connected piece can be homologous to one of $A$ and $B$. In this case, which we will call the \emph{disconnected phase}, the mutual information $S(A) + S(B) - S(A \cup B)$ vanishes. The second phase has the RT surface connect the subregions $A$ and $B$ to each other. Consequently, the mutual information does not vanish. We will call this the \emph{connected} phase. We will then construct the entanglement wedges $\mathcal W_E$ for the regions $A$, $B$, and $A \cup B$.
As we will argue below, while the definitions of $\mathcal W_E(A)$ and $\mathcal W_E(B)$ are straightforward and agree with the intersection of entanglement wedge in the full geometry cut off by the brane, the definition of the connected entanglement wedge is more subtle and requires us to be careful about the location and direction at which the RT surface ends on the cutoff surface. 

We find that, in general, and unlike AdS/CFT, spacelike separation of $A$ and $B$ is not enough to ensure EWN. As we will argue, this happens because spacelike separateness of $A$ and $B$, while sufficient to make sure that each connected piece of the associated RT surfaces is spacelike, is not enough to guarantee that the disconnected pieces of the RT surfaces are also spacelike separated. What we demonstrate is that the non-trivial condition EWN imposes precisely ensures that all the disconnected pieces of the RT surfaces are spacelike separated. Requiring that EWN is satisfied imposes a bound on the relative separation of $A$ and $B$ which is stronger than spacelike separation through the bulk.

It is well known \cite{Lewkowycz:2019xse,Geng:2019ruz,Grado-White:2020wlb,Neuenfeld:2023svs} that naively extending the HRT formula for cutoff surfaces can yield situations where SSA and EWN is violated. This motivated the authors of \cite{Grado-White:2020wlb} to propose a modification of the RT formula for cutoff surfaces, called \emph{restricted maximin}. The main idea of their proposal is to only consider bulk extremal surfaces which lie on a Cauchy slice that intersects $U$, which guarantees SSA and EWN. However, as we will argue below, applying their construction to a configuration of disjoint subregions $A$ and $B$ which violate our bounds will produce RT surfaces which disagree with naive expectations. For example, given a static geometry and subregions $A$ and $B$ on constant but different time-slices, the two disjoint components of the RT surface in the disconnected phase will have a non-trivial profile in time. Since this happens whenever $\mathcal W_E(A)$ and $\mathcal W_E(B)$ are timelike separated, this suggests that a deviation between the naive RT surfaces defined in this paper and restricted maximin surfaces indicates that $A$ and $B$ are not independent, possibly due to the non-local nature of the theory at the cutoff.

The issue of defining the subregion, $B$, on the brane/cutoff can be subtle depending on the types of boundary condition (BC) we impose on the brane. When discussing the brane/cutoff there appear to be at least two distinct ways in which we might interpret the physics on the brane which depends on whether we choose Dirichlet or Neumann BCs on the brane \cite{Geng:2023qwm,Deng:2023pjs,Geng:2024xpj}. When using Neumann BCs the way the brane is embedded in the higher dimensional AdS geometry is determined dynamically by analyzing equations of motion of the following Lorentzian action\footnote{Note in our setup in Poincare AdS the boundary of the bulk cutoff by the brane consists of the conformal boundary as well as the ETW brane so we should also include a Gibbons-Hawking term of the form $\int_{\text{bdry}}d^dy\sqrt{-h_{\text{bdry}}}K_{\text{bdry}}$. For simplicity we omit explicitly writing it this in the action and focus on the brane and bulk terms.} \cite{Takayanagi:2011zk}
\begin{equation}
    \label{Stotal}
    \frac{1}{16\pi G_N}\int_{\text{Bulk}}d^{d+1}x\sqrt{-g}(R-2\Lambda)+\frac{1}{8\pi G_N}\int_{\text{Brane}}d^d x\sqrt{-h}(K-T).
\end{equation}
The parameter $T$ denotes the tension of the ETW brane. Varying Eq.~\eqref{Stotal} with respect to the metric gives an equation that needs to be satisfied by the brane,
\begin{equation}
    K_{ab}-Kh_{ab}+Th_{ab}=0,
\end{equation}
where $h_{ab}$ is the induced metric on the brane and $K_{ab}$ is the extrinsic curvature. We can rewrite the equation above into a more simple form by substituting its trace,
\begin{equation}
    \label{EOWBEOM}
        K_{ab}=\frac{T}{d-1}h_{ab},
\end{equation}
which implies the brane/cutoff is a co-dimension one surface in the bulk of constant extrinsic curvature characterized by a tension $T$. Usually one can also consider dynamical gravity on the brane as is often done in double holography \cite{Takayanagi:2011zk,Almheiri:2019hni,Geng:2020fxl,Geng:2020qvw,Chen:2020uac,Chen:2020hmv,Hernandez:2020nem,Grimaldi:2022suv,Neuenfeld2021DoubleHA,Neuenfeld:2021wbl,Omiya:2021olc,Geng:2023iqd,Geng:2023qwm} and brane-world gravity \cite{Karch:2000ct,Geng:2023qwm}. This naively poses the questions of how to define cutoff subregions in a diffeomorphism invariant way. One possible way to address this by going back to the subregion $A$ which lies on the non-gravitating conformal boundary. Then we consider RT surfaces associated to $A$ which extend to the brane. In the case of Neumann BCs the location the RT surfaces end on the brane is dynamically determined and this in turn can be used to associate a subregion $B$ on the brane given some $A$ on the conformal boundary (i.e. $B$ is not independently chosen from $A$). Since we are considering dynamical gravity on the brane one can change the gravitational dynamics on the brane by adding terms to Eq. (\ref{Stotal}) (e.g, higher derivative corrections, DGP gravity, etc. on the brane \cite{Dvali:2000hr,Chen:2020hmv,Chen:2020uac,Geng:2022tfc,Geng:2022slq,Geng:2023qwm,Lee:2022efh}) these would generally affect where the RT surface ends on the brane and thereby shift the relative location of $A$ and $B$. One might speculate that it is possible to end up in configurations that violate EWN, in this case the constraints we find may be regarded as providing constraints to gravity on the brane. On the other hand, imposing Dirichlet boundary conditions on the brane one can no longer view the embedding of the brane in AdS as arising from the equation of motion given in Eq. (\ref{EOWBEOM}). In this case we simply view the brane as a cutoff surface of constant extrinsic curvature where all bulk fields satisfy Dirichlet boundary conditions on the brane, this is more akin to an interpretation of a cutoff in the $T\bar{T}$ context where gravity is effectively turned off on the brane. In such a case we can fix some gauge and naively choose spacelike separated subregions $A$ and $B$ ``independently'' of each other. In this case what we constrain through EWN is not gravity on the brane as we did in the Neumann case but rather the relative location of $A$ and $B$. The central point here is that regardless of the interpretation we place on the physics of the brane our results simply rely on having two subregions; $A$ on a conformal boundary and $B$ on the brane/cutoff along with a prescription to associate entanglement wedges to these subregions. With this we can derive constraints from EWN and such an investigation on its own will yield interesting results which we summarize in the following section.

\begin{comment}
    However, the theory on the cutoff surface might in general be gravitational as is the case in doubly holography \cite{Karch:2000ct}. This naively poses the question of how to define cutoff subregions in a diffeomorphism invariant way. In this paper, we restrict ourselves to the strict $N \to \infty$ limit, such that gravity is effectively turned off in the bulk. In this case subregions can be defined uniquely.\footnote{The authors of \cite{Bousso:2022hlz} proposed the notion of generalized entanglement wedges for gravitational theories. Their proposal only works for subregions of a time-reflection symmetric Cauchy slice and thus is not applicable to our configurations. 
} Extending this discussion to $1/N$ corrections, while interesting, will not be pursued here.
\end{comment}

\subsection{Overview}
This paper is roughly divided into two parts. The first part consists of Sections \ref{Section2Poincare} and \ref{ChapterDerivingConstraintForEWN} where we explore EWN in Poincare AdS$_3$ cutoff by an end-of-the-world (ETW) brane. In the second part of this work, given in Section \ref{EWNBTZChapterOfPaper}, we study EWN for a two-sided planar BTZ black hole geometry with the right exterior cutoff by an ETW brane.  

More specifically, in Section \ref{Section2Poincare} we consider two disjoint spacelike separated constant time intervals $A$ and $B$ on the conformal boundary and brane respectively. Identifying the RT surfaces that we will use in our prescription which are homologous to $A$, $B$, and $A\cup B$ we define entanglement wedges $\mathcal{W}_E(A)$, $\mathcal{W}_E(B)$, and $\mathcal{W}_E(A\cup B)$ respectively. We address the subtle aspects of correctly identifying the entanglement wedges for brane subregions, which involve finding spacelike separated regions from the relevant RT curve line segments. When we explicitly construct $\mathcal{W}_E(B)$ we show that the RT surface that is homologous to $B$ can generally be extended behind the brane and will be homologous to a subregion on the ``imaginary'' or ``virtual'' conformal boundary behind the brane, which we denote as $\text{Vir}(B)$. By defining an ``entanglement wedge'', denoted as $\mathcal{W}_E(\text{Vir}(B))$, in the extended spacetime we demonstrate that $\mathcal{W}_E(B)=\mathcal{W}_E(\text{Vir}(B))\cap\mathcal{M}_{\text{phys}}$, where $\mathcal{M}_{\text{phys}}$ is the physically relevant portion of the spacetime in front of the brane. However, when constructing $\mathcal{W}_E(A\cup B)$ in the connected phase, we demonstrate that it is generally not possible to view $\mathcal{W}_E(A\cup B)$ as the intersection of some kind of extended wedge with $\mathcal{M}_\text{phys}$. 

In Section \ref{ChapterDerivingConstraintForEWN} we explore whether EWN (i.e., requiring that $\mathcal{W}_E(A),\mathcal{W}_E(B)\subseteq\mathcal{W}_E(A\cup B)$) places non-trivial constraints on the relative placement of the spacelike separated intervals $A$ and $B$. We derive the necessary and sufficient condition on the set of parameters that define the location of $A$ and $B$ to ensure that the EWN is satisfied when $\mathcal{W}_E(A\cup B)$ is in the connected phase (see Eq. (\ref{FinalEWNCondition})). We show that the condition to satisfy EWN amounts to ensuring that the RT surfaces homologous to $A$, $B$, $A\cup B$ in the connected phase are all spacelike separated from each other. We conclude our exploration by emphasizing that in our setup we must non-trivially require that $\mathcal{W}_E(A)$ and $\mathcal{W}_E(B)$ be spacelike separated, which is a stronger condition in cutoff holography than simply requiring that $A$ and $B$ be spacelike separated. We connect this back to the expectation that holographic theories on cutoff surfaces are generally expected to exhibit a certain degree of non-locality and that our results corroborate this expectation. We also comment on how our results and prescription for defining entanglement wedges of subregions on cutoff surfaces compare to the restricted maximin prescription, which necessarily gives rise to entanglement wedges that respect EWN. We argue that the RT surfaces used in our prescription and the maximin prescription agree within the subset of parameter space where we EWN is satisfied (i.e. when all the relevant RT surfaces are spacelike separated).

In Section \ref{EWNBTZChapterOfPaper} we study EWN in a two-sided planar BTZ black hole background whose right exterior is cutoff by an ETW brane. We introduce the intervals $A_{\text{Br}}$ and $A_{\text{Bdry}}$ on the brane in the right exterior and on the boundary in the left exterior, respectively. We discuss the RT surfaces that will be homologous to the subregions $A_{\text{Br}}$, $A_{\text{Bdry}}$, and $A_{\text{Bdry}}\cup A_{\text{Br}}$ and use them to define the entanglement wedges $\mathcal{W}_E(A_{\text{Bdry}})$, $\mathcal{W}_E(A_{\text{Br}})$, and $\mathcal{W}_E(A_{\text{Bdry}}\cup A_{\text{Br}})$ respectively. Using similar methods employed in the first part of the paper we can write down the condition that needs to be placed on the intervals $A_{\text{Br}}$ and $A_{\text{Bdry}}$ so that $\mathcal{W}_E(A_{\text{Br}})\subset \mathcal{W}_E(A_{\text{Bdry}}\cup A_{\text{Br}})$ in the connected phase. We find that in the limit where the brane and boundary interval lengths diverge the condition has a simple geometric interpretation in the bulk. This interpretation is that requiring EWN in the connected phase prevents the connected RT surfaces from touching the black hole ``singularities'' at Kruskal time $\tau=\pm \pi/2$. This provides an interesting result in the context of \cite{Lee:2022efh} which also analyzed the same BTZ black hole setup to constrain DGP couplings (more specifically the gravitational coupling set by the value of a JT dilaton) on the brane. The result we obtain serves as another reason beyond the ones discussed in \cite{Lee:2022efh} to ignore RT surfaces that cross the ``singularities'' at $\tau=\pm \pi/2$.     

We conclude the work in Section \ref{ConclusionDiss} where we summarize the main results and ideas of the paper and discuss future avenues of exploration.

\section{Constructing Entanglement Wedges in AdS\texorpdfstring{$_3$}{3} cutoff by ETW Brane}
\label{Section2Poincare}
\begin{figure}
\centering
\includegraphics[width=120mm]{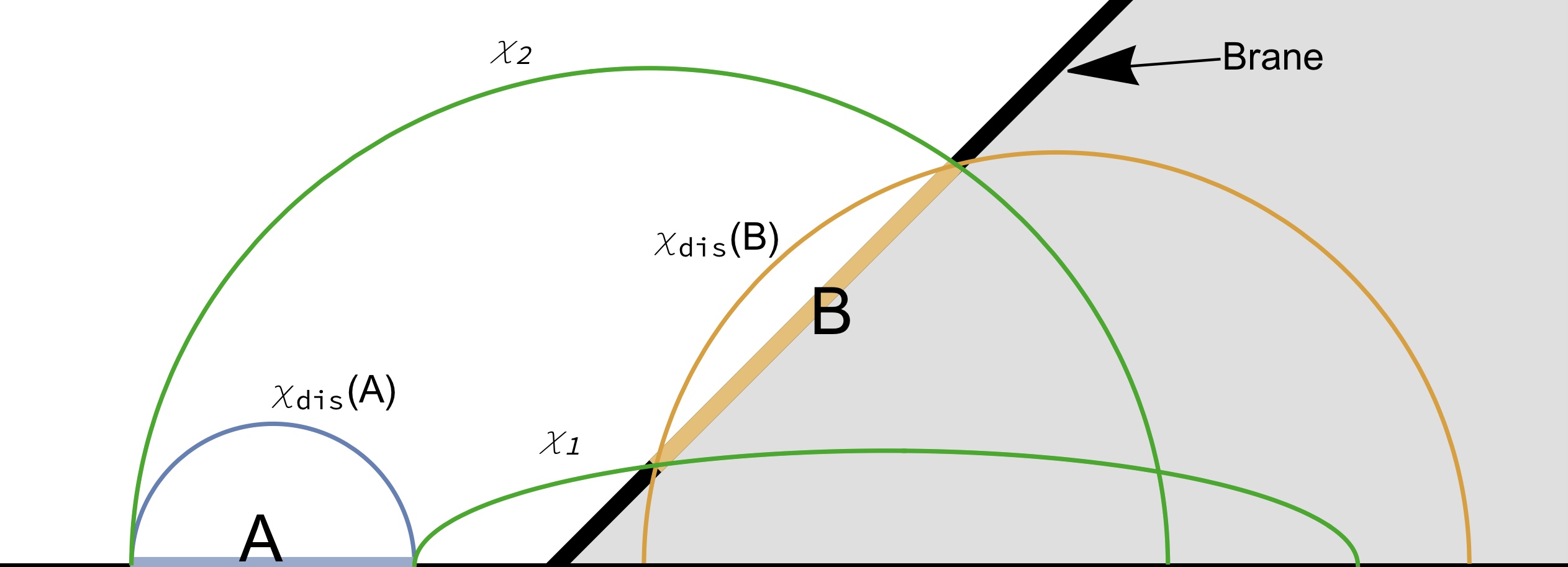}
    \caption{
    This plot showcases an example configuration of intervals $A$ and $B$, with $a_1=2, a_2=6, b_1=2, b_2=8$. The ETW brane with $\theta_0=\pi/4$ cuts off the shaded part of the spacetime. We have set $|t_b-t_a|=0.97(2+\sqrt{2})\approx 3.3$. The blue curve is $\chi_{\text{dis}}(A)$ and is anchored to the conformal boundary subregion $A$ and lies on the constant time slice $t=t_a$. The constant-time, solid yellow curve is $\chi_{\text{dis}}(B)$ and is anchored to the subregion $B$ on the brane. The two green curves show a projection of $\chi_{\text{con}}(A\cup B)=\chi_1\cup \chi_2$ and connect $A$ and $B$. For our particular choice of parameters we have $|k_1|=0.97$ and $|k_2|=0.97\left[\frac{2-\sqrt{2}}{2}\right]\approx 0.28$ which implies that both green line segments are ellipses in the $xz$-plane. Note that all curves defined can be extended behind the brane (in the shaded region), but our constructions will only rely on the portion of these line segments strictly in unshaded part of the spacetime.
    \label{EntangSurfPoincarePlot}}
\end{figure}

In this section, we will present a prescription to construct entanglement wedges in AdS$_3$ with an ETW brane acting as a cutoff surface. We will fix spacelike subregions $A$ and $B$ on the boundary and brane, respectively, and then consider candidate extremal surfaces that are homologous to $A$, $B$, and $A\cup B$, which we denote as $\chi_{\text{dis}}(A)$, $\chi_{\text{dis}}(B)$ and $\chi_{\text{con}}(A \cup B)=\chi_1\cup \chi_2$, which we depict in Figure \ref{EntangSurfPoincarePlot}. These extremal surfaces are then used to construct entanglement wedges $\mathcal{W}_E(A)$, $\mathcal{W}_E(B)$, and $\mathcal{W}_E(A\cup B)$ which will later be used to study EWN.

Section \ref{AdS3PoincareSetup} briefly reviews ETW branes in Poincare AdS$_3$. Explicit expressions for relevant RT surfaces are given in Section \ref{ExplicitPoincareRTSurfaces}. In Section \ref{ClassificationSectionRT} a classification of the extremal surfaces is given in terms of conic sections and we show that due to the presence of the ETW brane a larger set of surfaces than in standard holography must be considered. In Section \ref{EntanglementWedgeDefGenericSec} we describe how our entanglement wedges are constructed and also provide explicit examples. Through our construction we demonstrate that while $\mathcal{W}_E(B)$ can be thought of as naively being cut out of a larger wedge in the extended spacetime, this is generally not possible for $\mathcal{W}_E(A\cup B)$, since our prescription and the naive extended wedge prescription will differ near the brane.

\subsection{AdS\texorpdfstring{$_3$}{3} Poincare Patch with an ETW Brane}
\label{AdS3PoincareSetup}

Our starting point is AdS$_3$ in Poincare coordinates,
\begin{equation}
    ds^2=\frac{L^2\left(dz^2-dt^2+dx^2\right)}{z^2},
\end{equation}
where $L$ is the AdS radius, $z > 0$ is the radial bulk coordinate and $z=0$ is the conformal boundary which is parameterized by $t,x\in\mathbb{R}$. We consider the brane/cutoff as a co-dimension one surface in the bulk which intersects the conformal boundary at $x=0$ and is given by 
\begin{align}
\label{ETWsol}
z=z(x)=\cot\theta_0 x, & & x>0,
\end{align}
where $\theta_0\in (0,\pi/2]$. As we already discussed in the introduction, when one imposes Neumann boundary conditions on the brane, the brane is a surface which will satisfy Eq. (\ref{EOWBEOM}) with a tension given by $TL=\sin\theta_0$. When the tension is zero the ETW brane cuts off half of the AdS space and when $T=L^{-1}$ the ETW brane coincides with the conformal boundary. The limit of $T \lesssim \frac 1 L$, i.e., the limit in which the brane approaches the ``would be conformal boundary'' is referred to as the \emph{critical limit}. The critical limit is physically relevant in the discussion of gravity induced on the brane, particularly in higher dimensions. It is a limit in which bulk gravitons become localized on the brane and one can consider a conventional lower dimensional gravity theory on the world volume of the brane \cite{Karch:2000ct, Geng:2020qvw, Neuenfeld:2021wbl}. In the three-dimensional case discussed here, a theory of dilaton gravity gets induced on the brane, and the critical limit is the limit in which the cutoff scale of the brane theory becomes small \cite{Chen:2020uac,Geng:2022slq,Geng:2022tfc,Neuenfeld:2023svs}. In the case of Dirichlet BCs we simply interpret the brane as some cutoff surface of constant extrinsic curvature $K=2L^{-1}\sin\theta_0$, in this perspective there is no dynamical gravity on the brane.\footnote{With Dirichlet BCs the brane is more similar to a cutoff surface discussed in the $T\bar{T}$ context, whereas Neumann BCs is more appropriate in the discussion of islands in double holography and dynamical gravity on the brane.} In either perspective the key parameter which dictates how the brane/cutoff is embedded in the bulk is given by $\theta_0$ and all our results will be formulated in terms of $\theta_0$.

Throughout this work we will denote $\mathcal{M}_{\text{phys}}$ as the remaining spacetime after the cutoff is introduced. In the case of AdS$_3$ cut off by an ETW brane whose position is given by Eq.~\eqref{ETWsol} we can explicitly express $\mathcal{M}_{\text{phys}}$ as
\begin{equation}
    \mathcal{M}_{\text{phys}}=\{(t,x,z)|\, z\geq \Theta(x)\cot\theta_0 x\},
\end{equation}
where $\Theta(x)$ is the Heaviside step function. The metric on $\mathcal{M}_{\text{phys}}$ is identical to AdS$_3$ with the only difference being that the ETW brane cuts off a the portion of AdS$_3$ which is ``behind'' the brane.

\subsection{Candidate RT Surfaces in AdS\texorpdfstring{$_3$}{3} with an ETW Brane}
\label{ExplicitPoincareRTSurfaces}
We consider two constant time subregions, $A$ and $B$ living on the boundary of this spacetime. $A$ is an interval $[-a_2, -a_1]$ with $0 < a_1 < a_2$ at time $t_a$ on the conformal boundary at $z=0$. Subregion $B$ lives on the brane at time $t = t_b$. It is located at $x\in [b_1\sin\theta_0,b_2\sin\theta_0]$ for $0 < b_1 < b_2$ and $x = z \tan \theta_0$ such that $z\in[b_1\cos\theta_0,b_2\cos\theta_0]$. The proper length of this interval is given by $\ell_{\text{prop}}=\frac{L}{\cos\theta_0}\ln\left(1+\frac{b_2-b_1}{b_1}\right)$.

Since in our approach the RT surfaces by definition are the minimal extremal surfaces homologous to regions $A$ and $B$, we now need to consider the different extremal surfaces which are anchored to those subregions. Since the bulk is three dimensional the extremal surfaces are curves that extend through the bulk. We parameterize these curves in terms of the Poincare coordinate $x$ and extremize the area functional
\begin{equation}
\label{PoincareAdS3RTFunctional}
    \mathcal{A}[z,\dot{z},t,\dot{t};x]=L\int dx\frac{\sqrt{1+\dot{z}^2-\dot{t}^2}}{z(x)},
\end{equation}
where $\dot{z}=\frac{dz}{dx}$ and $\dot{t}=\frac{dt}{dx}$. The general solutions to the equations of motion associated with the functional in Eq. (\ref{PoincareAdS3RTFunctional}) are given by (see Appendix \ref{AppDerivingGenSolRTSurfAdS3})
\begin{equation}
\label{GenSolPoincareAdS3RtSurf}
        t(x)=kx+c_1, \qquad  z(x)=\sqrt{\left(1-k^2\right)\left(\frac{L^2}{H_x^2(1-k^2)^2}-(x+c_2)^2\right)}.
\end{equation}
The four undetermined constants, $k, H_x, c_1, c_2$ can be expressed in terms of the coordinates $a_1$, $a_2$, $b_1$, $b_2$, which determine the entangling surfaces.

There are two possible configurations of the extremal surfaces. Either, the extremal surface consists of two disconnected pieces which are separately homologous to $A$ and $B$. We will denote those surfaces by $\chi_{\text{dis}}(A)$ and $\chi_{\text{dis}}(B)$, respectively, where the subscript stands for \emph{disconnected}.
Explicitly, $\chi_{\text{dis}}(A)$ is given by
\begin{align}
\label{chidisAexp}
    t(x)=t_a &&  z(x)=\sqrt{-(a_1+x)(a_2+x)} && x\in[-a_2,-a_1].
\end{align}
This is simply a semi-circle of radius $\frac{a_2-a_1}{2}$ centered at $(x=-\frac{a_1+a_2}{2},z=0)$ on the constant $t=t_a$ plane. Similarly, $\chi_{\text{dis}}(B)$ is given by
\begin{align}
\label{chidisBexp}
        t(x)=t_b && z(x)=\sqrt{-x^2+\frac{b_1+b_2}{\sin\theta_0}x-b_1b_2} && x\in [b_1\sin\theta_0,b_2\sin\theta_0].
\end{align}
This curve is a section of the semi-circle whose radius is given by $R_b=\frac{\sqrt{b_1^2+b_2^2+2b_1b_2\cos(2\theta_0)}}{2\sin\theta_0}$ centered at $(x=\frac{b_1+b_2}{2\sin\theta_0},z=0)$ on the constant $t=t_b$ plane. In particular, we can think of it as an RT surface which ends on an interval of the asymptotic region in the nonphysical region of spacetime behind the brane, which intersects the ETW brane at the boundary of $B$.

The other configuration consists of two line segments which connect $(t=t_a,x=-a_1,z=0)$ to $(t=t_b,x=b_1\sin\theta_0,z=b_1\cos\theta_0)$ and $(t=t_a,x=-a_2,z=0)$ to $(t=t_b,x=b_2\sin\theta_0,z=b_2\cos\theta_0)$. We will denote those two surfaces by $\chi_1$ and $\chi_2$, respectively, and call their union the \emph{connected} surface\footnote{The fact that the \emph{connected} RT surface consists of two \emph{disconnected} segments is special to $AdS_3$ and we hope that this naming convention will not be the source of confusion.} $\chi_{\text{con}}(A\cup B)$. The explicit trajectories are given by
\begin{align}
\begin{split}
\label{Z1RT}
        t_i(x) &=t_a+k_i\left(x+a_i\right) \\ 
        z_i(x) &=\sqrt{\frac{(x+a_i)\left(b_i^2\cos^2\theta_0-(1-k_i^2)(x-b_i\sin\theta_0)(a_i+b_i\sin\theta_0)\right)}{a_i+b_i\sin\theta_0}},
\end{split}
\end{align}
where $x\in\left[-a_i,b_i\sin\theta_0\right]$ and $k_i=\frac{t_b-t_a}{a_i+b_i\sin\theta_0}$.

\subsection{General Classification of Extremal Surfaces in Poincare AdS\texorpdfstring{$_3$}{3}}
\label{ClassificationSectionRT}
In our setup, RT surfaces are conic sections in the $xz$-plane. In the special case where the endpoints lie on the same constant time-slice, they are half circles. However, more generally RT surfaces have a slope in time direction $k \neq 0$, c.f. Eqs.~\eqref{GenSolPoincareAdS3RtSurf} and \eqref{Z1RT}, such that the projection on the $xz$-plane is one of the three conic sections (ellipse, parabola, or hyperbola). Which of the three sections is realized depends on the value of $k$,
\begin{center}
\begin{tabular}{ r||c|c|c|c }
$|k|$ & $k=0$ & $0<|k|<1$ & $|k|=1$ & $|k|>1$ \\
\hline
$z(x)$ & circle & ellipse & parabola & hyperbola \\
\end{tabular}
\end{center}

In usual AdS/CFT, i.e.\ when the spacetime boundary is at asymptotic infinity, one only considers RT curves that have $|k|< 1$ (i.e. circles and ellipses), since all the extremal surfaces have to end on the $z=0$ boundary. RT surfaces with $|k| \geq 1$ may start on the asymptotic boundary at $z = 0$ but never return.\footnote{At least in the case where we restrict ourselves to a single Poincare patch as we will be doing in this paper.} However, in the presence of a cutoff, such as the brane under consideration, we can now attach one endpoint of the extremal surface to the brane for which generally $z>0$. This allows for novel configurations where the RT curve has $|k|\geq 1$. 

\begin{figure}
\centering
\includegraphics[width=120mm]{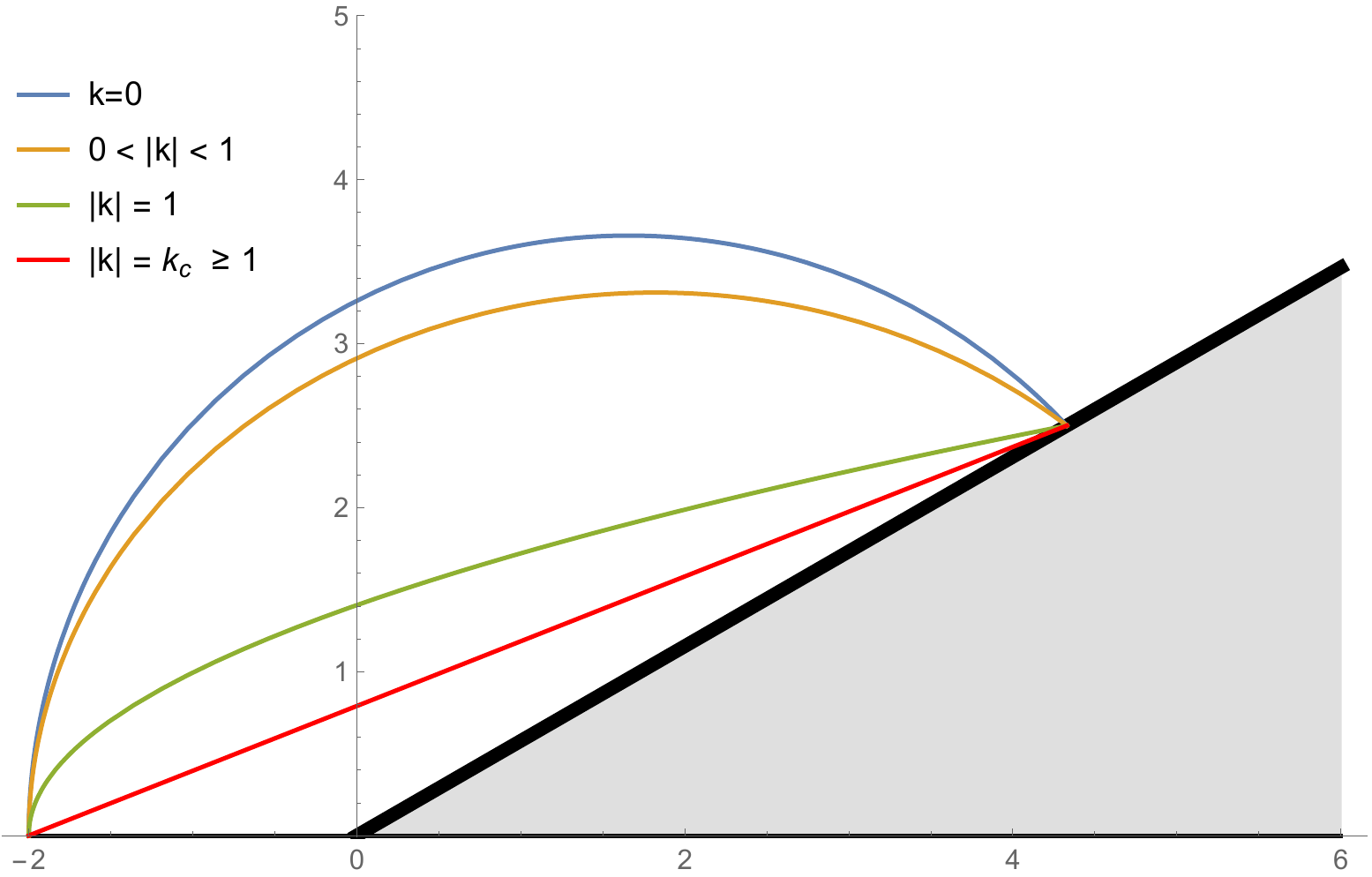}
    \caption{ 
    This plot shows extremal surfaces for various values of $|k|$ with $a_1=2$, $b_1=5$, $\theta_0=\pi/3$.  In particular, in our example we have $k_c \approx 1.075 > 1$ which can be computed from the formula given in Eq. (\ref{kcDefinitionMaintext}). The red line corresponds to setting $|k|=k_c>1$ which produces a straight line in the projection to the $xz$-plane.
    \label{RTVaryK}}
\end{figure}

Nonetheless, for $|k|$ above a certain threshold, the extremal surface becomes timelike.
Here, we consider only spacelike separated points with respect to the bulk metric, as expected in computations of entanglement entropy for spacelike separated subregions.\footnote{For work on timelike entanglement entropy see \cite{Doi:2022iyj,Liu:2022ugc,Doi:2023zaf,Li:2022tsv,Narayan:2022afv,Narayan:2023zen}.} In this case we have that $|\Delta t| <\sqrt{\Delta x^2 + \Delta z^2}$, where $\Delta x$ and $\Delta z$ are the coordinate distances between the RT endpoints. This places an upper bound on the parameter $|k| < k_c$, with $k_c$ given by
\begin{align}
\label{kcDefinitionMaintext}
    k_c = \sqrt{1 + \frac{\Delta z^2}{\Delta x^2}} \leq k_\text{max} = \frac{1}{\sin \theta_0}.
\end{align}
The value $k_\text{max}$ is special to our model and arises from maximizing $\frac{\Delta z^2}{\Delta x^2}$ over all boundary and brane points. Restricting the RT surface to end on the asymptotic boundary of course reproduces the situation in standard AdS/CFT with $|k| < 1$. In the limit where $|k|\to k_{c}$ the line segment becomes a null geodesic (i.e. a straight line in the $xz$-plane). We depict several configurations in Figure \ref{RTVaryK}.

Before continuing, it is important to note that in the definitions of $\chi_{\text{dis}}(A)$, $\chi_{\text{dis}}(B)$, and $\chi_{\text{con}}(A\cup B)$, Eqs.~\eqref{chidisAexp} -- \eqref{Z1RT}, we have restricted the range of the parameter $x$ along the curves so that they would lie within $\mathcal{M}_{\text{phys}}$. For our future discussions it will also be useful to define the extension of these extremal curves so they are not only defined in $\mathcal{M}_{\text{phys}}$ but also in AdS$_3$ which includes the region ``behind'' the brane, shaded in gray in Figs.~\ref{EntangSurfPoincarePlot} and \ref{RTVaryK}. This is achieved by simply lifting the restrictions on the parameter $x$ and continuing the expressions into the region behind the brane. We will distinguish these extended extremal curves with a bar on top of the $\chi$. For example, the extension of $\chi_{\text{dis}}(B)$ is given by $\bar{\chi}_{\text{dis}}(B)$ and it is straightforward to see in this definition that $\bar{\chi}_{\text{dis}}(B)\cap \mathcal{M}_{\text{phys}}=\chi_{\text{dis}}(B)$. 

Fig. \ref{EntangSurfPoincarePlot} shows an example of the boundary and brane subregions along with the associated extremal surfaces in the $xz$-plane. The shaded gray region is the portion of AdS that is cut off by the ETW brane and is not physically relevant in our discussion. The blue curve $\chi_{\text{dis}}(A)$ and the yellow curve $\chi_{\text{dis}}(B)$ are half-circles. The union of the green line segments form the connected surface, $\chi_{\text{con}}(A\cup B)$. In this particular case the two line green line segments correspond to  ellipses in the $xz$-plane.  

\subsection{Constructing the Entanglement Wedges}
\label{EntanglementWedgeDefGenericSec}
Once the candidate RT surfaces are known, we can construct the associated entanglement wedges.
In standard holography, the entanglement wedge is defined by taking a boundary subregion $U$ together with its associated RT surface $\chi(U)$. One then considers a partial Cauchy surface, $\Sigma_{U}$, whose boundary is given by $\partial \Sigma_{U}=U\cup \chi(U)$. The entanglement wedge of the boundary subregion $U$, denoted $\mathcal{W}_E(U)$, is then given by the domain of dependence of the partial Cauchy slice $\Sigma_{U}$.
It is tempting to suggest that to construct the entanglement wedge $\mathcal{W}_E(U)$ to an associated boundary region $U$ one can simply define an extended entanglement wedge using $\bar{\chi}(U)$ and then cut off the extended wedge with brane and identify the remaining portion in $\mathcal{M}_\text{phys}$ as $\mathcal{W}_E(U)$. As we shall see in our discussions, such a prescription is generally not valid in cases where our boundary subregions lie on cutoff surfaces such as the brane.

\begin{figure}[t]
\centering
\includegraphics[width=70mm]{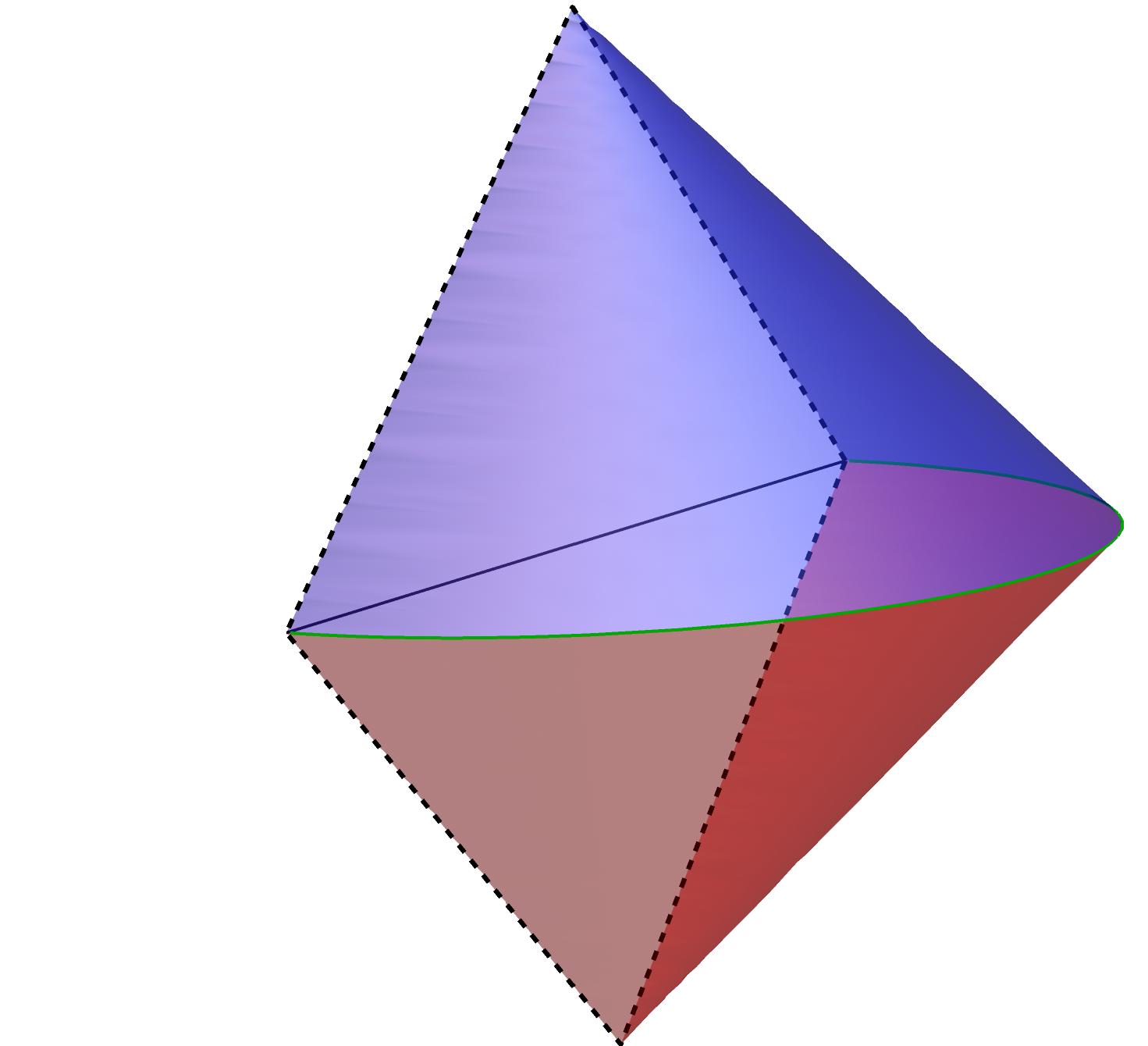}
    \caption{ An example of $\mathcal{W}_E(A)$. The intersection of the past cone and future cone is the RT surface $\chi_{\text{dis}}(A)$ shown in green. The dotted ``diamond'' is the causal diamond on the boundary associated with the interval $A$ which horizontally splits the diamond to the future and past of the interval.   \label{EntangWedgeBdry}}
\end{figure}

The explicit construction of the entanglement wedge of an interval on the conformal boundary can be understood by introducing a null congruence orthogonal to the RT surface, which is
directed towards the past and future towards the boundary. In the case of a simple geometry like Poincare AdS$_3$ it is straightforward to show that the null sheets generated by this family of null geodesics are light cones whose coordinate description is very simple.
In particular, in our setup for the interval $A$, one finds that $\mathcal{W}_E(A)$ is given by the points $(t,x,z)$ which satisfy
\begin{equation}
    \frac{a_2-a_1}{2}-|t-t_a|>\sqrt{z^2+\left(x+\frac{a_2+a_1}{2}\right)^2},
\end{equation}
see Fig. \ref{EntangWedgeBdry} for an illustration. Now let us define $\Sigma_A$ as a partial Cauchy slice whose boundary is given by $\partial \Sigma_A=A\cup \chi_{\text{dis}}(A)$. Any future(past) oriented causal curve going through a point $p$ to the past(future) of $\Sigma_A$ within the wedge enclosed by the light cones will eventually intersect $\Sigma_A$ and for any point outside the light cones one can find a causal curve that does not go through $\Sigma_A$. This means that the region enclosed by the light cones really is the domain of dependence of $\Sigma_A$ which by definition means it is $\mathcal{W}_E(A)$.

In our setup we also have spatial boundary subregions that (at least partially) live on the brane, namely $B$ and $A\cup B$. We already know that the RT curves associated with $B$ and $A\cup B$ are given by $\chi_{\text{dis}}(B)$ and $\chi_{\text{con}}(A\cup B)$ respectively. We define the entanglement wedge of the subregion $B$ denoted, $\mathcal{W}_E(B)$, as the domain of dependence of the partial Cauchy slice $\Sigma_B$ whose boundary is given by $\partial \Sigma_B=B\cup \chi_{\text{dis}}(B)$. In a similar manner we define the entanglement wedge of subregion $A\cup B$, denoted $\mathcal{W}_E(A\cup B)$, as the domain of dependence of the partial Cauchy slice $\Sigma_{A\cup B}$ whose boundary is given by $\partial \Sigma_{A\cup B}=A\cup B\cup \chi_{\text{con}}(A\cup B)=A\cup B\cup \chi_1\cup \chi_2$, where we remind the reader that $\chi_1$ and $\chi_2$ are the extremal surfaces which connect the boundaries of $A$ and $B$.

The challenge now is to characterize the set of points in $\mathcal{W}_E(B)$ and $\mathcal{W}_{E}(A\cup B)$. In the case of the boundary interval $A$ we can find $\mathcal{W}_E(A)$ by simply considering the set of spacelike separated points from $\chi_{\text{dis}}(A)$ towards the boundary and this naturally gives rise to the light cones bounding the entanglement wedge. In a similar manner $\mathcal{W}_E(B)$ is the set of spacelike separated points from $\chi_{\text{dis}}(B)$ going towards the brane. 

As for $\mathcal{W}_E(A\cup B)$ we will consider the region which is given by the intersection of points that are spacelike separated from $\chi_1$ towards $\chi_2$ with the points that are spacelike separated from $\chi_2$ towards $\chi_1$ in $\mathcal{M}_{\text{phys}}$. This will trace out a tube-like region through the bulk which will connect interval $A$ to interval $B$. In standard AdS/CFT this would give the domain of dependence of the partial Cauchy slice $\Sigma_{A\cup B}$ in the connected phase. As we will see in the next section, the situation is more complicated in the presence of a cutoff.

To explicitly characterize the spacelike regions of interest in $\mathcal{M}_{\text{phys}}$ we need to identify spacelike separated points from curve segments in AdS$_3$. Understanding the subtleties of doing this will be the primary aim of the discussions in Subsection \ref{SubSecSLPointsinAdS3}. We will then explicitly construct $\mathcal{W}_E(B)$ in Subsection \ref{ConstructWEBSec} and $\mathcal{W}_E(A\cup B)$ in Subsection \ref{ConstructWEAUBSec}. Note that much of the explicit construction that is discussed in this subsection is not required to formulate the conditions for EWN in Section \ref{ChapterDerivingConstraintForEWN}. Here we present the construction for the sake of completeness of the discussion of the various entanglement wedges that are involved in our setup and to illustrate subtleties in defining entanglement wedges in the connected phase.

\subsubsection{Spacelike Separated Points from RT Curves in Poincare AdS\texorpdfstring{$_3$}{3}}
\label{SubSecSLPointsinAdS3}
To understand the entanglement wedge of various brane-boundary subregions it is essential to characterize which points in the bulk are spacelike and timelike separated from the RT surfaces. We can identify the regions for extremal line segments through the methods discussed in Appendix \ref{SpacelikePointsFromCurve}. We summarize the procedure in the following. 

Suppose we have an extremal line segment, $\chi$, parameterized in terms of the Poincare coordinate $x\in[x_L,x_R]$ (i.e. we have a curve $z=z_{RT}(x)$ and $t=t_{RT}(x)$). At the endpoints of the line segment we introduce planes $\sigma_{L,R}$ that are orthogonal to the tangent vectors of the line segment,
\begin{equation}
\begin{split}
    &\sigma_{L,R}=\{(t,x,z)| 0=-\dot{t}_{RT}(x_{L,R})(t-t_{RT}(x_{L,R}))+(x-x_{L,R})+\dot{z}_{RT}(x_{L,R})(z-z_{RT}(x_{L,R}))\},\\
\end{split}
\end{equation}
where $\dot{t}_{RT}(x_{L,R})=\frac{dt_{RT}}{dx} \big\vert_{x=x_{L,R}}$ and $\dot{z}_{RT}(x_{L,R})=\frac{dz_{RT}}{dx} \big\vert_{x=x_{L,R}}$. The subscripts $L(R)$ indicate that the planes correspond to the endpoints at smaller $x$ or bigger $x$ which are left and right in our figures. As we work through explicit examples these regions will become more clear. The normalized unit tangent vector to the left and right endpoints, which we denote as $\mathcal{T}_L$ and $\mathcal{T}_R$ respectively, take the form
\begin{equation}
\label{TechnicalDefTangentVectors}
    \begin{split}
        &\mathcal{T}_{L,R}=\mathcal{N}\left[\dot{t}(x_{L,R})\partial_t+\partial_x+\dot{z}(x_{L,R})\partial_z\right], \qquad
        \mathcal{N}=\frac{z(x_{L,R})}{L\sqrt{-\dot{t}(x_{L,R})^2+1+\dot{z}(x_{L,R})^2}}.\\
    \end{split}
\end{equation}
Since $\mathcal{T}_{L,R}$ is a unit normal vector to the plane $\sigma_{L,R}$ we can use it to characterize which ``side'' of the plane $\sigma_{L,R}$ we are on. For any point $p_0=(t_0,x_0,z_0)$ we can take a line segment orthogonal to $\sigma_{L,R}$ that connects a point on $\sigma_{L,R}$ denoted as $p_{\sigma_{L,R}}$ to $p_0$ that satisfies $p_0^\mu-p_{\sigma_{L,R}}^\mu=\lambda\mathcal{T}^\mu_{L,R}$. Based on this, we say $p_0$ is to the in the direction of $\mathcal{T}_{L,R}$ ($-\mathcal{T}_{L,R}$) from $\sigma_{L,R}$ if $\lambda>0$ ($\lambda<0$). Using this terminology, we define $\mathcal{V}_\pm(\sigma_{L,R})$ to be the set of bulk points that are in the direction of $\pm\mathcal{T}_{L,R}$ from $\sigma_{L,R}$. We define $\mathcal{R}_M=\mathcal{V}_+(\sigma_L)\cap\mathcal{V}_-(\sigma_R)$, $\mathcal{R}_{L}=\mathcal{V}_-(\sigma_L)$, and $\mathcal{R}_{R}=\mathcal{V}_+(\sigma_R)$.

In the region $\mathcal{R}_M$ the set of spacelike separated points from the extremal curve can be understood by emitting null geodesics orthogonally from each point on the curve. We will refer to the null sheet generated by emitting null geodesics orthogonally from $\chi$ the ``null evolution'' of $\chi$ denoted $\text{NullEvo}(\chi)$. The points outside the null evolution of $\chi$ are spacelike separated from $\chi$.

In the region $\mathcal{R}_{L,R}$ the set of spacelike separated points can be understood as being outside the light cones emitted by the endpoints of the extremal surface. These are the points $(t,x,z)$ that will satisfy the following inequality $(t-t_{RT}(x_{L,R}))^2<(z-z_{RT}(x_{L,R}))^2+(x-x_{L,R})^2$. 

Importantly, in general, if we have an extremal surface $\bar{\chi}$ that lives in AdS$_3$ and another extremal surface $\chi=\bar{\chi}\cap \mathcal{M}_{\text{phys}}$, the set of points in $\mathcal M_\text{phys}$ which are spacelike separated from $\chi$ will generally differ from the set of spacelike separated points from $\bar{\chi}$ due to the finite extent of the extremal surfaces. In certain cases it is possible to correctly identify the entanglement wedge by naively defining a larger ``entanglement wedge'' using $\bar{\chi}$ and then simply take the portion of the larger wedge which lies in $\mathcal{M}_{\text{phys}}$. However in general such a naive approach can fail. 

\begin{comment}
    In particular, one could naively extended all extremal surfaces behind the brane, $\chi \to \bar{\chi}$, and define spacelike separated points to $\bar{\chi}$ in the physical spacetime $\mathcal M_\text{phys}$. In general, one will find that the set of spacelike separated points to $\bar{\chi}$ in $\mathcal{M}_{\text{phys}}$ can be different from the set of spacelike separated points in $\mathcal{M}_{\text{phys}}$ to $\chi$ itself.
\end{comment} 
As we will see in Section \ref{ConstructWEBSec}, our approach of using $\chi$ and this naive approach which uses $\bar{\chi}$ will correctly identify $\mathcal{W}_E(B)$. However, when constructing $\mathcal{W}_E(A\cup B)$ in the connected phase in Section \ref{ConstructWEAUBSec} the naive approach will not correctly give the entanglement wedge. It is also worth mentioning that if one defines an extended wedge $\bar{\mathcal{W}}$ in the extended spacetime using $\bar{\chi}$ we know by construction that null sheet enclosing $\bar{\mathcal{W}}$ will always fully enclose a co-dimension 0 region between $\bar{\chi}\cap\mathcal{M}_{\text{phys}}$ and the brane. However, it is possible for portions of the null sheet in front of the brane to be generated by geodesics that originate from $\bar{\chi}$ behind the brane. Our prescription relies only on null rays emitted from $\bar{\chi}\cap\mathcal{M}_{\text{phys}}$. In fact, we can say that the entanglement wedge can be correctly identified using the naive prescription iff the null sheet which encloses $\bar{\mathcal{W}}\cap\mathcal{M}_{\text{phys}}$ can be completely generated by null congruences emitted from $\bar{\chi}\cap\mathcal{M}_{\text{phys}}$. Such is the case for $\mathcal{W}_E(B)$ but not for $\mathcal{W}_E(A\cup B)$.

\subsubsection{Construction of \texorpdfstring{$\mathcal{W}_E(B)$}{W(B)}}
\label{ConstructWEBSec}
By definition $\mathcal{W}_E(B)$ is the set of points that are spacelike separated from $\chi_{\text{dis}}(B)$ towards the brane. In our case, it is relatively straightforward to explicitly construct $\mathcal{W}_E(B)$. We can define the regions $\mathcal{R}_{L,M,R}$ by introducing the planes $\sigma_{L}$ given by, 
\begin{equation}
    \label{LeftBraneLine}
        z_{\sigma_L}(t,x)=-\frac{b_1\sin(2\theta_0)x}{b_2+b_1\cos(2\theta_0)}+\frac{b_1(b_1+b_2)\cos\theta_0}{b_2+b_1\cos(2\theta_0)},
    \end{equation}
    and $\sigma_R$ given by,
  \begin{equation}
    \label{RightBraneLine}
        z_{\sigma_R}(t,x)=-\frac{b_2\sin(2\theta_0)x}{b_1+b_2\cos(2\theta_0)}+\frac{b_2(b_1+b_2)\cos\theta_0}{b_1+b_2\cos(2\theta_0)}.
    \end{equation}
Making use of the notation discussed immediately after Eq. (\ref{TechnicalDefTangentVectors}) we can define and describe $\mathcal{R}_{L,M,R}$. In particular, $\mathcal{R}_M=\mathcal{V}_+(\sigma_L)\cap \mathcal{V}_-(\sigma_R)$ can be intuitively understood as the region wedged between $\sigma_R$ and $\sigma_L$. The region $\mathcal{R}_{L}=\mathcal{V}_-(\sigma_L)$ can be understood as being between the planes $x=-\infty$ and $\sigma_{L}$ (i.e. to the ``left'' of $\mathcal{R}_M$). In a similar manner $\mathcal{R}_R=\mathcal{V}_+(\sigma_R)$ is the region between the planes $x=+\infty$ and $\sigma_{R}$ (i.e. to the ``right'' of $\mathcal{R}_M$). We adopt the notation $SL_{Q}(\chi_{\text{dis}}(B))$ to denote the set of spacelike separated points to $\chi_{\text{dis}}(B)$ in the subregion $Q=\{L,M,R\}$. Using this notation we can write,\footnote{We obtain the definition of $SL_M(\chi_{\text{dis}}(B))$ from Eq. (\ref{SLTowardsBdry}) and Eq. (\ref{SLAwayBdry}) by setting $x_\pm=\frac{b_1+b_2}{2\sin\theta_0}$ and $t_{\pm}=t_b\pm R_b$. The ``$+$'' corresponds to the spacelike points towards the boundary and the ``$-$'' corresponds to the set of spacelike points expanding away from the boundary.} 
\begin{align}
        SL_L(\chi_{\text{dis}}(B))&=\Bigl\{(t,x,z)\in\mathcal{R}_L \Big| (t-t_b)^2<(x-b_1\sin\theta_0)^2+(z-b_1\cos\theta_0)^2 \Bigr\} \\
        SL_R(\chi_{\text{dis}}(B))&=\Bigl\{(t,x,z)\in\mathcal{R}_R \Big| (t-t_b)^2<(x-b_2\sin\theta_0)^2+(z-b_2\cos\theta_0)^2 \Bigr\} \\
        SL_M(\chi_{\text{dis}}(B))&=\Bigl\{(t,x,z)\in\mathcal{R}_M \Big| |t-t_b|<\pm\left(R_b-\sqrt{z^2+\left(x-\frac{b_1+b_2}{2\sin\theta_0}\right)^2}\right) \Bigr\}.
\end{align}

\begin{figure}[th]
\centering
\includegraphics[width=150mm]{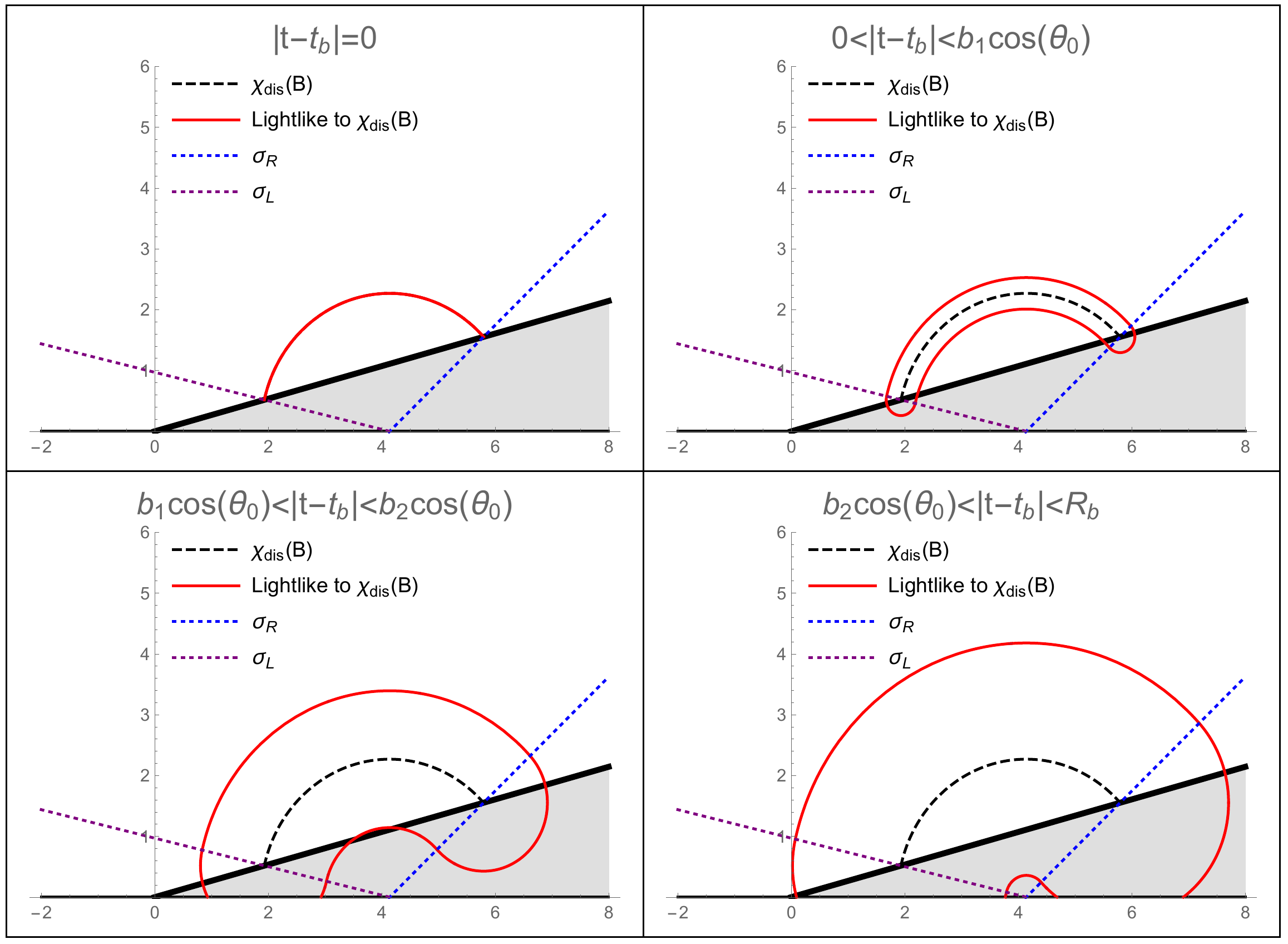}
    \caption{The figures above illustrate the set of spacelike, timelike, and null separated points from the RT curve, $\chi_{\text{dis}}(B)$, on various constant time slices. The projection of $\chi_{\text{dis}}(B)$ on each time slice is the dashed black line. The purple and blue dotted rays represent the planes $\sigma_L$ and $\sigma_R$ respectively which allow us to define the subregions $\mathcal{R}_{L,R,M}$. The solid red curve is the set of points that are lightlike separated from the RT curve. The set of points outside the region enclosed by the red curve are spacelike separated from the RT curve. For this particular plot we use the parameters: $\theta_0=\frac{5\pi}{12}$, $b_1=2$, $b_2=6$. The unshaded region is $\mathcal{M}_{\text{phys}}$. We discard everything in the shaded region behind the brane in our constructions.  \label{TimeSlicesWEB}}
\end{figure}

Due to the simplicity of out setup we can easily visualize all these regions together on various constant time slices. We depict them in Figure \ref{TimeSlicesWEB}. 
As time evolves, we can see that on each time slice a red curve encloses the RT curve line segment. This red curve is exactly the set of points that will be lightlike separated from $\chi_{\text{dis}}(B)$ at that particular time slice, and therefore the region outside the red curve is spacelike separated from $\chi_{\text{dis}}(B)$. Another interesting point to make in this case is that the planes $\sigma_L$ (the purple line in Figure \ref{TimeSlicesWEB}) and $\sigma_R$ (the blue line in Figure \ref{TimeSlicesWEB}) intersect along a line on the conformal boundary behind the brane. By construction, this line is the axis along which the tips of the future and past cones intersect along $\bar{\chi}_{\text{dis}}(B)$. In fact, we can see as we approach $|t-t_b|=R_b$ the ``shrinking'' part of the null surface will converge at the apex of the boundary cones. What will be physically relevant for our discussion is the restriction to $\mathcal{M}_{\text{phys}}$ (i.e. the unshaded portion of Figure \ref{TimeSlicesWEB} above the thick black line representing the brane). 

\begin{figure}[th]
\centering
\includegraphics[width=80mm]{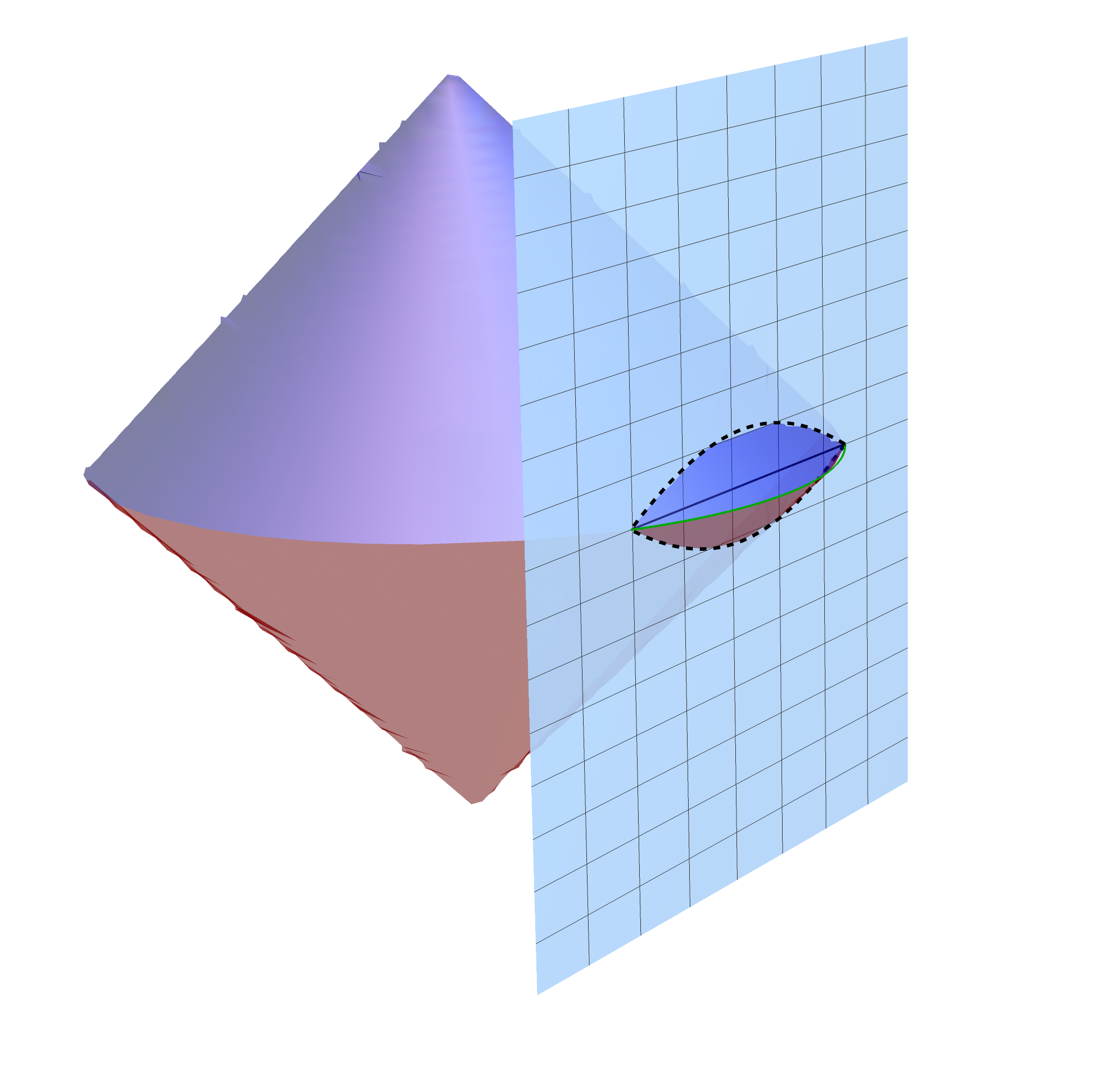}
    \caption{Depicted in the figure above is an example of $\mathcal{W}_E(B)$. There are actually two entanglement wedges in the figure. The very large wedge is what we called $\mathcal{W}_{E}(\text{Vir}(B))$ in Eq. (\ref{WEBDefVir}). The actual entanglement wedge $\mathcal{W}_E(B)$ is the small piece that is cut out of the larger wedge by the slanted plane which is the ETW brane. We can see that the intersection of $\mathcal{W}_E(B)$ with the ETW brane is not the naive causal diamond associated with the interval $B$ on the brane.  \label{EntangWedgeBrane}}
\end{figure}

The points spacelike separated from $\chi_{\text{dis}}(B)$ towards the brane define $\mathcal{W}_E(B)$. 
It is straightforward to see from Figure \ref{TimeSlicesWEB} that $\mathcal{W}_E(B)\subset \mathcal{R}_M$ and that
\begin{equation}
    \mathcal{W}_E(B)=\Biggl\{(t,x,z)\in\mathcal{M}_{\text{phys}} \Bigg|  R_b-|t-t_b|>\sqrt{z^2+\left(x-\frac{b_1+b_2}{2\sin\theta_0}\right)^2}\Biggr\}.
\end{equation}
In fact, one can understand the region $\mathcal{W}_E(B)$ as being cut from a larger entanglement wedge of a constant time slice interval that lives on the imagined, or virtual, asymptotic boundary behind the brane.
This interval can be found by analyzing where $\bar{\chi}_{\text{dis}}(B)$ intersects with the boundary this interval will be called $\text{Vir}(B)$. Using $\text{Vir}(B)$ we can express the entanglement wedge of the brane interval, $B$, as the intersection of the entanglement wedge of $\text{Vir}(B)$ with the region of $\mathcal{M}_{\text{phys}}$,
\begin{equation}
\label{WEBDefVir}
    \mathcal{W}_E(B)=\mathcal{W}_E(\text{Vir}(B))\cap \mathcal{M}_{\text{phys}}.
\end{equation}
The visualization of the points contained in the intersection is depicted in Figure \ref{EntangWedgeBrane}, where the brane cuts out a small section of the larger wedge. The smaller piece the brane cuts is precisely $\mathcal{W}_E(B)$. 

To conclude the discussion of the construction of $\mathcal{W}_E(B)$ we should emphasize that we did not use the extension $\bar{\chi}_{\text{dis}}(B)$ to define $\mathcal{W}_E(B)$. The statement $\mathcal{W}_E(B)=\mathcal{W}_E(\text{Vir}(B))\cap \mathcal{M}_{\text{phys}}$ followed from our prescription. So in this special case, the naive approach of defining the wedge by cutting a piece out of a larger wedge and our approach match. However, we will see that when we construct the connected wedge in the next section, this nice coincidence between our approach and the naive approach fails.

\subsubsection{Construction of \texorpdfstring{$\mathcal{W}_E(A \cup B)$}{W(AB)}}
\label{ConstructWEAUBSec}
Now that we have identified the entanglement wedges associated with the boundary interval $A$ and brane interval $B$ we can consider the construction of the entanglement wedge associated with $A\cup B$. In this case there are two possible phases that can exist for the entanglement wedge of $A\cup B$. The disconnected phase, which is given by the union of the entanglement wedges discussed thus far (i.e. $\mathcal{W}_E(A\cup B)=\mathcal{W}_E(A)\cup \mathcal{W}_E(B)$) and the connected phase. In the connected phase it is no longer true that the entanglement wedge is the union of the disjoint wedges(i.e. $\mathcal{W}_E(A\cup B)\neq \mathcal{W}_E(A)\cup \mathcal{W}_E(B)$). To determine $\mathcal{W}_E(A\cup B)$ in the connected phase we will adopt the procedure of finding the entanglement wedge through the prescription of finding the spacelike region away from the boundary from $\chi_1$ and the spacelike region toward the boundary for $\chi_2$ and then taking the intersection of those regions and restricting to $\mathcal{M}_{\text{phys}}$. Roughly speaking, the region we will define will look like a ``tube'' that connects $A$ and $B$ regions on the brane-boundary system. 

For each line segment we must begin by computing the relevant orthogonal planes. To begin, we note that both line segments $\chi_1$ and $\chi_2$ have their left endpoint on the conformal boundary so the planes for these points coincide for the conformal boundary and the region $\mathcal{R}_L=\varnothing$.\footnote{As we showed in Appendix \ref{SpacelikePointsFromCurve} the plane ``$\sigma_L$'' coincides with the boundary which implies that $\mathcal{R}_L=\varnothing$.} We only need to worry about $\sigma_R$ for $\chi_{1,2}$.

The orthogonal plane to $\chi_i$ at the point where $\chi_i$ intersects the brane is denoted $\sigma_{R,i}$, where $i=1,2$. The points on the plane satisfy
\begin{equation}
\label{OrthoGPlanechi1Br}
\begin{split}
    &0=-\frac{\Delta t}{\Delta x_i}(t-t_b)+(x-b_i\sin\theta_0)+\left(\frac{-\Delta x_i^2+\Delta z_i^2+\Delta t^2}{2\Delta z_i\Delta x_i}\right)(z-b_i\cos\theta_0)\\
    &\Rightarrow z(x,t)=\frac{2\Delta z_i}{\Delta t^2-\Delta x_i^2+\Delta z_i^2}\left(-x\Delta x_i+t\Delta t\right)+\frac{\Delta z_i\left[b_i^2-a_i^2+t_a^2-t_b^2\right]}{\Delta t^2-\Delta x_i^2+\Delta z_i^2},\\
\end{split}
\end{equation}
where $\Delta t=t_b-t_a$, $\Delta x_i=a_i+b_i\sin\theta_0$, and $\Delta z_i=b_i\cos\theta_0$. Using $\sigma_{R,i}$ we can define $\mathcal{R}_{M,i}$ and $\mathcal{R}_{R,i}$ using the notation/conventions introduced in Eq. (\ref{TechnicalDefTangentVectors}). In this particular scenario, we will find $\mathcal{R}_{M,i}$ as the region between the planes $x=-\infty$ and $\sigma_{R,i}$ and $\mathcal{R}_{M,i}$ as the region between the planes $x=+\infty$ and $\sigma_{R,i}$. Now we can define the set of spacelike points in each region
\begin{align}
    SL_R(\chi_i)&=\Biggl\{(t,x,z)\in \mathcal{R}_{R,i} \Bigg| (t-t_b)^2<(x-b_i\sin\theta_0)^2+(z-b_i\cos\theta_0)^2 \Biggr\}, \\
    SL_M(\chi_i)&=\Biggl\{(t,x,z)\in \mathcal{R}_{M,i} \Bigg| \text{Outside the region enclosed by NullEvo}(\chi_i)\Biggr\}.
\end{align}
We can then express the set of spacelike separated points in AdS$_3$ from $\chi_i$ as $SL_{AdS_3}(\chi_i)=SL_R(\chi_i)\cup SL_M(\chi_i)$. 

\begin{figure}[th!]
\centering
\includegraphics[width=150mm]{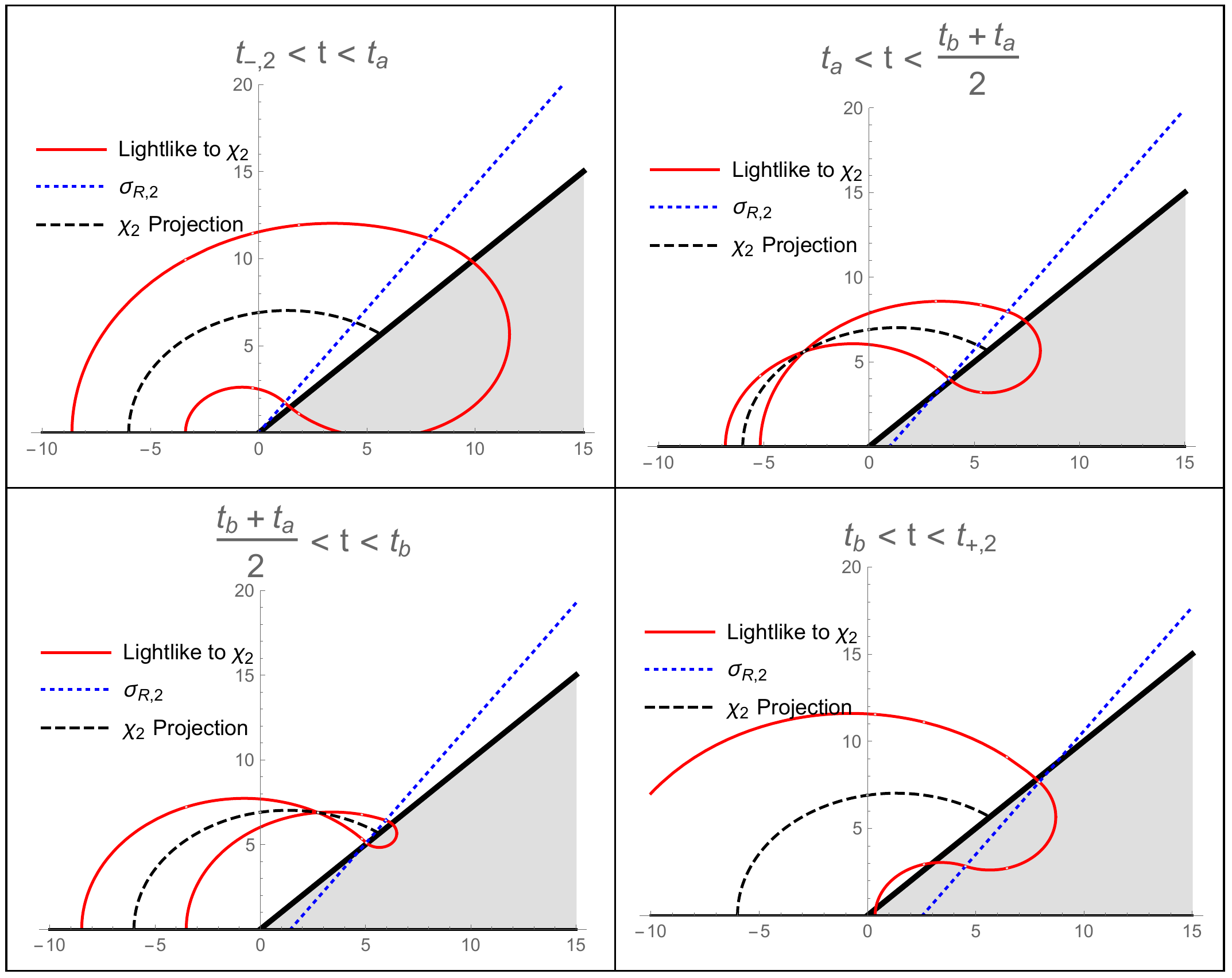}
    \caption{The figures above illustrate the set of spacelike, timelike, and null separated points from the RT curve, $\chi_2$, on various constant time slices. The brane is the thick black line and the projection of $\chi_2$ on each time slice is the dashed black line. The blue dotted ray represents the time slice of plane, $\sigma_{R,2}$ respectively which allow us to define the subregions $\mathcal{R}_{R,M}$. The solid red curve is the set of points that are lightlike separated from RT curve. The set of points outside the region which is enclosed by the red curve are spacelike separated from the RT curve. The restriction to $\mathcal{M}_{\text{phys}}$ can be obtained by simply ignoring everything behind the brane which is shaded. For this particular plot we use the parameters: $\theta_0=\frac{\pi}{4}$, $a_2=6$ $b_2=8$,  $t_b-t_a=0.97(2+\sqrt{2})\approx 3.3 \Rightarrow k_2=0.97\left(1-\frac{1}{\sqrt{2}}\right)\approx0.28<1$.  \label{chi2SLTimeSlice}}
\end{figure}

In the special case where $\chi_i$ is a line segment with $|k_i|<1$ we can write $SL_M(\chi_i)$ more explicitly. This is because we can express $\text{NullEvo}(\chi_i)$ in terms of light cones whose apexes are located at $(t,x,z)=(t_{\pm,i},x_{\pm,i},0)$ where
\begin{equation}
\label{tipConechi2}
\begin{split}
    t_{\pm,i} & =\frac{(a_i\pm t_a)^2+2b_i(a_i\pm t_a)\sin\theta_0+b_i^2-t_b^2}{2\left[(t_a-t_b)\pm(a_i+b_i\sin\theta_0)\right]},\\
    x_{\pm,i} & =\frac{[t_b-t_a\pm(b_i-a_i)][a_i+b_i\pm(t_a-t_b)]}{2\left[(t_a-t_b)\pm(a_i+b_i\sin\theta_0)\right]}.
\end{split}
\end{equation}
Using the discussion in Appendix \ref{NullEvoEquationschij} along with the expressions given in Eq. (\ref{OrthoGPlanechi1Br}) and Eq. (\ref{tipConechi2}) we can write down a formula for the null sheets emitted from $\chi_i$ in the special case where $|k_i|<1$, which is given by Eq. (\ref{DefOfFullNullEvolutionchi1}). Using this we can generate Figure \ref{chi2SLTimeSlice}, which plots various time slices of the null sheets/congruences originating from the $\chi_2$ we depicted in Figure \ref{EntangSurfPoincarePlot}.

As in Figure \ref{TimeSlicesWEB}, the red lines enclose the projection of $\chi_2$ (given by the dashed black line), although due to the nontrivial profile of $\chi_2$ over time, the red curve that surrounds the projection of $\chi_2$ takes on a more complicated shape. The blue dotted line is a time slice of the plane, $\sigma_{R,2}$, whose position changes over time, as we discussed. To the left of the dotted blue line the red line is obtained by the null evolution of $\chi_2$, while to the right of the dotted blue line the red line is generated by a light cone centered at the point where $\chi_2$ ends on the brane. 
We can see how the red line segments are continuously glued along the plane in every time slice, as they should be through our construction. 

In Figure \ref{chi2SLTimeSlice}, we consider 4 time regimes. The regime where $t_{-,2}<t<t_a$ the time slice is below $\chi_2$ the spacelike region on that slice is given by what is outside the red curve. The next two frames occur at time scales $t_a<t<t_b$. In these regimes the constant time slice intersects with $\chi_2$ this occurs exactly at the point on the dotted black line (representing $\chi_2$) where the red lines appear to intersect along a ``lobe''. We can see that this intersection point (``lobe'') moves along the $\chi_2$ as we shift the time slice in the window $t_a<t<t_b$. The last frame depicts the time window when $t_b<t<t_{+,2}$. Similar plots can also be obtained for $\chi_1$ as well, but they contain the same features, so we will not explicitly include them here.

It is worth noting that dotted blue line also allows us to identify the differences in our prescription for defining entanglement wedges and the naive prescription. In the naive prescription one would extend the RT surface behind the brane and compute the set of spacelike points to that extended curve and then restrict that to $\mathcal{M}_{\text{phys}}$. On a given time slice the spacelike regions defined in the two prescriptions will match to the left of the dotted blue lines in Figure \ref{chi2SLTimeSlice}, but to the right of the dotted blue lines the spacelike regions will differ and so will the associated wedges. 

Now that we have illustrated how we can find the set of spacelike regions from $\chi_{1,2}$ we can use this to construct the $\mathcal{W}_E(A\cup B)$ in the connected phase. We start by considering $SL^+_{AdS_3}(\chi_1) \cap SL^-_{AdS_3}(\chi_2)$ where $ SL^+_{AdS_3}(\chi_1)$ is the set of spacelike points from $\chi_1$ expanding ``outward'' towards $\chi_2$ and $ SL^-_{AdS_3}(\chi_2)$ is the set of spacelike points from $\chi_2$ expanding ``inward'' towards $\chi_1$. The intersection of the regions in $\mathcal{M}_{\text{phys}}$ will trace a smooth ``tube-shaped'' codimension 0 region enclosed by null sheets emitted from $\chi_{1,2}$ as long as $\chi_1$ and $\chi_2$ are spacelike separated from each other.\footnote{Unlike in standard holography, when we have a cutoff $\chi_1$ is not generally guaranteed to be spacelike separated from $\chi_2$ whenever $A$ and $B$ are spacelike separated. This fact will reappear in our discussions of EWN in Section \ref{SigmaABCauchyExploreSection}.} Furthermore, we can postulate the existence of a specific partial Cauchy slice $\Sigma_{A\cup B}$ with boundary $\partial \Sigma_{A\cup B}=A\cup B\cup \chi_1 \cup \chi_2$ and we can identify the domain of dependence of $\Sigma_{A\cup B}$ with the set of points $SL^+_{AdS_3}(\chi_1) \cap SL^-_{AdS_3}(\chi_2)\cap\mathcal{M}_{\text{phys}}$. For this reason we identify
\begin{equation}
    \mathcal{W}_E(A\cup B)=SL^+_{AdS_3}(\chi_1) \cap SL^-_{AdS_3}(\chi_2)\cap\mathcal{M}_{\text{phys}}.
\end{equation} 

\begin{figure}
\centering
\includegraphics[width=150mm]{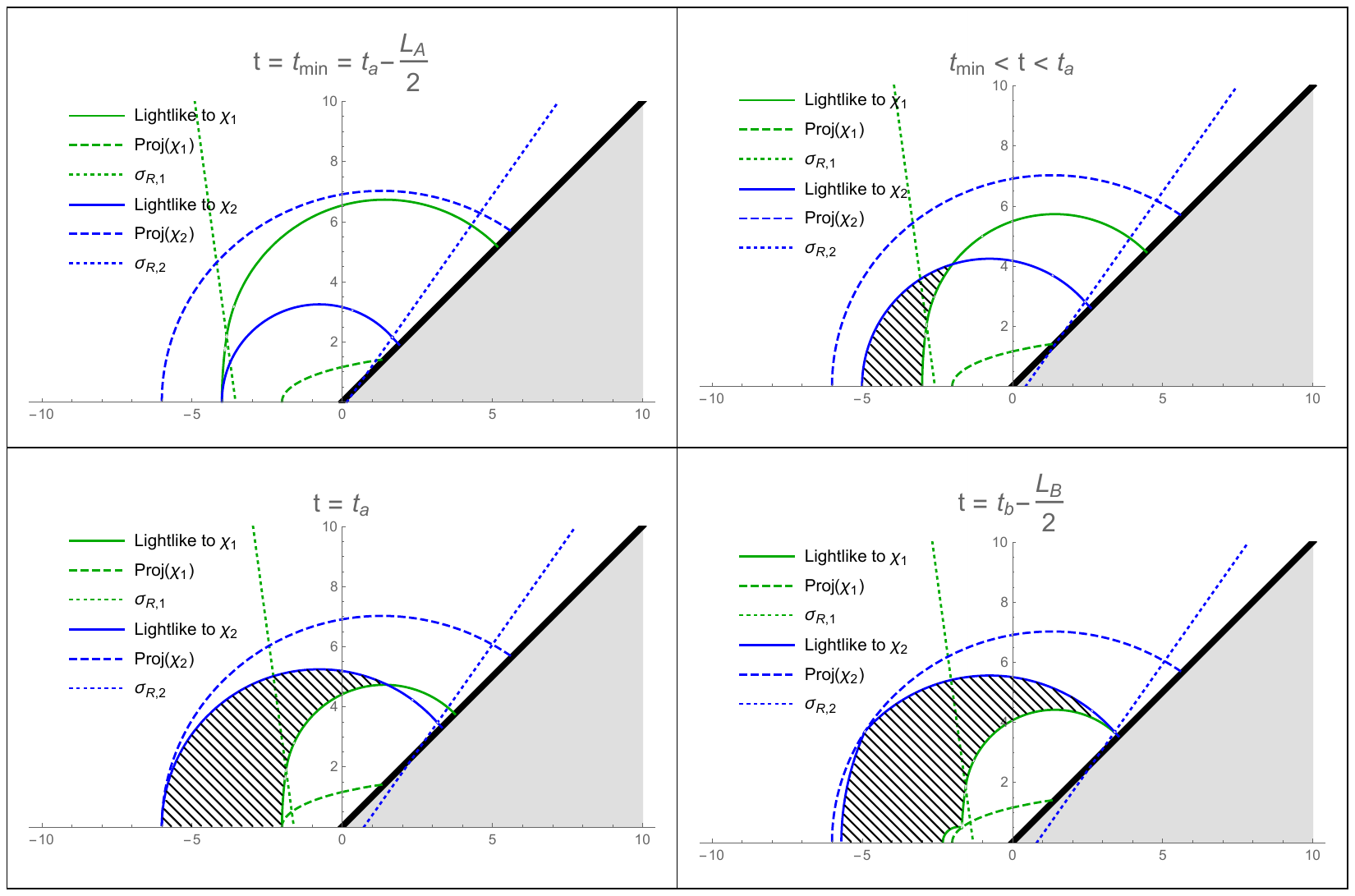}
\caption{Constant time slices of $\mathcal{W}_E(A\cup B)$ in the connected phase. With the following hierarchy of time scales $t_\text{min}=t_a-\frac{L_A}{2}<t_a<t_b-\frac{L_B}{2}<t_a+\frac{L_A}{2}<t_b<t_{\text{max}}\approx t_b+\frac{L_B}{2}-0.305$ ($L_A=a_2-a_1$, $L_B=b_2-b_1$). The actual region that is part of $\mathcal{W}_E(A\cup B)$ is shaded with lines. Green colored lines are relevant for $\chi_1$ and the blue colored lines are relevant for $\chi_2$. Dashed lines are projections of the RT curve line segments $\chi_{1,2}$. Dotted lines represent the planes $\sigma_{R,1,2}$. Solid lines represent lightlike separated points from the RT curve line segments (spacelike points to $\chi_2$ are below solid blue line and spacelike points to $\chi_1$ are above the solid green line). Specific parameters for the plot are: $\theta_0=\frac{\pi}{4}$, $a_1=b_1=2$, $a_2=6$, $b_2=8$, and $t_b-t_a=0.97(2+\sqrt{2})$. See the next set of slices in Figure \ref{AUBEntWegTimeslicesSet2}.   \label{AUBEntWegTimeslicesSet1}}
\end{figure}

\begin{figure}
\centering
\includegraphics[width=150mm]{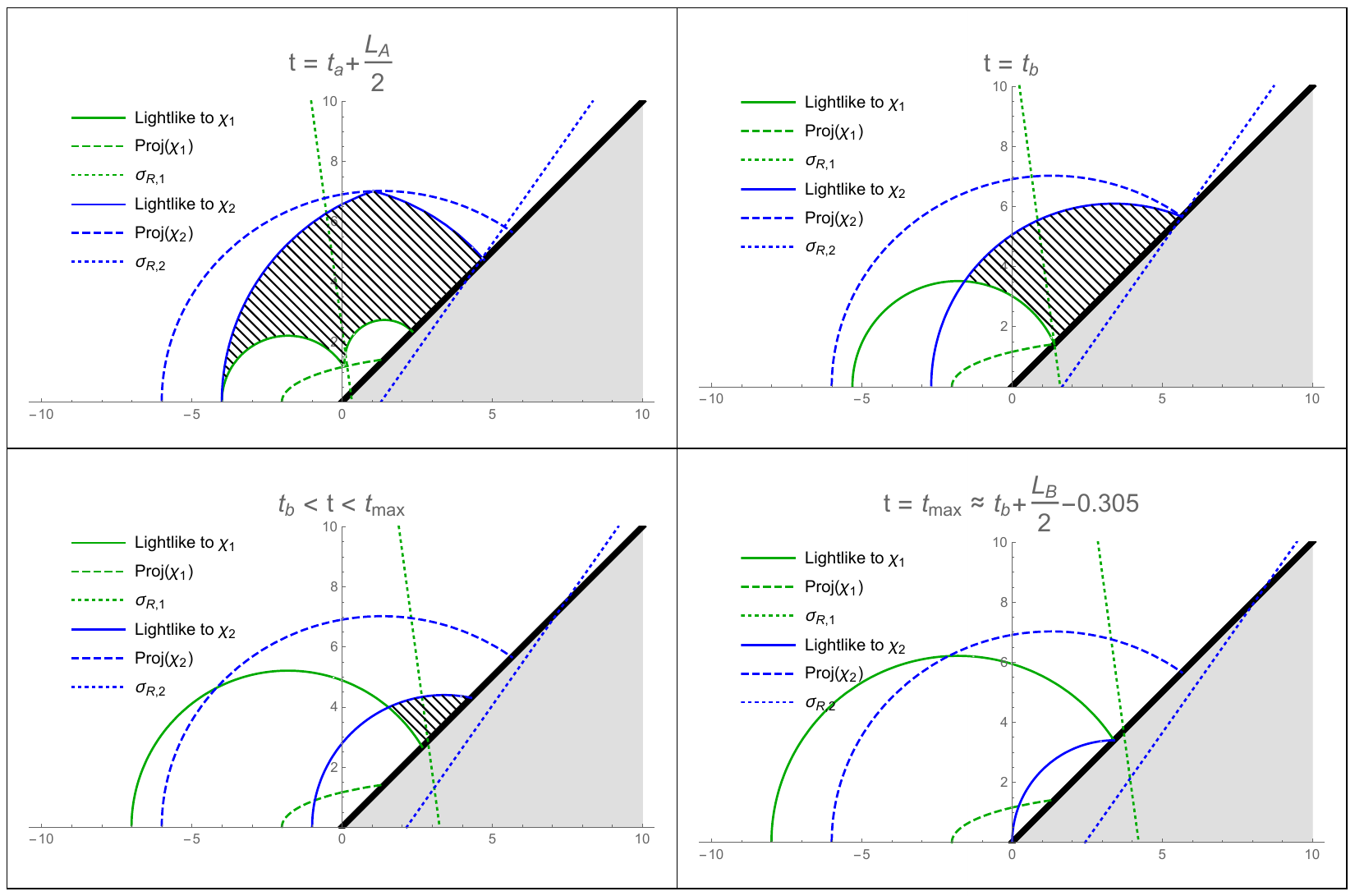}
\caption{Next set of time slices after the slices shown in Figure \ref{AUBEntWegTimeslicesSet1}.    \label{AUBEntWegTimeslicesSet2}}
\end{figure}

In Figures \ref{AUBEntWegTimeslicesSet1} and \ref{AUBEntWegTimeslicesSet2}, we consider an example to illustrate what the wedge will look like on various constant time slices. For each frame (constant time step) the brane is the thick black line. The remaining curves are color coded. Curves related to the line segment $\chi_1$ are green, and curves related to the line segment $\chi_2$ are blue. The dotted lines are the constant time slice of the planes which define the regions $\mathcal{R}_{R,M}$ for each line segment. The dashed lines are the projections of $\chi_{1,2}$ on the constant time slices. The projection of the partial Cauchy slice $\Sigma_{A\cup B}$ is contained between the dashed lines. The solid green line is the set of lightlike points from $\chi_1$ evolving towards $\chi_2$ (spacelike points to $\chi_1$ are above the solid green line) and the solid blue line is the set of lightlike points from $\chi_2$ evolving towards $\chi_1$ (spacelike points to $\chi_2$ are below the solid blue line). On each time slice we line shade the region that is spacelike to both $\chi_1$ and $\chi_2$ and represents a constant time slice of $\mathcal{W}_E(A\cup B)$.

To get intuition on what the various frames in Figures \ref{AUBEntWegTimeslicesSet1} and \ref{AUBEntWegTimeslicesSet2} represent it is useful to remember the basic ideas behind the construction of $\mathcal{W}_E(A\cup B)$. We take $\chi_1$ and emit null sheets outward towards $\chi_2$ and we consider another set of null sheets that are emitted from $\chi_2$ towards $\chi_1$. The sheets emitted towards the ``past'' will intersect along a curve which we will call $\mathcal{C}_{-}$ and the sheets emitted towards the future will intersect along another curve which we will denote $\mathcal{C}_+$. The region enclosed by these null sheets is precisely $\mathcal{W}_E(A\cup B)$. and it will trace out a ``tube.'' 

At time $t_{\text{min}}=t_a-\frac{a_2-a_1}{2}$ (Figure \ref{AUBEntWegTimeslicesSet1}, top left frame) we see that there is no shaded region, this is because that time slice only contains the past tip of the causal diamond of $A$ on the conformal boundary which is represented by the intersection of the green and blue solid lines at the conformal boundary. In fact, this past apex is precisely where $\mathcal{C}_-$ intersects with the conformal boundary. 

In the next time frame (Figure \ref{AUBEntWegTimeslicesSet1}, top right frame) we consider time scales in the window $t_{\text{min}}<t<t_a$, we will begin to obtain slices of the connected wedge indicated by the line-shaded region. The intersection of the solid green and blue lines which pinch off the line-shaded region is the ``past apex'' of the entanglement wedge tube and will lie on the line $\mathcal{C}_-$.  

At $t=t_a$ (Figure \ref{AUBEntWegTimeslicesSet1}, bottom left frame) we can see that the connected wedge on the conformal boundary spans the entire subregion $A$ we can also see that the line-shaded region describing the connected wedge grows and the ``past apex'' (indicated by the intersection between green and blue solid lines) moves further along the curve $\mathcal{C}_-$ towards the brane.

At time $t=t_b-\frac{b_2-b_1}{2}$ (Figure \ref{AUBEntWegTimeslicesSet1}, bottom right frame), the line-shaded region becomes larger and develops ``kinks'' where it touches the projections of $\chi_1,\chi_2$. These kinks were also seen in the discussion of Figure \ref{chi2SLTimeSlice} and occur when the time slice intersects with $\chi_{1,2}$ and will persist at all times $t_a<t<t_b$. Another important feature to mention on this time slice is that the intersection of the solid blue and green curve now occurs on the brane, and this is where the curve $\mathcal{C}_-$ will terminate on the brane and this point represents the ``past tip of the causal diamond'' on the brane. We put ``past tip of the causal diamond'' in quotation marks because, in general, the intersection of the entanglement wedge with the brane will not be the naive causal diamond of the brane interval $B$. So for all $t>t_b-\frac{b_2-b_1}{2}$ the curve $\mathcal{C}_-$ will not appear.

At a later time, $t=t_a+\frac{a_2-a_1}{2}$ (Figure \ref{AUBEntWegTimeslicesSet2}, top left frame), we see the appearance of $\mathcal{C}_+$ at the conformal boundary which represents the future tip of the causal diamond of $A$ on the conformal boundary where the blue and green lines intersect. Similar to $\mathcal{C_-}$, the curve $\mathcal{C}_+$ will eventually terminate on the brane at some later time, which we will calculate explicitly.

At time $t=t_b$ (Figure \ref{AUBEntWegTimeslicesSet2}, top right frame), we can see that the time slice intersects with $\mathcal{C}_+$ closer to the brane indicated by the intersection of the green and blue lines, and the line-shaded region intersects with the brane along the brane interval $B$. 

As time progresses past $t=t_b$ into the regime where $t_b<t<t_{\text{max}}$ (Figure \ref{AUBEntWegTimeslicesSet2}, bottom left frame), we will see that the line-shaded region will shrink and the intersection of the green and blue lines will become closer to the brane. 

The last frame of Figure \ref{AUBEntWegTimeslicesSet2} occurs at $t=t_{\text{max}}$ which is the time where $\mathcal{C_+}$ terminates on the brane and we have fully moved through the connected wedge through time. Naively, one might guess that the time scale $t=t_{\text{max}}$ should be given by $t_b+\frac{b_2-b_1}{2}$. However, this is not the case. The reason for this can be clearly seen upon closer inspection of the final frame in Figure \ref{AUBEntWegTimeslicesSet2}. Start by noting the intersection occurs to the left of both the dotted green and blue lines. This implies the point of intersection should be understood by projecting the null evolution of $\chi_{1}$ and $\chi_{2}$ in the region $\mathcal{R}_{M,1}$  and $\mathcal{R}_{M,2}$ respectively onto the brane. Referring to Appendix \ref{NullEvoEquationschij} we can find expressions for these sheets in the appropriate regions given by $z_{\text{Null},\mathcal{R}_{M,i}}^{(\pm)}(x,t)$. We need to find intersections of $z_{\text{Null},\mathcal{R}_{M,2}}^{(-)}(x,t)$ and $z_{\text{Null},\mathcal{R}_{M,1}}^{(+)}(x,t)$. Using the fact that $t_i(x)<t_b<t$ we can write
\begin{equation}
    \begin{split}
        &z_{\text{Null},\mathcal{R}_{M,2}}^{(-)}(x,t)=\sqrt{(t-t_{+,2})^2-(x-x_{+,2})^2},\\
        &z_{\text{Null},\mathcal{R}_{M,1}}^{(+)}(x,t)=\sqrt{(t-t_{-,1})^2-(x-x_{-,1})^2}.\\
    \end{split}
\end{equation}
Setting $z_{\text{Null},\mathcal{R}_{M,2}}^{(-)}(x,t)=z_{\text{Null},\mathcal{R}_{M,1}}^{(+)}(x,t)$ will describe a curve. We know this curve has to terminate on the brane at the intersection point we are interested so restricting $z$ to the brane we can actually set $z_{\text{Null},\mathcal{R}_{M,2}}^{(-)}(x,t)=z_{\text{Null},\mathcal{R}_{M,1}}^{(+)}(x,t)=x\cot\theta_0$ which leads to a non-linear 2 by 2 system of equations. 
\begin{equation}
    \begin{split}
        (t-t_{+,2})^2 & =x^2\cot^2\theta_0+(x-x_{+,2})^2,\\
        (t-t_{-,1})^2 & =x^2\cot^2\theta_0+(x-x_{-,1})^2.\\
    \end{split}
\end{equation}
We can find solutions to these equations and for our particular case we will numerically find
\begin{equation}
    t_{\text{Br}\cap\mathcal{C}_+}\approx 2.695=\frac{b_2-b_1}{2}-0.305, 
\end{equation}
where $t_a=-0.97(2+\sqrt{2})$ and $t_b=0$ and $t_{\pm,i}$ and $x_{\pm,i}$ are explicitly given by Eq. (\ref{tipConechi2}).
It is also possible to find solutions in closed form expressions but for the sake of simplicity of the presentation we just resort to numerically determining the time scale.

\begin{figure}[h!]
\centering
\includegraphics[width=95mm]{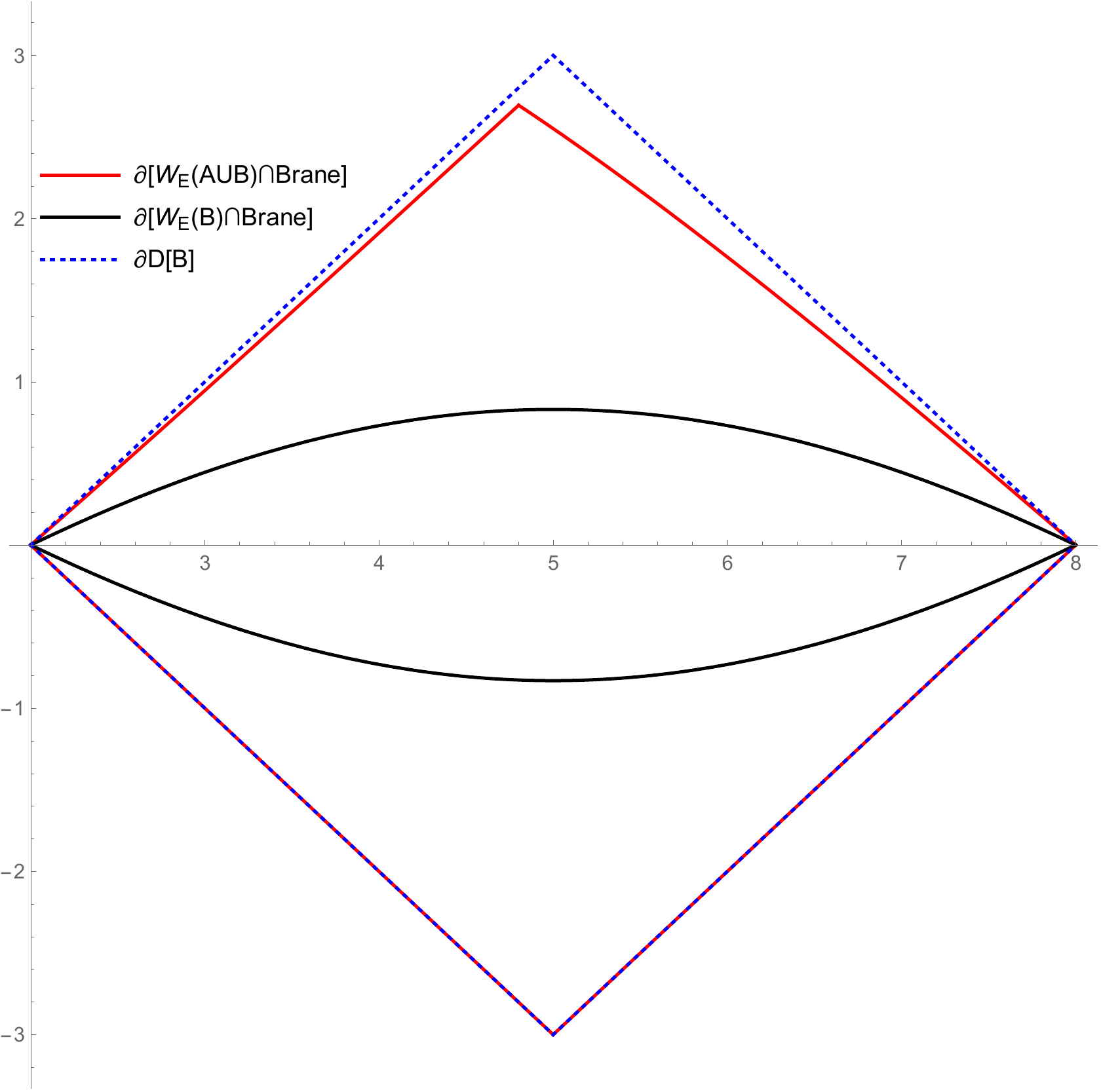}
\caption{This plot (which uses same parameters described in Figure \ref{AUBEntWegTimeslicesSet1}) shows the intersection of $\mathcal{W}_E(B)$ with the brane which is the region enclosed by the solid black line. The intersection of the connected wedge $\mathcal{W}_E(A\cup B)$ with the brane is enclosed by the solid red line. The region enclosed by the dotted blue lines is the naive causal diamond that one would associate with the interval $B$ just based on the induced metric on the brane. In general the intersection of the entanglement wedges with the brane are not the naive causal diamond, but rather a proper subset of the naive causal diamond.}
\label{WedgeIntBrane}
\end{figure}

To conclude this section we will discuss the plot in Figure \ref{WedgeIntBrane} which is the intersection of $\mathcal{W}_E(A\cup B)$ in the connected phase and $\mathcal{W}_E(B)$ with the brane for the same set of parameters considered in Figures \ref{AUBEntWegTimeslicesSet1} and \ref{AUBEntWegTimeslicesSet2}.
The intersection of $\mathcal{W}_E(B)$ with the brane is contained in the intersection of the connected wedge with the brane. This will always be the case in our construction as long as EWN is satisfied. We compare the intersection of the connected wedge with the brane with the naive domain of dependence of $B$ denoted $D(B)$. We can see that the connected wedge is contained in $D(B)$. Generally, whenever EWN is respected, we will have $\mathcal{W}_E(B)\subseteq\mathcal{W}_E(A\cup B)\subseteq D(B)$.

As we have seen, the precise analytic understanding of the region $\mathcal{W}_E(A\cup B)$ is relatively complicated compared to understanding the entanglement wedges $\mathcal{W}_E(A)$ and $\mathcal{W}_E(B)$. In the next set of Sections (Section \ref{ChapterDerivingConstraintForEWN}) we will derive constraints in terms of the parameters $a_{1,2},b_{1,2},\theta_0,|\Delta t|$ which need to be satisfied to ensure that our entanglement wedges are well defined and satisfy EWN. 

\newpage

\section{Entanglement Wedge Nesting in Poincar\'e AdS}
\label{ChapterDerivingConstraintForEWN}
In this section, we will study conditions on the relative location of constant time intervals $A$ and $B$, which are located on the asymptotic boundary and on the cutoff brane respectively, such that EWN is satisfied, that is, $\mathcal{W}_E(A), \mathcal{W}_E(B)\subseteq\mathcal{W}_E(A\cup B)$. Clearly, this condition is only non-trivial if $\mathcal{W}_E(A\cup B) \neq \mathcal{W}_E(A) \cup \mathcal{W}_E(B)$, ie., whenever we are in the connected phase. 
Assuming the latter, in Section \ref{PoincareAdS3SuffCondSec} we derive a condition, presented in Eq.~\eqref{FinalEWNCondition}, on the time separation between $A$ and $B$ such that EWN holds. 
In Section \ref {DominanceOfConnectedPhaseSection} we study when the connected phase dominates and our conditions for EWN should be applied. We find that there are nontrivial configurations in which the RT surfaces are in the connected phase and EWN fails, even though the intervals $A$ and $B$ are spacelike separated with respect to both the bulk and boundary metric. 
In Section \ref{SigmaABCauchyExploreSection}, we argue that these violations appear whenever the RT surfaces in the connected or disconnected phase are not spacelike separated. In Section \ref{DiscussionEWNConstrintChapt}, we discuss how insisting that $A$ and $B$ be spacelike separated in the bulk is generally not enough to ensure EWN will be satisfied. We connect this to the idea that the holographic theory on the cutoff surface is non-local which leads to a stronger requirement than spacelike separation of $A$ and $B$ to define independent subregions in non-local holographic theories. Finally, in Section \ref{RestMaximinSec}, we compare our prescription for defining entanglement wedges/RT surfaces in cutoff holography with the restricted maximin prescription of \cite{Grado-White:2020wlb}.
Our results demonstrate that while restricted maximin is designed such that EWN is satisfied under the condition that $A$ and $B$ are spacelike separated, the results obtained from restricted maximin disagree with a naive application of the RT formula if the ``naive'' RT surfaces corresponding to $A$ and $B$ are in causal contact. Even in the simplest cases, like the one we discuss, this can lead to RT surfaces which significantly differ from the naive expectation. 

\subsection{Condition for EWN}
\label{PoincareAdS3SuffCondSec}
We will now derive conditions under which EWN holds. 
To this end consider two subregions $A$ and $B$, located at constant times $t_a$ and $t_b$, respectively. We choose $A$ to be a subset of the asymptotic boundary, while $B$ is a subset of the cutoff surface, i.e., the brane.
Recall that EWN in our setup can only be violated when the connected phase dominates. Therefore, conditions that we derive in this section are only necessary and sufficient if the connected phase dominates. We will return to this point later in Section \ref{DominanceOfConnectedPhaseSection}.

As a warm-up, let us begin with the special case of $t_a=t_b$. In this situation the RT surfaces $\chi_{\text{dis}}(A)$, $\chi_{\text{dis}}(B)$, and $\chi_{\text{con}}(A\cup B)$ which were defined in the previous section all lie on the same constant time slice. It is clear that since the partial Cauchy slices $\Sigma_A,\Sigma_B$ are contained in $\Sigma_{A\cup B}$, the domain of dependence of $\Sigma_A \cup\Sigma_B$ is contained in the domain of dependence of $\Sigma_{A\cup B}$; in other words, $\mathcal{W}_E(A),\mathcal{W}_E(B)\subseteq \mathcal{W}_E(A\cup B)$. To see this, take any $p\in \mathcal{W}_{E}(A)$. Any causal curve going through $p$ must also go through $\Sigma_A$. But since $\Sigma_A\subset \Sigma_{A\cup B}$ all causal curves passing through $p$ also intersect $\Sigma_{A\cup B}$ Thus, $p\in \mathcal{W}_{E}(A\cup B)$ and $\mathcal{W}_E(A)$ must be a subset of $\mathcal{W}_E(A\cup B)$. The proof for $\mathcal{W}_E(B)$ follows analogously.

Now, let us take the more general case in which $t_a\neq t_b$. In this case, the strategy will be to ensure that the null congruences originating from $\chi_1$ or $\chi_2$ that enclose $\mathcal{W}_E(A \cup B)$ cannot puncture $\mathcal{W}_E(B)$ and $\mathcal{W}_E(A)$. This means that causal signals originating from $\chi_{\text{con}}(A\cup B)$ and, in particular, the complement of $\mathcal{W}_E(A \cup B)$ cannot influence events in $\mathcal{W}_E(A)$ and $\mathcal{W}_E(B)$. As long as such a condition holds, we are guaranteed $\mathcal{W}_E(A),W_{E}(B)\subseteq W_{E}(A\cup B)$. 

The condition that $\chi_i$ is unable to influence events in $\mathcal{W}_E(A)$ can be expressed as 
\begin{equation}
\label{SuffEWNA}
    \left(\frac{a_2-a_1}{2}+|t_{i}(x)-t_a|\right)^2 \leq z_{i}(x)^2+\left(x+\frac{a_2+a_1}{2}\right)^2,
\end{equation}
where $z_{i}(x)$ and $t_{i}(x)$ are given in Eq. \eqref{Z1RT}, and the index, $i=1,2$, indicates whether we consider $\chi_1$ or $\chi_2$. This condition arises from requiring that on every time-slice the spatial distance of the light congruence, which emanates from $\chi_{\text{dis}}(A)$, to the center of the region $A$ is less than the distance of the curve $\chi_i$ to the center of region $A$. As shown in Appendix \ref{AppendixNiaveConAdS3}, it turns out that among the two conditions, the most stringent one is given by analyzing the constraint for $\chi_1$, that is, the curve closer to the defect (c.f.~Fig.~\ref{EntangSurfPoincarePlot}). This condition can be rewritten as
\begin{equation}
\label{MostStringentEWNACondd}
    |\Delta t|\leq-\frac{a_2-a_1}{2}+\sqrt{b_1^2\cos^2\theta_0+\left(b_1\sin\theta_0+\frac{a_2+a_1}{2}\right)^2}\equiv \Delta t_{A,EWN}.
\end{equation}
The weaker condition we obtain from the analysis of $\chi_2$ takes the same form as Eq.~\eqref{MostStringentEWNACondd} with the replacement $b_1\to b_2$. It is straightforward to see that Eq.~\eqref{MostStringentEWNACondd} is not only a sufficient condition for EWN of $\mathcal{W}_E(A)$ in the connected phase, but also necessary, since violations of the condition necessarily lead to violations in EWN of $\mathcal{W}_E(A)$ because null congruences from $\chi_1$ will puncture $\mathcal{W}_E(A)$.

\begin{figure}
\centering
\includegraphics[width=120mm]{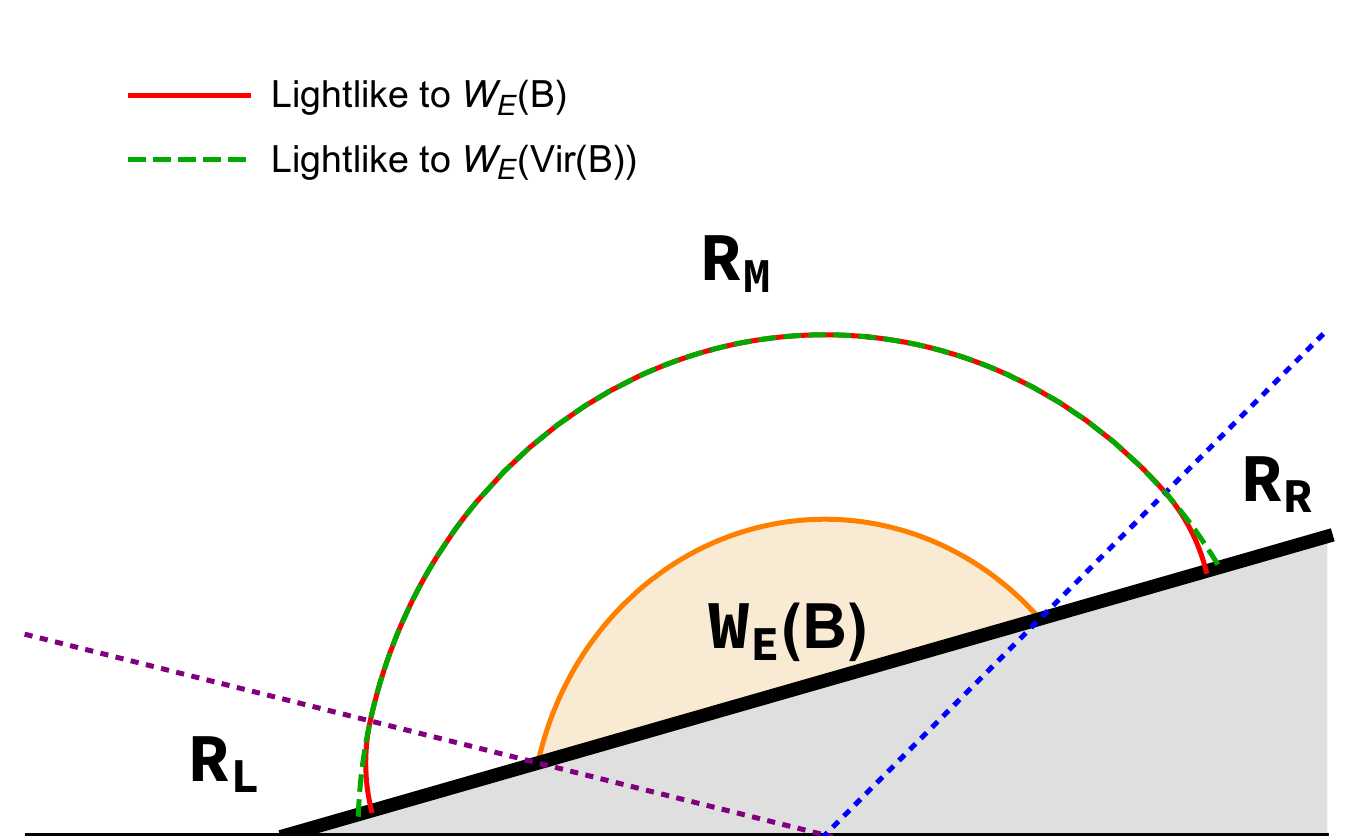}
\caption{We depict different notions of spacelike separation along some time slice $t\neq t_b$. Points outside the region enclosed by solid red are spacelike separated to $\mathcal{W}_E(B)$. Points outside the dashed green line are spacelike separated from $\mathcal{W}_E(\text{Vir}(B))$. Enforcing that $\chi_i$ satisfies Eq. (\ref{NaiveSLSeparationPoincareCond}) throughout $\mathcal{M}_{\text{phys}}$ is equivalent to being outside the region enclosed by dotted green line. In particular, being spacelike to $\mathcal{W}_E(\text{Vir}(B))$ implies being spacelike to $\mathcal{W}_E(B)$.}
\label{IllustratingSuffCondFig}
\end{figure}

Now we turn to the more complicated and involved case of deriving sufficient and necessary conditions to ensure $\mathcal{W}_E(B)\subseteq \mathcal{W}_E(A\cup B)$. To obtain sufficient conditions for EWN we require that $\chi_i$ be spacelike separated from $\mathcal{W}_E(B)$. However, following our discussion in Section \ref{ConstructWEBSec} we know that we need to analyze the condition in a piecewise manner for each region $\mathcal{R}_{L}$, $\mathcal{R}_{R}$, and $\mathcal{R}_{M}$, defined with respect to the extremal surface $\chi_{\text{dis}}(B)$. The condition that the points $(t_{i}(x),x,z_{i}(x))$ on $\chi_i$ which are spacelike separated from $\mathcal{W}_E(B)$ in $\mathcal {R}_L$ or $\mathcal {R}_R$ need to satisfy reads
\begin{equation}
    \label{RLConstraint}
        (t_{i}(x)-t_b)^2< (x-b_{j}\sin\theta_0)^2+(z_{i}(x)-b_{j}\cos\theta_0)^2,
\end{equation}
where $j=1$ if the point $(t_{i}(x), x, z_i(x))$ lies in $\mathcal{R}_L$ and $j=2$ if it lies in $\mathcal{R}_R$, while, as above, $i = 1,2$ enumerates the two RT surfaces that connect $A$ and $B$. This condition comes from requiring that $\chi_1$ and $\chi_2$ need to be spacelike separated to the entangling surface on the cutoff brane. For the points of $\chi_i$ that are located in $\mathcal{R}_{M}$, we impose the usual condition, familiar from the intervals at the asymptotic boundary, c.f., Eq.~\eqref{SuffEWNA}
\begin{equation}
\label{NaiveSLSeparationPoincareCond}
|t_{i}(x)-t_b|+R_b< \sqrt{\left(x-\frac{b_1+b_2}{2\sin\theta_0}\right)^2+z_{i}(x)^2}.
\end{equation}
Out of the conditions written in Eqs. (\ref{RLConstraint}) and (\ref{NaiveSLSeparationPoincareCond}), we claim that it is sufficient for $\chi_i$ to satisfy Eq. (\ref{NaiveSLSeparationPoincareCond}) in $\mathcal{M}_{\text{phys}}$. The reason for this is easily seen from Figure \ref{IllustratingSuffCondFig}. 
It depicts the various notions of spacelike separation to $\mathcal{W}_E(B)$ on some time slice $t\neq t_b$. The collective conditions in Eq. (\ref{RLConstraint}) and Eq. (\ref{NaiveSLSeparationPoincareCond}) applied to the relevant regions can be visually understood as requiring $\chi_i$ to be outside the region enclosed by the solid red line. Moreover, satisfying only Eq. \eqref{NaiveSLSeparationPoincareCond} in $\mathcal{M}_{\text{phys}}$ is equivalent to $\chi_i$ staying outside the region enclosed by the dashed green line. What is clear from Figure \ref{IllustratingSuffCondFig} is that a point being in the region outside the dashed green line implies that the same point is also outside the region enclosed by the solid red line. In Appendix \ref{AppendixNiaveConAdS3} we analyze the constraints in Eq. (\ref{NaiveSLSeparationPoincareCond}) for $\chi_i$ and show that it is satisfied inside $\mathcal{M}_{\text{phys}}$ whenever
\begin{equation}
\label{SuffConEWNNaivedelt}
   |\Delta t|<  a_{i}+\frac{b_1+b_2-\sqrt{b_1^2+b_2^2+2b_1b_2\cos(2\theta_0)}}{2\sin\theta_0} \equiv \Delta t_{i,EWN},
\end{equation}
where the right hand side defines the additional bounds $\Delta t_{i,EWN}$.\footnote{Among the two bounds the $i=1$ bound is more stringent. However, as we will soon show, satisfying $i=1$ bound is not necessary to satisfy EWN. Although the $i=1$ constraint is not relevant when formulating necessary and sufficient conditions for EWN it will reappear in our discussion of trying to understand non-locality in cutoff holography in Section \ref{DiscussionEWNConstrintChapt}.}

Let us now investigate whether, under which circumstances, Eq.~\eqref{SuffConEWNNaivedelt} is necessary. We begin by focusing on $\chi_1$ (that is, $i=1$). Suppose that we fix $a_{1,2},b_{1,2},\theta_0$. As we vary $\Delta t$ or equivalently $|k_1|$ (cf., Eq.~\eqref{Z1RT}) we will change the profile of the projection of $\chi_1$ onto a constant time slice without changing the endpoints. We know that since $\chi_1$ eventually intersects the boundary, a portion of $\chi_1$ always has to lie in $\mathcal{R}_L$. This just leaves us with the question of whether $\chi_1$ can lie in $\mathcal{R}_M$ and $\mathcal{R}_R$. It is useful to note that if no portion of $\chi_1$ lies in $\mathcal{R}_M$ then no portion of $\chi_1$ can lie within $\mathcal{R}_R$ either. Therefore, showing that $\chi_1\cap \mathcal{R}_M=\varnothing$ implies that $\chi_1\subset\mathcal{R}_L$. In Appendix \ref{chi1RMCondAppendix} we show that $\chi_1\cap\mathcal{R}_M=\varnothing$ whenever $|k_1|\geq k_{1,*}$, where $k_{1,*}$ is given by
 \begin{equation}
     k_{1,*}=\sqrt{1-\frac{(b_1\cos\theta_0)^2}{(a_1+b_1\sin\theta_0)^2}\left(1+\frac{4\sin\theta_0(a_1+b_1\sin\theta_0)}{b_2+b_1\cos(2\theta_0)}\right)}.
 \end{equation} 
To understand where this quantity comes from we define the point $p_1=\chi_1\cap\text{Brane}$. When $|k_1|=k_{1,*}$ one can show that the tangent line to $\chi_1$ at $p_1$ will coincide with the ray in the $xz$-plane that separates the regions $\mathcal{R}_L$ and $\mathcal{R}_M$. When $|k_1|>k_{1,*}$, one can further show that the tangent line lives in $\mathcal{R}_{L}$. This implies that within some neighborhood of $p_1$, $\chi_1\subset\mathcal{R}_L$. With some additional work detailed in Appendix \ref{chi1RMCondAppendix} we can prove that this is enough to show that $\chi_1\subset \mathcal{R}_L$ throughout $\mathcal{M}_\text{phys}$. An astute reader may notice that $k_{1,*}$ can actually become imaginary and worry about how our discussion will apply in such a scenario. We also cover such cases in Appendix \ref{chi1RMCondAppendix}. Imaginary $k_{1,*}$ simply indicates that we have chosen a combination of $a_{1,2},b_{1,2},\theta_0$ such that for every choice of $|k_1|$ the tangent to $\chi_1$ at point $p$ will always lie in $\mathcal{R}_L$ which means in such cases $\chi_1\subset \mathcal{R}_L$. So technically our condition should be $\chi_1\subset \mathcal{R}_L$ iff $|k_1|\geq \text{Re}[k_{1,*}]$. This implies that when $k_{1,*}$ is imaginary, satisfying the bound $|\Delta t|<\Delta t_{1,EWN}$ is not necessary to ensure EWN. This leaves us with the case where $k_{1,*}$ is real. To analyze this we define
 \begin{equation}
\begin{split}
    k_{1,EWN}&=\frac{\Delta t_{1,EWN}}{a_1+b_1\sin\theta_0}\\
    &=1+\frac{b_2+b_1\cos(2\theta_0)-\sqrt{\left(b_2+b_1\cos(2\theta_0)\right)^2+b_1^2\sin^2(2\theta_0)}}{2(a_1+b_1\sin\theta_0)\sin\theta_0}.\\
\end{split}
\end{equation}
Now consider the following expression,
\begin{equation}
\begin{split}
&\left[\frac{2(b_2+b_1\cos(2\theta_0))\sin^2\theta_0(a_1+b_1\sin\theta_0)^2}{b_1+b_2+2a_1\sin\theta_0}\right](k_{1,EWN}^2-k_{1,*}^2)\\
&=\sqrt{b_1^2+b_2^2+2b_1b_2\cos(2\theta_0)}\left[-(b_2+b_1\cos(2\theta_0))+\sqrt{(b_2+b_1\cos(2\theta_0))^2+b_1^2\sin^2(2\theta_0)}\right]\geq 0.
\end{split}
\end{equation}
It thus follows that $k_{1,EWN}^2-k_{1,*}^2\geq 0$ and we have shown that $k_{1,EWN}\geq k_{1,*}$. Using $k_1=\frac{\Delta t}{a_1+b_1\sin\theta_0}$ we can rewrite the condition $|\Delta t|<\Delta t_{1,EWN}$ as $|k_1|<k_{1,EWN}$. Thus, whenever $|k_1|\geq k_{1,EWN}$, it follows that $|k_1|\geq k_{1,*}$. In other words, if the sufficient condition \eqref{SuffConEWNNaivedelt} is violated, we are in a regime where it is not necessary, since the extremal surface $\chi_1$ is completely contained in $\mathcal R_L$. Instead, the sufficient and necessary condition is simply the condition in Eq.~\eqref{RLConstraint} with $i=j=1$. We show in Appendix \ref{AppendixNiaveConAdS3} that this is trivially satisfied when
\begin{align}
    |\Delta t|<\sqrt{\Delta x_1+\Delta z_1}=\sqrt{a_1^2+b_1^2+2a_1b_1\sin\theta_0} \equiv \Delta t_c, 
\end{align}
which is equivalent to saying that $A$ and $B$ are spacelike separated.

A similar analysis can also be performed for $\chi_2$, although with some differences. It is straightforward to convince oneself that there will always be portions of $\chi_2$ that live in $\mathcal{R}_M$ and $\mathcal{R}_L$ so it really only leaves us to understand under which conditions $\chi_2$ might live in $\mathcal{R}_R$. In Appendix \ref{chi1RMCondAppendix} we prove that $\chi_2\cap\mathcal{R}_{R}=\varnothing$ iff either $b_1+b_2\cos(2\theta_0)\leq 0$ or $b_1+b_2\cos(2\theta_0)>0$ and $|k_2|\geq \text{Re}[k_{2,*}]$, where 
\begin{equation}
     k_{2,*}=\sqrt{1-\frac{(b_2\cos\theta_0)^2}{(a_2+b_2\sin\theta_0)^2}\left(1+\frac{4\sin\theta_0(a_2+b_2\sin\theta_0)}{b_1+b_2\cos(2\theta_0)}\right)},
\end{equation}
is analogous to $k_{1,*}$ we defined in the analysis of $\chi_1$ with similar subtleties and interpretations we already discussed.\footnote{The reader may be wondering why there are two distinct cases here. The reason for this is that unlike the ray that separates $\mathcal{R}_L$ and $\mathcal{R}_M$, which always has a non-positive slope; the ray that separates $\mathcal{R}_R$ and $\mathcal{R}_M$ can have a positive or negative slope depending on the sign of $b_1+b_2\cos(2\theta_0)$. In the case where $b_1+b_2\cos(2\theta_0)>0$ the ray has a negative slope and we can do a similar analysis as we did for $\chi_1$ by defining $k_{2,*}$. In the case where $b_1+b_2\cos(2\theta_0)\leq 0$ the slope of the ray is non-negative and in such a case it is not possible for $\chi_2$ to be in $\mathcal{R}_R$.} For the case of $b_1+b_2\cos(2\theta_0)\leq 0$ we can see that a violation of $|\Delta t|<\Delta t_{2,EWN}$ will necessarily lead to intersections of $\mathcal{W}_E(B)$ with null congruences originating from $\chi_2$ in arbitrarily small neighborhoods of $p_2=\chi_2\cap\text{Brane}$. So in this case $|\Delta t|<\Delta t_{2,EWN}$ is necessary and sufficient. The remaining case of $b_1+b_2\cos(2\theta_0)>0$ can be analyzed in an analogous manner as the case of $\chi_1$. In fact, we can also show that $k_{2,EWN}=\frac{\Delta t_{2,EWN}}{a_2+b_2\sin\theta_0}\geq k_{2,*}$ which implies that anytime $|k_2|\geq k_{2,EWN}$ then $\chi_2\cap\mathcal{R}_{R}=\varnothing$ which again will imply $|\Delta t|<\Delta t_{2,EWN}$ is both sufficient and necessary to ensure EWN in the connected phase.

Combining our results from the analysis of $\chi_{1,2}$, we conclude that the necessary and sufficient condition to ensure $\mathcal{W}_E(B)\subseteq\mathcal{W}_E(A\cup B)$ in the connected phase is $|\Delta t|<\min\{\Delta t_{c},\Delta t_{2,EWN}\}$. We have to augment the two bounds with a minimum because the ordering between the two quantities changes depending on the specific choices of $a_{1,2},b_{1,2},\theta_0$. Finally, we can formulate the condition for EWN of the entire setup (that is, ensuring that $\mathcal{W}_E(A),\mathcal{W}_E(B)\subseteq\mathcal{W}_E(A\cup B)$) by requiring that $|\Delta t|<\min\{\Delta t_{2,EWN},\Delta t_{A,EWN},\Delta t_c\}$. We can eliminate $\Delta t_c$ from the result by considering
\begin{equation}
\label{OrderbetweentAEWNandtc}
    \begin{split}
        \frac{2(\Delta t_{c}^2-\Delta t_{A,EWN}^2)}{a_2-a_1}
        &=-(a_1+a_2+2b_1\sin\theta_0)+\sqrt{(a_1+a_2+2b_1\sin\theta_0)^2+4b_1^2\cos^2\theta_0}.
    \end{split}
\end{equation}
It is easy to see that the right-hand side is positive from which it follows that $(\Delta t_{c}^2-\Delta t_{A,EWN}^2)\geq 0$. Since both $\Delta t_c$ and $\Delta t_{A,EWN}$ are positive, this implies that $\Delta t_{A,EWN} \leq \Delta t_c$. This yields the final necessary and sufficient condition to ensure EWN of our setup in the event a connected phase dominates,
\begin{equation}
\label{FinalEWNCondition}
\begin{split}
    &|\Delta t|<\Delta t_{EWN}=\min\{\Delta t_{A,EWN},\Delta t_{2,EWN}\},\\
\end{split}
\end{equation}
where $\Delta t_{A,EWN}$ and $\Delta t_{2,EWN}$ are defined in Eq. (\ref{MostStringentEWNACondd}) and Eq. (\ref{SuffConEWNNaivedelt}) respectively.

As a quick check it is useful to consider the limit as $\theta_0\to\pi/2$. In this limit the brane becomes a conformal boundary and we should recover standard holography results. Indeed, we can see that
\begin{equation}
    \lim_{\theta_0\to\frac{\pi}{2}}\Delta t_{EWN}=a_1+b_1=\Delta t_{A,EWN}|_{\theta_0=\pi/2}.
\end{equation}
Quick inspection of the expression in Eq. (\ref{OrderbetweentAEWNandtc}) reveals that $\Delta t_c|_{\theta_0=\pi/2}=\Delta t_{A,EWN}|_{\theta_0=\pi/2}$. In other words, in the critical limit we recover standard holography results that EWN trivially follows from requiring spacelike separation of the intervals $A$ and $B$.\footnote{One can also take the parameters used to generate Figures \ref{AUBEntWegTimeslicesSet1},\ref{AUBEntWegTimeslicesSet2},\ref{WedgeIntBrane} and verify that those choices of parameters do indeed satisfy Eq. (\ref{FinalEWNCondition}) which is consistent with the features we described in those example plots.}

\subsection{Dominance of the Connected Phase}
\label{DominanceOfConnectedPhaseSection}
Thus far, we have been exploring the issue of when EWN is violated under the assumption that the RT surfaces associated to the region $A \cup B$ are in the connected phase, i.e., $\mathcal{W}_E(A\cup B) \neq \mathcal{W}_E(A) \cup \mathcal{W}_E(B)$. This condition is necessary to have a non-trivial constraint from EWN. To finally show that EWN is violated, we need to discuss under which circumstances the condition
\begin{equation}
\label{ConnCondDirich}
    \mathcal{A}_{\chi_{\text{dis}}(A)}+\mathcal{A}_{\chi_{\text{dis}}(B)}-\mathcal{A}_{\chi_{\text{con}}(A\cup B)} > 0
\end{equation}
holds which implies that the RT surfaces are in the connected phase. Here, $\mathcal A$ denotes the area of an extremal surface, while the subscript specifies which surface we are considering.
In Appendix \ref{ConnectedConditionAppend} we analyze Eq.~\eqref{ConnCondDirich} and show that we can rephrase the condition of the RT surfaces being in the connected phase in terms of $\Delta t$ as $\Delta t^2>\Delta t_{\text{con}}^2$ with
\begin{equation}
\label{ConnectedConditionDominance}
\begin{split}
    &\Delta t_{\text{con}}^2=\frac{\Delta x_1^2+\Delta x_2^2+\Delta z_1^2+\Delta z_2^2}{2}-\sqrt{\left(\frac{\Delta x_1^2-\Delta x_2^2+\Delta z_1^2-\Delta z_2^2}{2}\right)^2+4a_-^2\Delta z_1\Delta z_2e^{\mathcal{A}_{\chi_{\text{dis}}(B)}/L}},
\end{split}
\end{equation}
where $\Delta x_{1,2}=a_{1,2}+b_{1,2}\sin\theta_0$, $\Delta z_{1,2}=b_{1,2}\cos\theta_0$, and $a_{\pm}=\frac{1}{2}(a_2\pm a_1)$.
The condition for a connected phase places a lower bound on $|\Delta t|$, which is intuitively clear since increasing the time between $A$ and $B$ makes the corresponding connected extremal surfaces more lightlike (i.e., decreases their area).
 
Recall that the condition for EWN, Eq. (\ref{FinalEWNCondition}), places an upper bound on $|\Delta t|$, while the condition for dominance of the connected phase places a lower bound on $|\Delta t|$. Therefore, in order to have a genuine connected phase for which our EWN conditions are nontrivial, it is sufficient to check that the connected phase occurs before $A$ and $B$ become timelike separated, i.e., $\Delta t_{c}-\Delta t_{\text{con}}>0$. This ensures that a connected phase dominates as $|\Delta t|\to\Delta t_{c}$ from below. Using the fact that $\Delta t_c=\sqrt{\Delta x_1+\Delta z_1}$ it is straightforward to see that generally, $\Delta t_{c}^2-\Delta t_{\text{con}}^2>0\Rightarrow\Delta t_c-\Delta t_{\text{con}}>0$.\footnote{This can be straightforwardly proved by noting that $\Delta t_c^2=\Delta x_1^2+\Delta z_1^2$. Using this in Eq. (\ref{ConnectedConditionDominance}) we can show that $\Delta t_{c}^2-\Delta t_{\text{con}}^2=-\frac{\Delta x_2^2-\Delta x_1^2+\Delta z_2^2-\Delta z_1^2}{2}+\sqrt{\left(\frac{\Delta x_1^2-\Delta x_2^2+\Delta z_1^2-\Delta z_2^2}{2}\right)^2+4a_-^2\Delta z_1\Delta z_2e^{\mathcal{A}_{\chi_{\text{dis}}(B)}/L}}$, which is clearly positive when $\theta_0\in(0,\pi/2)$. This implies that as $|\Delta t|\to\Delta t_c$ a connected phase will always dominate.}

It is also interesting to ask if a connected phase already exists as we approach $\Delta t_{EWN}=\min\{\Delta t_{A,EWN},\Delta t_{2,EWN}\}$ from below. To facilitate our discussion, we will find it useful to understand when $\Delta t_{EWN}=\Delta t_{2,EWN}$ and when $\Delta t_{EWN}=\Delta t_{A,EWN}$. Start by noting the following identity,
\begin{equation}
\label{TransitionCondition}
    \begin{split}
        &\Delta t_{2,EWN}-\Delta t_{A,EWN} =2a_- - \ell_*,
    \end{split}
\end{equation}
where we defined the length scale\footnote{The positivity of $\ell_*$ can be seen as follows. Define $\Delta t_{+,EWN}=a_++\frac{b_1+b_2-\sqrt{b_1^2+b_2^2+2b_1b_2\cos(2\theta_0)}}{2\sin\theta_0}\geq 0$ and $\Delta t_{+,c}=\sqrt{a_+^2+b_1^2+2a_+b_1\sin\theta_0}\geq 0$. Then $\ell_*=\Delta t_{+,c}-\Delta t_{+,EWN}$. Then it is a straightforward exercise to explicitly compute and see $\Delta t_{+,c}^2-\Delta t_{+,EWN}^2\geq 0\Rightarrow\ell_*\geq 0$.}
\begin{align}
\label{DefinitionOfellstar}
\ell_*=-\left(a_++\frac{b_1+b_2-\sqrt{b_1^2+b_2^2+2b_1b_2\cos(2\theta_0)}}{2\sin\theta_0}\right)+\sqrt{a_+^2+b_1^2+2a_+b_1\sin\theta_0}.
\end{align}
It follows that, $2a_-\leq \ell_*\Leftrightarrow\Delta t_{2,EWN}\leq \Delta t_{A,EWN}$ and $2a_-> \ell_*\Leftrightarrow\Delta t_{A,EWN}< \Delta t_{2,EWN}$. Using the length scale $\ell_*$ we numerically analyze the quantity $\Delta t_{EWN}-\Delta t_{\text{con}}$. The details of our numerical analysis is discussed near the end of Appendix \ref{ConnectedConditionAppend} after Eq. (\ref{DeltatForConnectedPhase}). We find that $\Delta t_{EWN}-\Delta t_{\text{con}}> 0$ whenever $2a_- > \ell_*$. Thus, if the subregion $A$ is sufficiently large, the connected phase will always dominate as we approach the EWN bound from below. This demonstrates that the constraint in Eq. (\ref{FinalEWNCondition}) is non-trivial. 

In the remaining regime where $2a_-\leq \ell_*$ we find the opposite results, namely that $\Delta t_{EWN}-\Delta t_{\text{con}}\leq 0$ (saturation occurs exactly when $2a_-=\ell_*$). This implies that as $|\Delta t|\to\Delta t_{EWN}$ the disconnected phase actually dominates. So it appears that for sufficiently small $A$ the condition for EWN we wrote only becomes relevant once $|\Delta t|>\Delta t_{A,EWN}$ and sufficiently close to $\Delta t_c$. We organize all these results in Table \ref{tab:my_label}. 
\begin{table}[ht!]
    \centering
   \begin{tabular}{ l|c|c } 
 %\hline
  & $2a_-\leq \ell_*$ & $2a_-> \ell_*$ \\ 
 %\hhline{|=|=|=|}
 \hline
 $|\Delta t|\to\Delta t_{2,EWN}$ & Disconnected & Connected \\ 
 %\hline
 $|\Delta t|\to\Delta t_{A,EWN}$ & Disconnected & Connected \\ 
 %\hline
 $|\Delta t|\to\Delta t_{c}$ & Connected & Connected \\ 
 %\hline
\end{tabular}
    \caption{This table shows the relation between the size of $A$ and the phase of the RT surface as we approach different bounds. Connected (Disconnected) means we satisfy (violate) Eq.~\eqref{ConnCondDirich}.} 
    \label{tab:my_label}
\end{table}

In summary, we have shown that spacelike separateness of two regions $A$ and $B$ with respect to the bulk, is not sufficient to ensure EWN, at least if one of those regions is located at a finite cutoff in the bulk. When EWN precisely fails depends on the size of the involved intervals. At least in our case and for sufficiently large intervals we can give a precise condition, Eq. \eqref{FinalEWNCondition}, which is equivalent to EWN. In the next section, we will demonstrate that the conditions we obtained for EWN can be reinterpreted as playing the role of ensuring that all extremal surfaces in our setup are spacelike separated from each other. 

\subsection{EWN and Spacelike Separation of Extremal Surfaces}
\label{SigmaABCauchyExploreSection}
To make more sense of the findings of Sections \ref{PoincareAdS3SuffCondSec} and \ref{DominanceOfConnectedPhaseSection} it is useful to understand how the various extremal surfaces are separated from each other, in particular, whether they are spacelike separated. 
The conditions we derived for EWN in previous sections certainly guarantee that the line segments, $\chi_1$ and $\chi_2$, generating the RT surface in the connected phase are always spacelike separated from the RT curves in the disconnected phase. However, we have not said anything about the separation between $\chi_{\text{dis}}(A)$ or $\chi_{\text{dis}}(B)$ or the separation between $\chi_1$ and $\chi_2$, we will explore this in what follows.       

Let us begin by investigating under which conditions the extremal surfaces $\chi_{\text{dis}}(A)$ and $\chi_{\text{dis}}(B)$ are spacelike separated. Start by noting that the congruence of null geodesics orthogonal to $\chi_{\text{dis}}(A)$ is simply given by
\begin{equation}
\label{NullCOngruchiAA}
     \left(\frac{a_2-a_1}{2}+|t-t_a|\right)^2=z^2+\left(x+\frac{a_2+a_1}{2}\right)^2.
\end{equation}
A point $(t,x,z)$ is spacelike separated from $\chi_{\text{dis}}(A)$ will satisfy Eq. (\ref{NullCOngruchiAA}) with the ``$=$'' sign replaced by a ``$<$'' sign. The RT surface $\chi_{\text{dis}}(B)$ is located at $t = t_b$ and $z=z(x)$, which is given in Eq. (\ref{chidisBexp}). 
Using this, we can write the condition for $\chi_{\text{dis}}(B)$ to be spacelike separated from $\chi_{\text{dis}}(A)$  as
\begin{equation}
\begin{split}
    \left(\frac{a_2-a_1}{2}+|\Delta t|\right)^2<&\left(-x^2+\frac{b_1+b_2}{\sin\theta_0}x-b_1b_2\right)+\left(x+\frac{a_2+a_1}{2}\right)^2\\
    &=\left(a_2+a_1+\frac{b_1+b_2}{\sin\theta_0}\right)x+\left(\frac{a_1+a_2}{2}\right)^2-b_1b_2.
\end{split}
\end{equation}
This has to be true for any choice of $x\in[b_1\sin\theta_0,b_2\sin\theta_0]$. Since the right hand side of the inequality is a linear function of $x$ with a positive slope it means that the inequality is satisfied for $x\in[b_1\sin\theta_0,b_2\sin\theta_0]$ if and only if it is satisfied at $x=b_1\sin\theta_0$. Plugging this in we find
\begin{equation}
\begin{split}
    \left(|\Delta t|+\frac{a_2-a_1}{2}\right)^2< b_1^2\cos^2\theta_0+\left(b_1\sin\theta_0+\frac{a_2+a_1}{2}\right)^2,
\end{split}
\end{equation}
which, after isolating for $|\Delta t|$, is precisely the condition that $|\Delta t|< \Delta t_{A,EWN}$.
This proves that $|\Delta t|< \Delta t_{A,EWN}$ is equivalent to requiring the spacelike separation of the disconnected RT surfaces. Importantly, this condition is actually stronger than simply requiring that $A$ and $B$ are spacelike separated in the bulk due to the computation we did in Eq. (\ref{OrderbetweentAEWNandtc}).

Now we turn to the issue of understanding when $\chi_1$ and $\chi_2$ are spacelike separated. We start by claiming that anytime $|\Delta t|\geq \Delta t_{2,EWN}$ then $\chi_1$ and $\chi_2$ cannot be spacelike separated. To prove this, it is useful to make use of the fact that we can express $\chi_2$ as the intersection of lightcones located at the boundary whose apexes are located at the points given in Eq. (\ref{tipConechi2}).\footnote{This statement is strictly speaking only true when $|k_2|\leq1$. However, note that if $|\Delta t|=\Delta t_{2,EWN}$ we have that $k_{2,EWN}=\frac{\Delta t_{2,EWN}}{a_2+b_2\sin\theta_0}=1+\frac{b_1+b_2\cos(2\theta_0)-\sqrt{(b_1+b_2\cos(2\theta_0))^2+b_2^2\sin^2(2\theta_0)}}{2\sin\theta_0(a_2+b_2\sin\theta_0)}$<1 when $\theta_0\in(0,\pi/2)$. For the purposes of our proof we need only show $\chi_1$ and $\chi_2$ fail to be spacelike separated from each other at $|\Delta t|=\Delta t_{2,EWN}$ and also very slightly above the bound as well so there is no issue here. Furthermore, the numerical calculations we did in Appendix \ref{AppendixProvingCasesWhenNotCauchy} make no such assumption and agree with what we have proved using this method.} This means that $\chi_1$ is spacelike separated to $\bar{\chi}_2$ (i.e. the extension of $\chi_2$ behind brane which we discussed near the end of Section \ref{ClassificationSectionRT}) whenever\footnote{Note the orientation of the inequality is correct. Since $\chi_1$ should be contained in the region enclosed by the lightcones whose apexes are at $(t_{\pm,2},x_{\pm,2},0)$ which makes $\chi_1$ timelike to $(t_{\pm,2},x_{\pm,2},0)$.}
\begin{equation}
\label{conditionForSLAuBCauchy}
    -(t_1(x)-t_{\pm,2})^2+\left(x-x_{\pm,2}\right)^2+z_1(x)^2<0.
\end{equation}
Without loss of generality we will consider the case when $\Delta t\geq 0$. Let us consider the situation when $\Delta t=\Delta t_{2,EWN}$ and we set $x=b_1\sin\theta_0-\epsilon$ with $\epsilon\geq0$,
\begin{equation}
\label{DeltasForchi2barTochi1}
\begin{split}
    &-(t_1(x)-t_{+,2})^2+\left(x-x_{+,2}\right)^2+z_1(x)^2\vert_{\Delta t=\Delta t_{2,EWN},x=b_1\sin\theta_0-\epsilon}\\
    &=-\frac{(a_2-a_1)\left(a_2+a_1+\frac{b_1+b_2}{\sin\theta_0}\right)}{a_1+b_1\sin\theta_0}\epsilon\leq 0.\\
\end{split}
\end{equation}
When $\epsilon=0$ we know that some subset of points on $\bar{\chi_2}$ are in fact null separated from the point where $p_1=\chi_1\cap\text{Brane}$. The next step is to determine which portion of $\bar{\chi}_2$ is null separated from $p_1$.\footnote{The reason this step is necessary is because it is important to ensure that $x=b_1\sin\theta_0\in\chi_1$ is null separated from $\chi_2\subset \bar{\chi}_2$ (i.e. the point on $p_1\in\chi_1$ needs to be null separated from some point on $\bar{\chi}_2$ in front of the brane).} This can be done by understanding where in the $xz$-plane the tip of the cone which defines what is spacelike separated from $\bar{\chi}_2$ is equal to when $\Delta t=\Delta t_{2,EWN}$. The answer is very simple; it is given by $x_{+,2}=\frac{b_1+b_2}{2\sin\theta_0}$. The important aspect of this is $x_{+,2}>0$. Now project our entire setup into the $xz$-plane and draw a straight line that goes from the tip of the cone located at $(z=0,x=x_{+,2}>0)$ to the point where $\chi_1$ intersects with the brane located at $(z=b_1\cos\theta_0,x=b_1\sin\theta_0)$. It is straightforward to show that the slope of the straight line we constructed is negative for any choice of parameters in our setup. If we follow this ray it will intersect with $\chi_2$ at $p_{\text{int}}=(t_2(x_{\text{int},2}),x_{\text{int},2},z_2(x_{\text{int},2}))$, where $x=x_{\text{int},2}<b_2\sin\theta_0$. It is precisely $p_{\text{int}}\in \chi_2$ which is null separated from $p_1\in\chi_1$. So far we understand that at pair of points ($p_1\in\chi_1$ and $p_{\text{int}}\in\chi_2$) the extremal surfaces become null separated when $|\Delta t|=\Delta t_{2,EWN}$. Our next task is to understand what happens, we slightly perturb away from $\Delta t_{2,EWN}$. To do this we consider the following,
\begin{equation}
\label{Timelikeincreasingt}
    \begin{split}
    &-(t_1(x)-t_{+,2})^2+\left(x-x_{+,2}\right)^2+z_1(x)^2\vert_{\Delta t=\Delta t_{2,EWN}+\delta t,x=b_1\sin\theta_0}\\
    &=\frac{2(b_2-b_1)\sin\theta_0\sqrt{b_1^2+b_2^2+2b_1b_2\cos(2\theta_0)}}{-b_1-b_2\cos(2\theta_0)+\sqrt{b_1^2+b_2^2+2b_1b_2\cos(2\theta_0)}}\delta t+\mathcal{O}(\delta t^2).
\end{split}
\end{equation}
When $\delta t>0$ the expression above is also greater than zero. This represents a violation of the spacelike condition in Eq. (\ref{conditionForSLAuBCauchy}) and implies that a portion of $\bar{\chi}_2$ is now timelike separated from $p_1\in\chi_1$. Due to our previous discussion below Eq. (\ref{DeltasForchi2barTochi1}) we can actually conclude that $p_1$ is timelike separated from $p_{\text{int}}\in\chi_2$ as well as an open neighborhood of points on $\chi_2$ centered around $p_{\text{int}}$. This proves our claim that $\chi_1$ cannot be spacelike to $\chi_2$ when $|\Delta t|\geq\Delta t_{2,EWN}$.

Let us now turn to $|\Delta t|<\Delta t_{2,EWN}$. In such a regime, there are a few comments that we can make based on the results of the computations in Eq. (\ref{DeltasForchi2barTochi1}) and Eq. (\ref{Timelikeincreasingt}). In particular, we see that when $\epsilon>0$ in Eq. (\ref{DeltasForchi2barTochi1}) the ``$+$'' bound given in Eq. (\ref{conditionForSLAuBCauchy}) is not violated. This suggests that, when $|\Delta t|=\Delta t_{2,EWN}$ only $p_{\text{int}}\in\chi_2$ and $p_1\in \chi_1$ fail to be spacelike separated from each other but all other points are indeed spacelike separated. The result in Eq. (\ref{Timelikeincreasingt}) in the case where $\delta t<0$ suggests that the same pair of points are spacelike separated to each other when $|\Delta t|<\Delta t_{2,EWN}$ and indeed we should also expect all other points on $\chi_1$ and $\chi_2$ to be spacelike separated also as we reduce $|\Delta t|$.\footnote{Here, we only considered the ``$+$'' bound of Eq. (\ref{conditionForSLAuBCauchy}). We should also perform a similar analysis for ``$-$'' bound, which however is harder to analyze. We thus resorted to numerical computations (see Appendix \ref{AppendixProvingCasesWhenNotCauchy}) to verify that claims we made for $|\Delta t|<\Delta t_{2,EWN}$. } 
This is because if only a pair of points fail to be spacelike separated from at $|\Delta t|=\Delta t_{2,EWN}$ reducing $|\Delta t|$ to be below the marginal case increases the curves' spacelike separation. 
Based on these observations, we expect that $|\Delta t|<\Delta t_{2,EWN}$ implies that $\chi_1$ and $\chi_2$ are spacelike separated. We have also verified this numerically in Appendix \ref{AppendixProvingCasesWhenNotCauchy}.

We have thus shown that the necessary and sufficient condition for EWN we derived in Eq. (\ref{FinalEWNCondition}) is equivalent to requiring that all the extremal surfaces we used to define our entanglement wedges (i.e. the surfaces $\chi_{\text{dis}}(A)$, $\chi_{\text{dis}}(B)$, $\chi_{1,2}$) are all spacelike separated from each other. In standard AdS this is guaranteed by spacelike separation of $A$ and $B$ in the bulk and boundary. However, as we have seen here, even in the simplest cases this is not true anymore in the presence of a cutoff.

Note that our construction of the entanglement wedge $\mathcal W_E(A \cup B)$ discussed in Section \ref{EntanglementWedgeDefGenericSec} only asked for points which were spacelike separated from and located between $\chi_1$ and $\chi_2$. However, as seen above, the volume obtained this way is not necessarily the domain of dependence of a partial Cauchy slice $\Sigma_{A\cup B}$. In the case of holography at a cutoff this enters as an additional requirement and in our case gives rise to the bound $\Delta t_{2,EWN}$. All the different wedges we consider can only be constructed iff $|\Delta t|<\min\{\Delta t_c,\Delta t_{2,EWN}\}$. The bound $|\Delta t| <\Delta t_{A,EWN}$ is logically independent of this and, as discussed below, ensures that the partial Cauchy slices $\Sigma_A$ and $\Sigma_B$ can lie on a single Cauchy slice.

\subsection{Consequences for Locality in Cutoff Holography}
\label{DiscussionEWNConstrintChapt}
From the analysis in Sections \ref{PoincareAdS3SuffCondSec} -- \ref{SigmaABCauchyExploreSection} we have learned that our condition for EWN, which was written in Eq. (\ref{FinalEWNCondition}) non-trivially applies in the regime where $2a_->\ell_*$. In this regime, the condition for EWN becomes $|\Delta t|<\Delta t_{A,EWN}$ which can be equivalently reinterpreted as the requirement that $\chi_{\text{dis}}(A)$ and $\chi_{\text{dis}}(B)$ are spacelike separated. This observation has implications for the discussion of non-locality in cutoff holography. 

In standard holography the boundary theory is local and two subregions are independent if there are not causally related, i.e., they are spacelike separated. However, in cutoff holography one generally expects the theory on the cutoff to be a non-local theory and it is less clear how independent subregions can be defined. 
It has been known in our setup involving ETW branes that spacelike separation in the bulk sense is stronger than spacelike separation in the boundary sense due to shortcuts that exist through the bulk between points on the brane and boundary \cite{Omiya:2021olc}.\footnote{As a note to the reader it was suggested in \cite{Omiya:2021olc} that the existence of such shortcuts was in tension with having a causal effective field theory (EFT) in the brane-boundary system (i.e. the ``intermediate description''). In the recent work \cite{Geng:2025yys}, it was shown that this is not problematic when one views the theory in the intermediate description as an EFT with a cutoff. It was argued that the geodesic shortcut is actually a UV effect from the perspective of the brane-boundary system (it is encoded by heavy modes KK modes). Since the EFT on the brane-boundary system comes with a cutoff it is safe from this UV effect. Coupling to the heavy KK modes in the UV was also explored as they could potentially lead to signatures of causality violation, but in this case it was shown that constraints from unitarity also prevented non-causal signatures from appearing in correlators.} An interesting open question in such setups is which notion of ``spacelike'' separation we need to define subregions on the holographic cutoff surface that are independent of each other. A naive candidate condition would be that the boundary subregions should be spacelike separated through the bulk as this ensures no causal signaling can occur between $A$ and $B$ either through the bulk or boundary. 

Our computation in Eq. \eqref{OrderbetweentAEWNandtc} proves that this requirement is not sufficient. In particular, as we have seen, it is possible to construct configurations in which $A$ and $B$ cannot influence each other causally through the bulk, but the entanglement wedges can. Such examples cannot be constructed in the case where $\theta_0=\pi/2$ (i.e. in standard holography). This implies that --- provided entanglement wedges work in cutoff holography in the same way they do in standard AdS/CFT --- $\mathcal{W}_E(A)$ and $\mathcal{W}_E(B)$, and thus $A$ and $B$, are not independent of each other even when $A$ and $B$ are spacelike separated through the bulk. We suggest that this might be regarded a geometric bulk manifestation of the non-local nature of the theory on the cutoff. In such cases our bound on entanglement wedge nesting disallows one to consider $A$ and $B$ as independent subregions. And clearly, the condition that $\mathcal{W}_E(A)$ cannot causally influence $\mathcal{W}_E(B)$ is a necessary condition for ``independent'' subregions in non-local holographic theories such as the one we have on the brane. Here we are not making any claims in contradiction/tension to the findings of \cite{Geng:2025yys}. The existence of such non-local correlations between $A$ and $B$ do not pose a problem to the causality of the EFT on the brane-boundary system. Since the effect described here is geometric and in the bulk, we  expect these to be UV effects which manifest well above the cutoff of the EFT, thereby leaving the local EFT structure at low energies intact. Similar to how the geodesic shortcut in bulk was argued in \cite{Geng:2025yys} to be encoded by heavy KK modes in the UV regime we might speculate a similar description also exists for the enhanced non-local correlations between $A$ and $B$ described here. We leave more detailed investigations in such a direction to future work. 

To conclude the discussion of this section it is also interesting to consider what the condition $|\Delta t|<\Delta t_{A, EWN}$ implies about the separation between $A$ and $\text{Vir}(B)$ (which was used in the discussion of Eq. (\ref{WEBDefVir})). Taking inspiration from recent works which think about a notion of induced causality on the brane in terms of subregions that exist behind the brane on the imaginary conformal boundary \cite{Mori:2023swn,Franken:2024wmh}, we will think of $B$ as a coarse grained version of $\text{Vir}(B)$ which lives on the imaginary conformal boundary behind the brane. Through this identification we might suggest a natural way to think about independent subregions on the brane/cutoff is to restrict $B$ to be the such that $\text{Vir}(B)$ is spacelike separated from $A$ in the full spacetime without the brane. Interestingly enough, this exactly gives the condition that $|\Delta t|<\Delta t_{1,EWN}$ which was derived as a sufficient condition for EWN in Eq. (\ref{SuffConEWNNaivedelt}).\footnote{Recall that in Section \ref{PoincareAdS3SuffCondSec}, this condition represented the most stringent sufficient condition to ensure $\mathcal{W}_E(B)\subseteq\mathcal{W}_E(A\cup B)$, we disregarded it because it was actually not a necessary condition but it reappears here as well.} Using the fact that $\Delta t_{2,EWN}-\Delta t_{1,EWN}=2a_-$ along with the identity in Eq. (\ref{TransitionCondition}) we can easily see that $\Delta t_{1,EWN}-\Delta t_{A,EWN}=-\ell_*\leq 0$, where saturation occurs when $\theta_0=\pi/2$. This implies that any time $A$ and $\text{Vir}(B)$ are spacelike separated then EWN is will be satisfied. However, due to the fact that $\Delta t_{1,EWN}-\Delta t_{A,EWN}\leq 0$ there also exist configurations where EWN is satisfied but $A$ and $\text{Vir}(B)$ fail to be spacelike separated and therefore would not be independent in this alternate notion of subregion independence formulated in terms of subregions behind the brane. 

\subsection{Relation to Restricted Maximin}
\label{RestMaximinSec}
Trying to define holography on a cutoff surface does not only cause problems for EWN, but also other inequalities which need to be obeyed by entanglement entropies, such as strong subadditivity (SSA). Based on ideas put forward in \cite{Wall:2012uf}, the authors of \cite{Grado-White:2020wlb} proposed a prescription, called \emph{restricted maximin}, which associates a quantity which obeys SSA to any set of achronal co-dimension two bulk regions. This quantity is thus a natural candidate for a holographic entanglement entropy. It was further shown in \cite{Grado-White:2020wlb} that their version of entanglement entropy also obeys monogamy of mutual information and the corresponding entanglement wedges satisfy EWN. It is interesting to contrast our construction of entanglement wedges to the construction using restricted maximin \cite{Grado-White:2020wlb}. 

In the restricted maximin prescription, one considers some subregion $U$ on the space-time boundary (which is allowed to be located at a finite cutoff) and then considers Cauchy surfaces $\Sigma$ in the bulk which end on the cutoff along a co-dimension 2 surface $\gamma$ (i.e. $\Sigma\vert_{\text{cutoff}}=\gamma$) such that $U\subset\gamma$. On each Cauchy slice one is instructed to find the minimal area surface homologous to $U$. Maximizing the area of the minimal area surfaces over the choice of all Cauchy slices with the property $U\subset\gamma$ yields the restricted maximin surface. In \cite{Grado-White:2020wlb} it was shown that the construction of entanglement wedges using the maximin prescription satisfies EWN anytime the subregions on the cutoff/boundary are spacelike separated through the bulk. This is quite different from our conclusion which stated that simply requiring $A$ and $B$ to be spacelike separated is not sufficient to ensure EWN. 

The key to resolving this apparent tension between our work and \cite{Grado-White:2020wlb} is to note that our prescription for defining RT surfaces is generally not the same as in the restricted maximin approach. To understand this, recall that our approach starts with choosing the ``naive'' RT surfaces homologous to the subregions in which we are interested. What our results in Section \ref{PoincareAdS3SuffCondSec}-\ref{SigmaABCauchyExploreSection} demonstrate is that such ``naive'' RT surfaces are generally not spacelike separated from each other. In our approach we need to enforce EWN non-trivially by restricting the placement of allowed boundary regions. This ensures that all the RT surfaces are spacelike separated. In the restricted maximin prescription, the construction always involves surfaces on restricted Cauchy slices, which ensures that all RT surfaces are already spacelike separated by construction. From this perspective, it is no surprise that EWN follows trivially from restricted maximin just by requiring $A$ and $B$ to be spacelike separated (i.e., EWN is ``baked'' into the maximin procedure through the ingredient of the restricted Cauchy slice). Furthermore, we expect that anytime EWN is respected in our prescription the ``naive'' and restricted maximin RT surfaces will coincide.

However, this also implies that there are configurations in which restricted maximin RT surfaces do not agree with the naive expectation. Consider the scenario used in this paper, where we have two spacelike separated subregions $A$ and $B$ at constant times $t_a$ and $t_b$ on a cutoff surface. As we have seen explicitly, it is possible to choose $t_a$ and $t_b$ such that $A$ and $B$ are spacelike separated, while their RT surfaces, which are located on constant time slices, are not. In particular, as demonstrated, this is possible for a static geometry. According to our prescription, such configurations should be ruled out, since EWN is violated. However, following restricted maximin it is indeed possible to assign an RT surface and an associated entropy. The price to pay is that even the disconnected, restricted maximin surfaces do not lie on constant time slices anymore, although the bulk geometry is static and both $A$ and $B$ are located on a constant time slice -- in stark contrast to expectations from AdS/CFT. Although we do not intend to make definitive statements, a deviation between the ``naive'' and ``restricted maximin'' RT surfaces might indicate that the chosen boundary regions are not truly independent. This would in particular imply that partial traces over subregions have to be handled with care: Given subregions $A$ and $B$ which violate our EWN bounds one might want to compute the entanglement entropy of the reduced system on $A$. However, when doing so using restricted maximin, one needs to remember to only consider Cauchy slices which are anchored on $B$, although $B$ is not part of the system under consideration anymore.

\begin{figure}[h!]
\centering
\includegraphics[width=120mm]{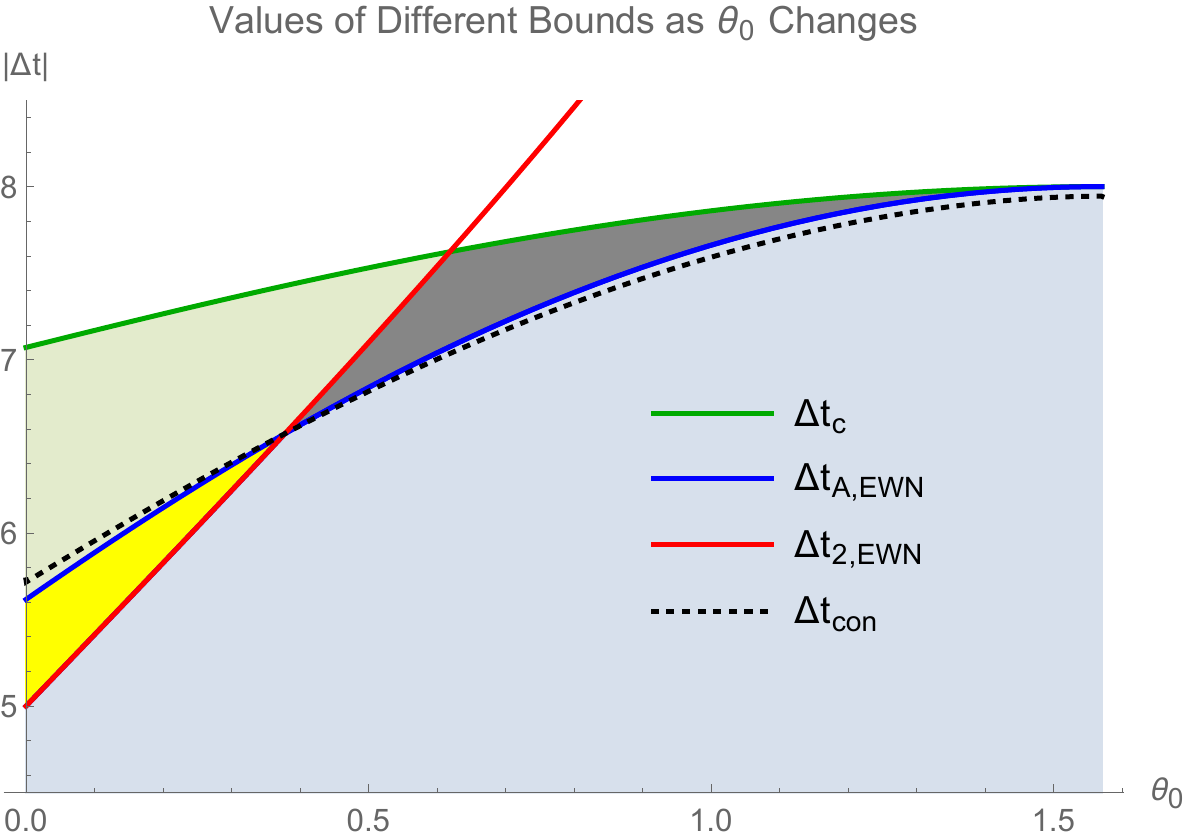}
\caption{We fixed $a_1 = 1, a_2 = 5, b_1 = 7, b_2 = 10$ and vary $\theta_0$ between $0$ and $\pi/2$. Our
construction of all entanglement wedges is well defined below $\min\{\Delta t_{2,EWN},\Delta t_c\}$ which is the union of the gray and blue shaded regions. Restricted maximin can be applied in the parameter space below the solid green line. Restricted maximin and our prescription are expected to give same RT surfaces in shaded blue region where EWN is respected. }
\label{BoundsVaryWiththetaplot}
\end{figure}

We can summarize the relation between restricted maximin and our construction with the help of Figure \ref{BoundsVaryWiththetaplot}, where we fix some values of $a_{1,2}$ and $b_{1,2}$ and plot the values of $\Delta t_{c}$ (solid green line), $\Delta t_{A,EWN}$ (solid blue line), $\Delta t_{2,EWN}$ (solid red line), and $\Delta t_{\text{con}}$ (dotted black line) as functions of $\theta_0$.
Start by noting that one can apply the restricted maximin procedure anywhere below the solid green line when $A$ and $B$ are spacelike separated, and one will always obtain entanglement wedges that satisfy the EWN (these wedges are defined using the maximin RT surfaces). In our prescription, we have learned that we can construct $\mathcal{W}_E(A\cup B)$ in the connected phase (as well as all other wedges) using the ``naive'' RT surfaces iff $|\Delta t|<\min\{\Delta t_c,\Delta t_{2,EWN}\}$. This corresponds to the union of the gray and blue regions. The gray region is always below the red line but above the blue line and this implies $\chi_{1,2}$ are spacelike separated to each other but the disconnected RT surfaces $\chi_{\text{dis}}(A)$, $\chi_{\text{dis}}(B)$ are not. In this portion of parameter space the disconnected RT surfaces in the restricted maximin prescription will not agree with our ``naive'' RT surfaces in the disconnected phase but we expect the connected RT surfaces to agree. The blue region ($|\Delta t|<\Delta t_{EWN}=\min\{\Delta t_{2,EWN},\Delta t_{A,EWN}\}$) is where all ``naive'' RT surfaces are spacelike separated and EWN is satisfied. This is precisely the portion of parameter space where we expect all the maximin and ``naive'' RT surfaces to agree. This just leaves us with the union of the shaded green and yellow regions. In the yellow region, we are below the blue line but above the red line. Here we expect the disconnected phase maximin RT surfaces to match the ``naive'' ones but the connected ones will not match. Finally, in the green region we are above both the red and blue lines so none of the naive RT surfaces are spacelike separated from each other. In this parameter regime we expect no agreement between the maximin RT surfaces and the ``naive'' ones.

\section{Entanglement Wedge Nesting in a Planar BTZ Black Hole Geometry}
\label{EWNBTZChapterOfPaper}
In this section, we will study conditions for entanglement wedge nesting for a two-sided planar BTZ black hole geometry with one exterior having an ETW brane. We will adapt the formalisms developed for AdS$_3$ in the previous sections and apply them to the BTZ black hole. The BTZ black hole geometry has been of great interest in a variety of contexts ranging from black hole microstates \cite{Hartman:2013qma,Kourkoulou:2017zaj,Miyaji:2021ktr}, brane world cosmologies \cite{Cooper:2018cmb,Antonini:2019qkt,Waddell:2022fbn}, as well as constraining DGP couplings in brane-world setups \cite{Lee:2022efh,Geng:2022slq,Geng:2022tfc,Geng:2023qwm}, which motivates their study using our techniques.

In Section \ref{BTZBHSecwithETW} we will introduce the two-sided BTZ black hole along with the ETW brane which will cut off a portion of the right exterior. We will also describe the configuration of intervals we will consider in the two exteriors. In particular, $A_{\text{Bdry}}$ is an interval defined on the conformal boundary in the left exterior and $A_{\text{Br}}$ is an interval defined in the right exterior on the brane. After doing this, in Section \ref{RTSurfBTZSection} we derive expressions for the extremal surfaces we will be considering which will be anchored to $A_{\text{Bdry}}$ and $A_{\text{Br}}$. In \ref{ThermalExtremalSurfacesubsec} we give explicit expressions for thermal RT surfaces anchored to $A_{\text{Bdry}}$ and $A_{\text{Br}}$ which remain in their respective exteriors. In \ref{ConnectedSurfaceBTZsection} we compute the connected RT surface which goes through the horizon and connects  $A_{\text{Bdry}}$ and $A_{\text{Br}}$. In Section \ref{BTZWedgeSection} we give explicit expressions characterizing the wedges associated with the thermal RT surfaces depicted in Figures \ref{EntangWedgeBdryBTZ} - \ref{EntangWedgeBraneBTZ}. In Section \ref{EWNSuffConBTZPlanarSec} we show that EWN of $\mathcal{W}_E(A_{\text{Br}})$ constrains the relative time shift between $A_{\text{Bdry}}$ and $A_{\text{Br}}$ in a way that prevents the RT surface in the connected phase from crossing the black hole ``singularity''.

\subsection{Two-Sided BTZ Black Holes with an ETW Brane}
\label{BTZBHSecwithETW}
\begin{figure}[t]
\centering
\includegraphics[width=80mm]{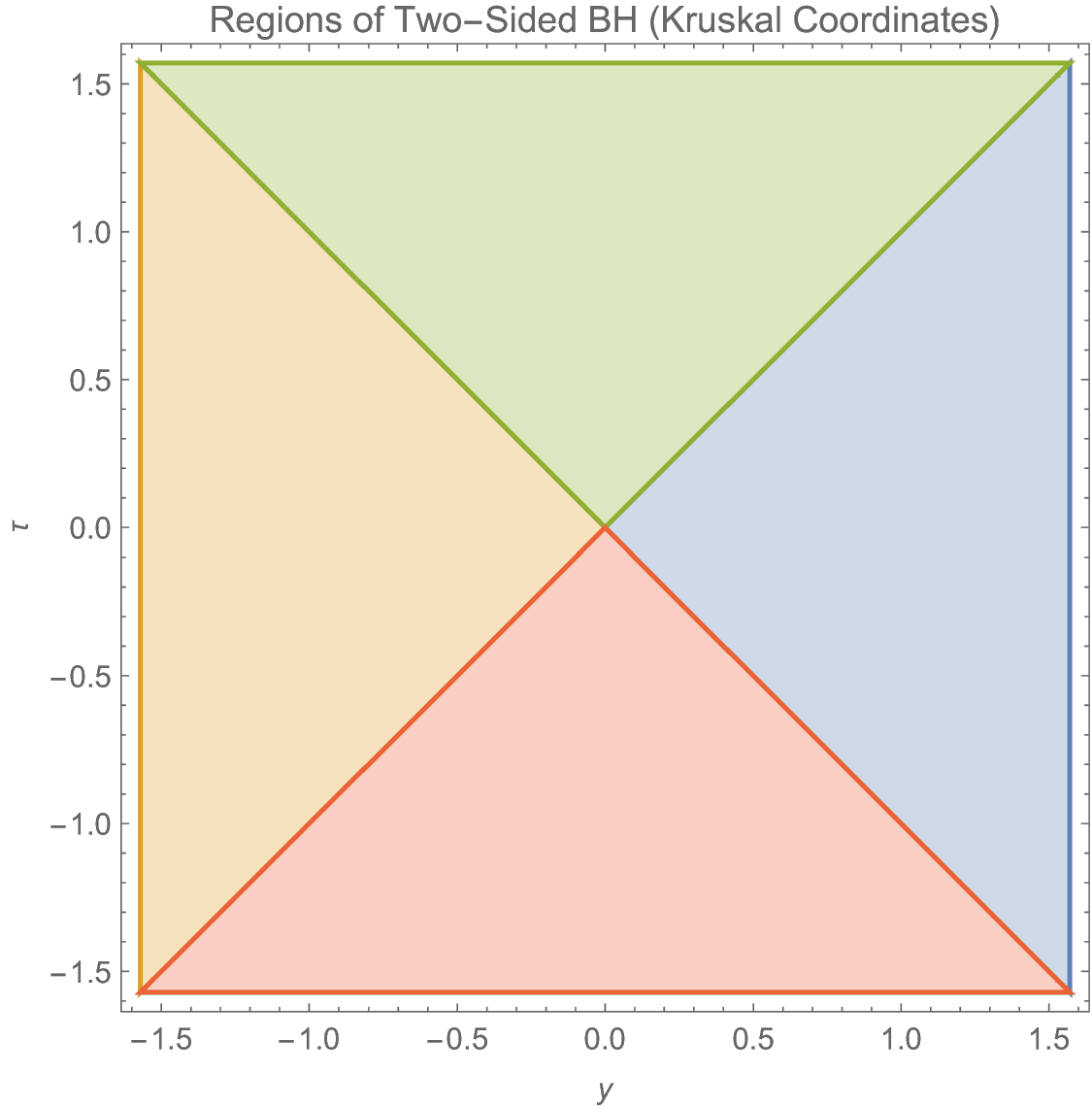}
    \caption{Depicted above is a Two-Sided Planar BTZ black hole spacetime in Kruskal coordinates (with the $x$ coordinate suppressed but one can imagine it coming out/into the page). The horizon of the black hole are the diagonals which split the Penrose diagram into four quadrants ($\text{Ext}_{L,R}$ and $\text{Int}_{Past.,Fut.}$). Constant time slices in Schwarzschild coordinates are straight lines which run through the exterior regions and pass through the bifurcate horizon $y = \tau = 0$. \label{TwoSidedKruskalBHDiagram}}
\end{figure}

We start by discussing the two-sided planar BTZ black hole with one side having an ETW brane. Nothing in what follows will depend on the fact that the black hole is planar, as long as the horizon size is sufficiently big relative to the brane and boundary intervals.
The line element for the two sided BTZ black hole written in Kruskal-Szekeres coordinates is
\begin{equation}
\label{KruSkLineElement}
    ds^2=\frac{1}{\cos^2y}\left[L^2\left(-d\tau^2+dy^2\right)+\frac{r_+^2}{L^2}\cos^2\tau dx^2\right],
\end{equation}
where $\tau,y\in[-\pi/2,\pi/2]$, $x\in\mathbb{R}$, $L$ is the AdS radius, and $r_+$ is the horizon radius in Schwarzschild coordinates. The right and left conformal boundaries are located at $y=\pi/2$ and $y=-\pi/2$, respectively. In these coordinates, the future and past event horizons are located at $\tau=\pm y$. They split the spacetime into four distinct regions: The \emph{right} and \emph{left exteriors},
\begin{align}
\begin{split}
    \text{Ext}_R &=\{(\tau,y,x):|\tau|<y,y\in(0,\pi/2]\} \\ \text{Ext}_L &=\{(\tau,y,x):|\tau|<-y,y\in[-\pi/2,0)\},
    \end{split}
\end{align}
as well as the \emph{future} and \emph{past interiors,}
\begin{align}
    \begin{split}
    \text{Int}_{\text{Fut}} &=\{(\tau,y,x):|y|<\tau,\tau\in(0,\pi/2]\}, \\ \text{Int}_{\text{Past}}&=\{(\tau,y,x):|y|<-\tau,\tau\in[-\pi/2,0)\},
    \end{split}
\end{align}
see Figure \ref{TwoSidedKruskalBHDiagram}.

The Schwarzschild coordinate description $(t,r,\bar{x})$ of the exterior regions, $\text{Ext}_{L,R}$, is related to the Kruskal coordinates through the change of coordinates
\begin{align}
\label{KruskaltoSchCoordTrans}
\begin{split}
    r=r_+\frac{\cos\tau}{\cos y},\qquad \qquad  \bar{x}=x, \\
    t=\frac{L^2}{2r_+}\ln\left[-\frac{\tan\left(\frac{\tau+y}{2}\right)}{\tan\left(\frac{\tau-y}{2}\right)}\right]=\frac{L^2}{2r_+}\ln\left[\frac{\sin y+\sin\tau}{\sin y-\sin\tau}\right].
\end{split}
\end{align}
Applying these coordinate transformations to the exterior region $\text{Ext}_{L}$ ($\text{Ext}_{R}$), the line element given in Eq. (\ref{KruSkLineElement}) takes the form of the (planar) BTZ line element in Schwarzschild coordinates which covers the left (right) exterior region,
\begin{equation}
\label{SchwarzBTZBH}
\begin{split}
    ds^2=-f(r)dt^2+\frac{dr^2}{f(r)}+\frac{r^2}{L^2}dx^2, \qquad \text{with } \qquad f(r)=\frac{r^2-r_+^2}{L^2},
\end{split}
\end{equation}
where $t\in \mathbb{R}$, $r>r_+$ and $x\in \mathbb{R}$. As we will see, using Schwarzschild coordinates is useful when discussing extremal surfaces which stay in the exterior regions, while Kruskal coordinates are beneficial to discuss extremal surfaces which start and end in the exterior regions, but cross through the interior regions.
A particularly useful redefinition of the radial coordinate is $r=\frac{L^2}{z}$ so that the exterior exists for $z\in(0,z_+)$. Using this new radial coordinate gives the following expression for the planar BTZ black hole line element,
\begin{equation}
\begin{split}
    &ds^2=\frac{L^2}{z^2}\left[-\tilde{f}(z)dt^2+\frac{dz^2}{\tilde{f}(z)}+dx^2\right]\qquad \text{with } \qquad\tilde{f}(z)=1-\frac{z^2}{z_+^2}.\\
\end{split}
\end{equation}

We now introduce an ETW brane in our setup. The ETW brane will be a timelike co-dimension one surface, which satisfies the following equations of motion involving the extrinsic curvature and induced metric\footnote{As we discussed in the introduction, this is when we impose Neumann BCs on the brane. In the Dirichlet case we simply view the brane as a cutoff with constant extrinsic curvature set by $y_{\text{Br}}$. All our results will be formulated in terms of $y_{\text{Br}}$.}
\begin{equation}
\begin{split}
    K_{ab}-h_{ab}+T_0h_{ab}=0.
\end{split}
\end{equation}
One can check that a hypersurface $y=y_{\text{Br}}=\text{constant}$, satisfies the equation of motion above with
\begin{equation}
    T_0=\sin y_{\text{Br}}.
\end{equation}
We will be interested in the case where the ETW brane is in the right exterior (i.e. $y_{\text{Br}}\in (0,\pi/2]$). With the help of the coordinate transformations in Eq. (\ref{KruskaltoSchCoordTrans}), we can also express the ETW brane $y=y_{\text{Br}}$ in the right exterior in Schwarzschild coordinates as
\begin{equation}
\label{BraneTraj}
    r(t)=r_+\sqrt{\frac{1-\sin^2y_{\text{Br}}\tanh^2\left(\frac{r_+t}{L^2}\right)}{\cos^2y_{\text{Br}}}}.
\end{equation}
We can see that in Schwarzschild coordinates the brane emerges from the past horizon at $t=-\infty$ and falls into the future horizon at $t=\infty$ with some nontrivial profile in the $tr$ plane in the right exterior.

\subsection{Subregions and RT Surfaces}
\label{RTSurfBTZSection}

For the sake of mathematical simplicity we will restrict ourselves to a very limited configuration of boundary and brane intervals. In particular, we will consider constant Schwarzschild time slice intervals of equal length on both the boundary and brane centered at $x=0$. We will only allow for a relative shift in time for the boundary and brane intervals. We will denote the constant time interval on the left boundary at Schwarzschild time $t=t_{\text{Bdry}}$ as $A_{\text{Bdry}}$. The constant time interval on the brane in the right exterior at Schwarzschild time $t=t_{\text{Br}}$ will be denoted by $A_{\text{Br}}$. For the configuration of intervals on the asymptotic boundary and brane described above there are two kinds of possible RT surfaces which are discussed in the next two subsections.

\subsubsection{Thermal Extremal Surfaces}
\label{ThermalExtremalSurfacesubsec}
Let us begin by finding extremal surfaces which stay within a single exterior region; we will call those \emph{thermal extremal surfaces}. Due to time-translation symmetry in each region, thermal extremal surfaces are located on a constant Schwarzschild time-slice. We denote the thermal surface in left exterior anchored to the boundary as $\chi_{\text{ther}}(A_{\text{Bdry}})$ and the thermal surface in the right exterior anchored to the brane as $\chi_{\text{ther}}(A_{\text{Br}})$. The extremal surfaces $\chi_{\text{ther}}(A_{\text{Bdry}})$ and $\chi_{\text{ther}}(A_{\text{Br}})$ are analogues to $\chi_{\text{dis}}(A)$ and $\chi_{\text{dis}}(B)$ in our previous AdS$_3$ setup which will be used to construct disconnected wedges in  our study of EWN. Thermal extremal surfaces extremize the area functional which in the BTZ geometry reads
\begin{equation}
\label{ThermalRTFunctional}
\begin{split}
    &\mathcal{A}[t,\dot{t},z,\dot{z};x]=\int dx\frac{L}{z}\sqrt{-\tilde{f}(z)\dot{t}^2+\frac{\dot{z}^2}{\tilde{f}(z)}+1}
\end{split}
\end{equation}
with the horizon location in Schwarzschild coordinates (with radial coordinate $z$). 
\begin{align}
    z_+=\frac{L^2}{r_+}, && \tilde{f}(z)=1-\frac{z^2}{z_+^2}.
\end{align}
With some work, detailed in Appendix \ref{ThermalRTSurfaceDeivationAppendix}, we obtain the following expression for a thermal RT surface anchored to a constant time slice ($t_L=t_{\text{Bdry}}$) interval $[-a, a]$ on the conformal boundary of $\text{Ext}_L$ (i.e. $\chi_{\text{ther}}(A_{\text{Bdry}})$)
\begin{align}
\label{ThermalRTSchBdryAnchor}
    t(x)=t_{\text{Bdry}} && z(x)=z_+\sqrt{1-\frac{\cosh^2\left(\frac{x}{z_+}\right)}{\cosh^2\left(\frac{a}{z_+}\right)}}.
\end{align}

Using the expression above for the thermal extremal surface anchored to a boundary interval we can also easily deduce the expression for the thermal extremal surface which is anchored to a constant time slice interval on the brane at time $t=t_{\text{Br}}$ in the right exterior (i.e. $\chi_{\text{ther}}(A_{\text{Br}})$). To do this we define the ``virtual interval'' associated to the interval on the brane denoted $\text{Vir}(A_{\text{Br}})$ which is defined to be an interval which is obtained by continuing the extremal surface anchored to the brane past the brane into the non-physical region. We can characterize this interval below,
\begin{equation}
\begin{split}
    \text{Vir}(A_{\text{Br}}) & =\{(t,x,z)|t=t_{\text{Br}},x\in [-a',a'],z=0\},\\
    a' & = z_+\operatorname{arccosh}\left[\frac{\cosh\left(\frac{a}{z_+}\right)}{\sqrt{1-\left(\frac{z_{Br}}{z_+}\right)^2}}\right]\geq a.
\end{split}
\end{equation}
The thermal extremal surface $\chi_{\text{ther}}(\text{Vir}(A_{\text{Br}}))$ is then described by
\begin{align}
    &t(x)=t_{\text{Br}}, && z(x)=z_+\sqrt{1-\frac{\cosh^2\left(\frac{x}{z_+}\right)}{\cosh^2\left(\frac{a'}{z_+}\right)}}.
\end{align}
It it straightforward to see that $\chi_{\text{ther}}(A_{\text{Br}})=\chi_{\text{ther}}(\text{Vir}(A_{\text{Br}}))\cap \mathcal{M}_{\text{phys}}$, where $\mathcal{M}_{\text{phys}}$ is the part of the Planar BTZ spacetime not cut off by the brane (this just amounts to restricting the range of $x$ from $[-a',a']$ to $[-a,a]$). This gives the following, final, expression for $\chi_{\text{ther}}(A_{\text{Br}})$, where $x\in[-a,a]$
\begin{align}
\label{ThermalRTAnchoredToBr}
t(x)=t_{\text{Br}}, && z(x)=z_+\sqrt{1-\frac{\cosh^2\left(\frac{x}{z_+}\right)}{\cosh^2\left(\frac{a}{z_+}\right)}\left(1-\frac{z_{Br}^2}{z_+^2}\right)}.
\end{align}

We can use our results to rewrite the same thermal extremal surface in Kruskal coordinates using the inverse of the transformation Eq.~\eqref{KruskaltoSchCoordTrans}. With a little bit of algebra and a careful distinction between the left and right exteriors the transformation from Schwarzschild to Kruskal coordinates reads
\begin{align}
\label{SchwToKruskalCoordTrans}
    \begin{split}
        \tan\left(v\right)=\pm e^{-\frac{t}{z_+}}\sqrt{\frac{z_+-z}{z_++z}}, & \qquad \tan\left(u\right)=\mp e^{\frac{t}{z_+}}\sqrt{\frac{z_+-z}{z_++z}}\\
        u=\frac{\tau+y}{2}, &  \qquad v=\frac{\tau-y}{2},
    \end{split}
\end{align}
where the different choice of signs give us the right and left exteriors. In particular, the left exterior is given when $-u,v>0$ and the right exterior is given when $-v,u>0$. With some work detailed in Appendix \ref{ThermalRTSurfaceDeivationAppendix} we can express $\chi_{\text{ther}}(A_{\text{Bdry}})$ as
\begin{equation}
\label{ThermalRTKruskalLeftExt}
    \begin{split}
        &\sin\left(\tau(x)\right)=-\frac{\cosh\left(\frac{x}{z_+}\right)\sinh\left(\frac{t_{\text{Bdry}}}{z_+}\right)}{\sqrt{\cosh^2\left(\frac{a}{z_+}\right)+\cosh^2\left(\frac{x}{z_+}\right)\sinh^2\left(\frac{t_{\text{Bdry}}}{z_+}\right)}},\\
        &\sin\left(y(x)\right)=-\frac{\cosh\left(\frac{t_{\text{Bdry}}}{z_+}\right)\cosh\left(\frac{x}{z_+}\right)}{\sqrt{\cosh^2\left(\frac{a}{z_+}\right)+\cosh^2\left(\frac{x}{z_+}\right)\sinh^2\left(\frac{t_{\text{Bdry}}}{z_+}\right)}}.
    \end{split}
\end{equation}
Similarly, $\chi_{\text{ther}}(A_{\text{Br}})$ is given by
\begin{equation}
\label{ThermalRTRightExtBr}
\begin{split}
    &\sin\left(\tau(x)\right)=\sin[v(x)+u(x)]=\frac{\sqrt{1-\frac{z_{Br}^2}{z_+^2}}\cosh\left(\frac{x}{z_+}\right)\sinh\left(\frac{t_{\text{Br}}}{z_+}\right)}{\sqrt{\cosh^2\left(\frac{a}{z_+}\right)+\left(1-\frac{z_{Br}^2}{z_+^2}\right)\cosh^2\left(\frac{x}{z_+}\right)\sinh^2\left(\frac{t_{\text{Br}}}{z_+}\right)}},\\
    &\sin(y(x))=\sin[u(x)-v(x)]=\frac{\sqrt{1-\frac{z_{Br}^2}{z_+^2}}\cosh\left(\frac{x}{z_+}\right)\cosh\left(\frac{t_{\text{Br}}}{z_+}\right)}{\sqrt{\cosh^2\left(\frac{a}{z_+}\right)+\left(1-\frac{z_{Br}^2}{z_+^2}\right)\cosh^2\left(\frac{x}{z_+}\right)\sinh^2\left(\frac{t_{\text{Br}}}{z_+}\right)}}.
\end{split}
\end{equation}
\subsubsection{Connected Extremal Surfaces}
\label{ConnectedSurfaceBTZsection}
\emph{Connected extremal surfaces} are extremal surfaces that connect the boundary and brane interval and pass through the interior of the black hole. We will denote them as $\chi_{\text{con}}(A_{\text{Bdry}}\cup A_{\text{Br}})$.

To find the connected extremal surfaces we extremize the area functional in Kruskal coordinates,
\begin{equation}
\label{ConnRTKruskFunc}
     \mathcal{A}[\tau,\dot{\tau},x,\dot{x};y]=\int dy \frac{\sqrt{L^2\left(-\dot{\tau}^2+1\right)+\frac{r_+^2}{L^2}\cos^2\tau\dot{x}^2}}{\cos y}.
\end{equation}
We can simplify the functional above by recalling that $A_{\text{Br}}$ and $A_{\text{Bdry}}$ are intervals with the same $x$-coordinate for the endpoints. This means that a connected surface that starts at $x=\pm a$ also needs to end at $x=\pm a$. This is only possible if $\dot{x}=0$. After setting $\dot{x}=0$ we arrive at the task of extremizing the functional
\begin{equation}
\label{ConnRTKruskSimpleVerFunc}
     \mathcal{A}[\tau,\dot{\tau},x,\dot{x};y]=\int dy \frac{\sqrt{L^2\left(-\dot{\tau}^2+1\right)}}{\cos y}.
\end{equation}
With the work detailed in Appendix \ref{ConnectedRTSurfaceDeivationAppendix}, we arrive at the following general expression for the extremal curve(s) that go through the horizon and connects the brane and boundary intervals
\begin{equation}
\label{ConnRTSurfKruskCord}
    x(y)=\pm a, \qquad 
     \tau(y)=\tau_{\text{Bdry}}+\text{arcsin}\left(\frac{c_\tau}{\sqrt{1+c_\tau^2}}\sin y\right)+\text{arcsin}\left(\frac{c_\tau}{\sqrt{1+c_\tau^2}}\right).
\end{equation}
For future convenience, we define $\tau_{\text{Br}}=\tau(y_{\text{Br}})$. We can see that for a given a fixed value of $\tau_{\text{Bdry}}$ (in the left exterior) the constant parameter $c_\tau\in \mathbb{R}$ will determine what $\tau_{\text{Br}}$ (in the right exterior) will be and vice-versa. In the special case of $c_\tau=0$ the connected surface has a trivial horizontal profile in Kruskal time coordinates. As $c_{\tau}$ tends to $\pm\infty$ the extremal curves become null.

When we do our analysis of entanglement wedge nesting we will find it useful to work in Schwarzschild coordinates. To find the expressions for the connected extremal curves we start with Eq. (\ref{ConnRTSurfKruskCord}) and make use of the coordinate transformation that takes exterior regions in Kruskal coordinates to Schwarzschild coordinates given by Eq. (\ref{KruskaltoSchCoordTrans}).
The result is the following expression (see Appendix \ref{ConnectedRTSurfaceDeivationAppendix} for details)
\begin{align}
\label{RightExtConnRTSch}
        x(t)=\pm a, &&
        \left(\frac{z(t)}{z_+}\right)^2=1-\frac{1-\frac{z_{Br}^2}{z_{+}^2}}{\left[\cosh\left(\frac{|\Delta t|}{z_+}\right)+\left|\frac{B}{A}\right|\sinh\left(\frac{|\Delta t|}{z_+}\right)\right]^2},
\end{align}
with constants given by
\begin{align}
\label{constantsBTZ}
\begin{split}
        A&=\cos\tau_*\sinh\left(\frac{t_{\text{Br}}}{z_+}\right)-c_t\cosh\left(\frac{t_{\text{Br}}}{z_+}\right)\\
        B&=\cos\tau_*\cosh\left(\frac{t_{\text{Br}}}{z_+}\right)-c_t\sinh\left(\frac{t_{\text{Br}}}{z_+}\right)\\
        \tau_*&=\tau_{\text{Br}}-\arcsin(c_t\sin y_{\text{Br}}), \qquad c_t=\frac{c_\tau}{\sqrt{1+c_\tau^2}}, \qquad \Delta t=t-t_{\text{Br}}.
        \end{split}
\end{align}
The absolute value signs in the expressions are a book-keeping tool which circumvents the issue of worrying about the specific signs we need to use for $\Delta t$ for a given $c_t$.

\subsection{Construction of Entanglement Wedges}
\label{BTZWedgeSection}
\begin{figure}[t]
\centering
\begin{subfigure}{0.48\textwidth}
\includegraphics[width=40mm]{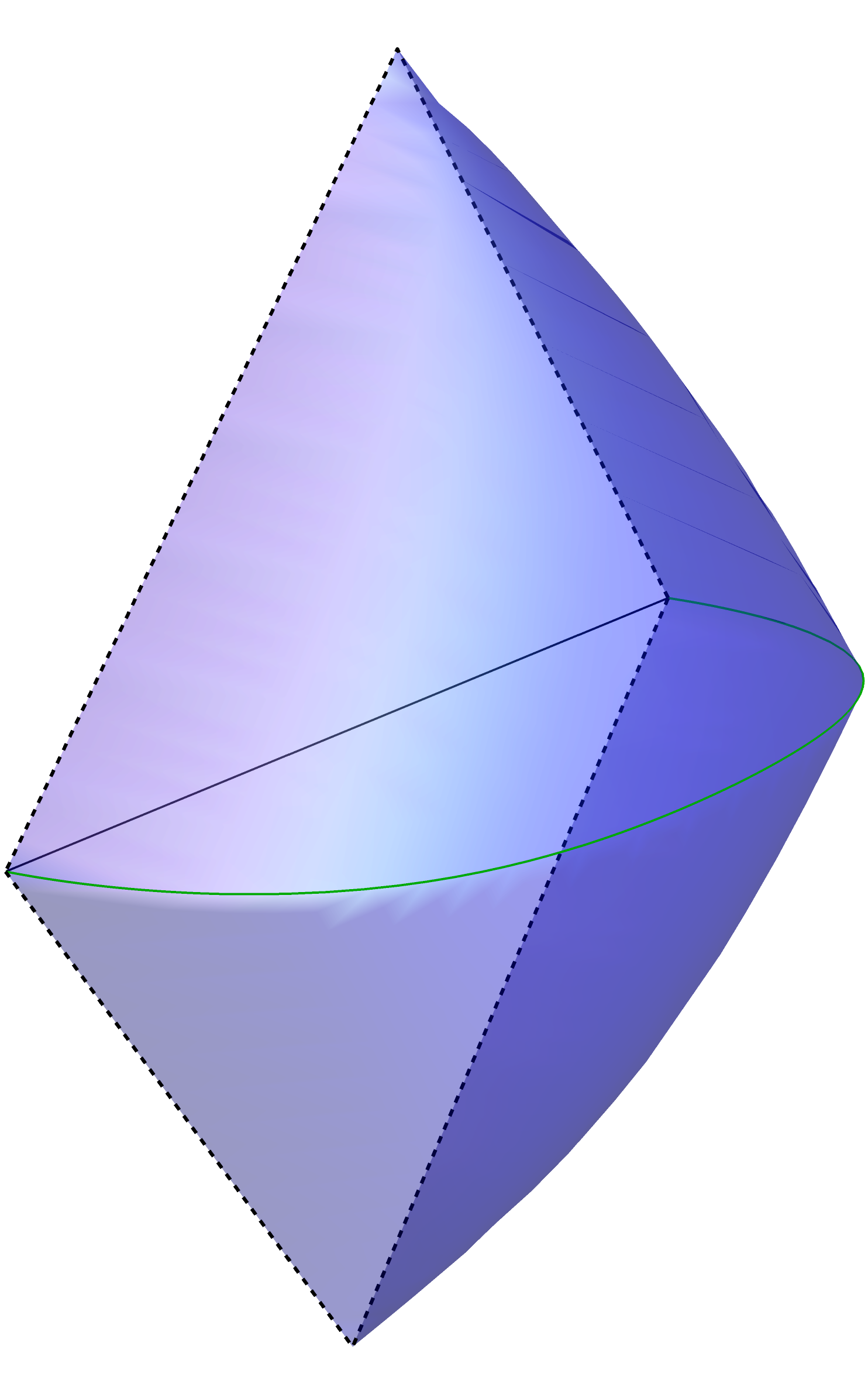}
\subcaption{\label{EntangWedgeBdryBTZ}}
\end{subfigure}
\begin{subfigure}{0.48\textwidth}
\includegraphics[width=48mm]{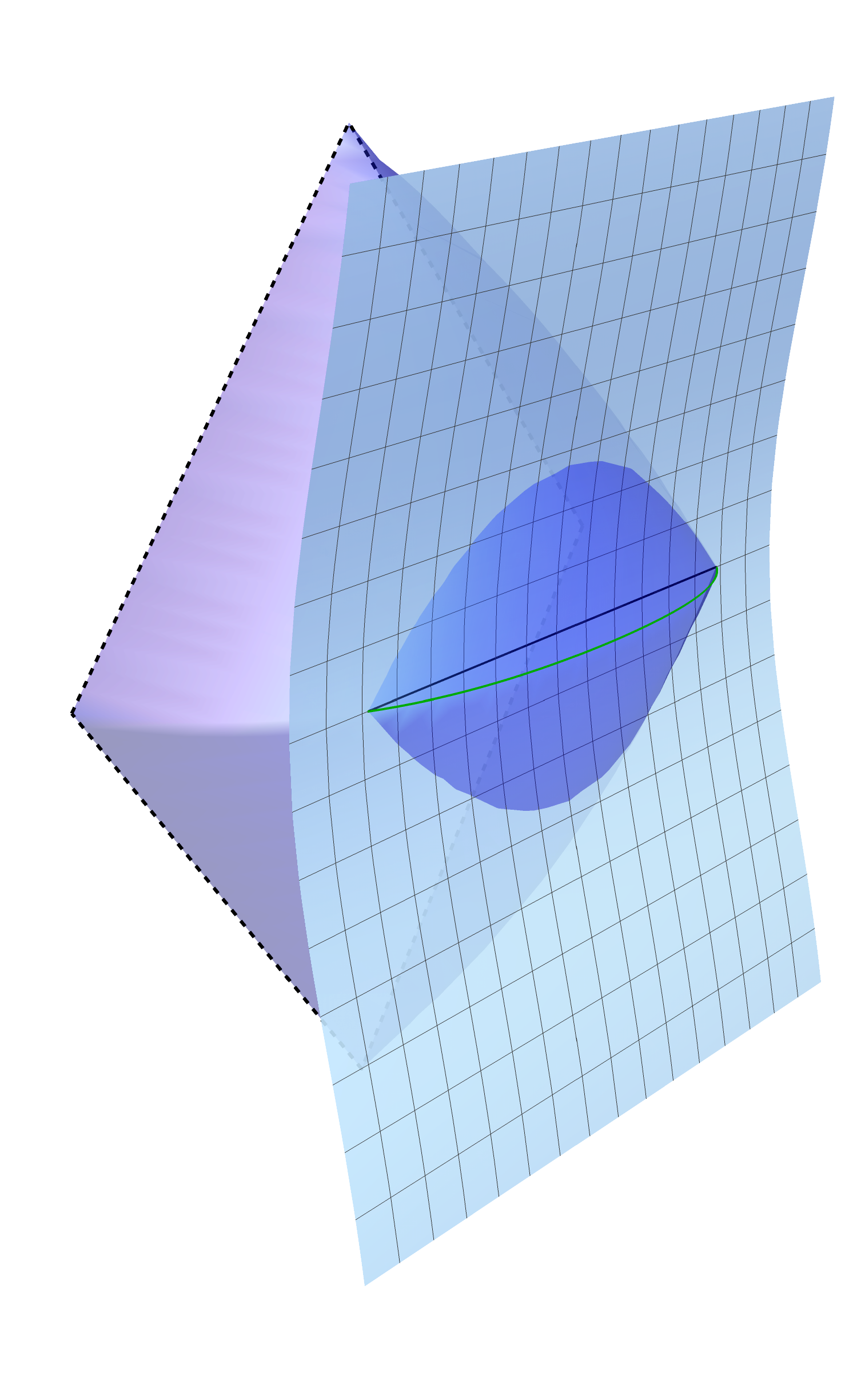}
\subcaption{\label{EntangWedgeBraneBTZ}}
\end{subfigure}
    \caption{\textbf{(a)} An example of $\mathcal{W}_E(A_{\text{Bdry}})$. The intersection of the past and future cones is the RT surface $\chi_{\text{ther}}(A_{\text{Bdry}})$ shown in green. The dotted ``diamond'' is the causal diamond on the boundary associated with the interval $A_{\text{Bdry}}$ which horizontally splits the diamond to the future and past of the interval. Unlike in AdS$_3$ vacuum, the RT surface deeper in the bulk is ``flattened'' due to the black hole geometry. 
    \textbf{(b)} An example of $\mathcal{W}_E(A_{\text{Br}})$ (purple) and an ETW brane (blue). The full wedge which extends behind the brane is what we called $\mathcal{W}_{E}(\text{Vir}(A_{\text{Br}}))$ in Appendix \ref{AppendixEntangleWedgesConstTimeIntervalsBTZ}. The actual entanglement wedge $\mathcal{W}_E(A_{\text{Br}})$ in the physical subregion is the smaller piece in front of the ETW brane. }
\end{figure}

We now turn to discussing the entanglement wedges associated to the regions $A_{\text{Bdry}}$, $A_{\text{Br}}$, and $A_{\text{Bdry}}\cup A_{\text{Br}}$. Starting with $A_{\text{Bdry}}$, we use $\Sigma_{A_{\text{Bdry}}}$ to denote a partial Cauchy surface with boundary $\partial \Sigma_{A_{\text{Bdry}}}=A_{\text{Bdry}}\cup \chi_{\text{ther}}(A_{\text{Bdry}})$. We define the entanglement wedge $\mathcal{W}_E(A_{\text{Bdry}})$ of $A_{\text{Bdry}}$ as the domain of dependence of $\Sigma_{A_{\text{Bdry}}}$. A convenient prescription which can be used to explicitly construct $\mathcal{W}_E(A_{\text{Bdry}})$ involves the family of null geodesics beginning on $\chi_{\text{ther}}(A_{\text{Bdry}})$ which are also orthogonal to $\chi_{\text{ther}}(A_{\text{Bdry}})$ and evolve them towards the boundary. The set of these null geodesics enclose a co-dimension 0 region which is exactly $\mathcal{W}_E(A_{\text{Bdry}})$ as shown in Figure \ref{EntangWedgeBdryBTZ}. 

The points $(t,x,z)\in\mathcal{W}_E(A_{\text{Bdry}})$ which belong to the entanglement wedge obey
\begin{equation}
\label{NullBdryOfEntangleWedgeBTZ}
     a-|t-t_{\text{Bdry}}|  > z_+\text{arctanh}\left[\sqrt{1-\frac{1-\frac{z^2}{z_+^2}}{\cosh^2\left(\frac{x}{z_+}\right)}}\right].
\end{equation}
where equality holds for the boundary of $\mathcal{W}_E(A_{\text{Bdry}})$ which is null (see Appendix \ref{AppendixEntangleWedgesConstTimeIntervalsBTZ}).

In an analogous manner we define the entanglement wedge associated with the region $A_{\text{Br}}$ on the brane. We define $\Sigma_{A_{\text{Br}}}$ as the partial Cauchy slice whose boundary is given by $\partial\Sigma_{A_{\text{Br}}}=A_{\text{Br}}\cup\chi_{\text{ther}}(A_{\text{Br}})$. Then the entanglement wedge of $A_{\text{Br}}$, denoted $\mathcal{W}_E(A_{\text{Br}})$, is simply the domain of dependence of the $\Sigma_{A_{\text{Br}}}$. Furthermore, a more explicit construction of the of the wedge can be made by considering the a family of null orthogonal geodesics beginning on $\chi_{\text{ther}}(A_{\text{Br}})$ and evolve towards the boundary. The set of these null geodesics enclose a co-dimension 0 region which is exactly $\mathcal{W}_E(A_{\text{Br}})$ as shown in Figure \ref{EntangWedgeBraneBTZ}.

As shown in Appendix \ref{AppendixEntangleWedgesConstTimeIntervalsBTZ} the set of points in $\mathcal{W}_E(A_{\text{Br}})=\mathcal{W}_E(\text{Vir}(A_{\text{Br}}))\cap \mathcal{M}_{\text{phys}}$, satisfy
\begin{equation}
\label{EntWedgeBTZtoBr}
    \begin{split}
    &a'-|t-t_{\text{Br}}|>z_+\text{arctanh}\left[\sqrt{1-\frac{1-\frac{z^2}{z_+^2}}{\cosh^2\left(\frac{x}{z_+}\right)}}\right], \qquad  a'=z_+\text{arccosh}\left[\frac{\cosh\left(\frac{a}{z_+}\right)}{\sqrt{1-\frac{z_{Br}^2}{z_+^2}}}\right],
    \end{split}
\end{equation}
where we make the implicit restriction on $(t,x,z)$ to be in $\mathcal{M}_{\text{phys}}$.

Finally, we turn to the definition of the entanglement wedge $\mathcal{W}_E(A_{\text{Bdry}}\cup A_{\text{Br}})$. As in the Poincare case, $\mathcal{W}_E(A_{\text{Bdry}}\cup A_{\text{Br}})$ can be in one of two possible phases which we will call the connected and disconnected phase, depending on whether the RT surfaces connect the asymptotic boundary with the brane or not. 
In the disconnected phase we have $\mathcal{W}_E(A_{\text{Bdry}}\cup A_{\text{Br}})=\mathcal{W}_E(A_{\text{Bdry}})\cup\mathcal{W}_E(A_{\text{Br}})$ which is simply the union of the separate entanglement wedges of $A_{\text{Bdry}}$ and $A_{\text{Br}}$ which we already discussed. 
In the connected phase we define $\Sigma_{A_{\text{Bdry}}\cup A_{\text{Br}}}$ as the partial Cauchy slice with a boundary given by $\partial\Sigma_{A_{\text{Bdry}}\cup A_{\text{Br}}}=\chi_{\text{con}}(A_{\text{Bdry}}\cup A_{\text{Br}})\cup A_{\text{Bdry}}\cup A_{\text{Br}}$. Then using this partial Cauchy slice we define the connected phase entanglement wedge as the domain of dependence of $\Sigma_{A_{\text{Bdry}}\cup A_{\text{Br}}}$.

The explicit construction of $\mathcal{W}_E(A_{\text{Bdry}}\cup A_{\text{Br}})$ in the connected phase can be done in a manner similar to what we outlined in Poincare AdS$_3$, although the technical details will be different. Since for the purposes of finding constraints using EWN we only need to explicitly characterize the disconnected wedges, we refrain from explicitly characterizing the connected wedge here.

\subsection{Condition for EWN}
\label{EWNSuffConBTZPlanarSec}
In this section we will derive constraints on the parameters defining the separation of boundary and brane subregions in Kruskal time such that $\mathcal{W}_E(A_{\text{Br}})\subset \mathcal{W}_E(A_{\text{Br}}\cup A_{\text{Bdry}})$ in the connected phase and also provide the geometric interpretation of the constraints. 

Just as in the Poincare AdS$_3$ analysis in Section \ref{PoincareAdS3SuffCondSec} a sufficient condition for EWN is given by requiring that all null geodesics originating from $\chi_{\text{con}}(A_{\text{Bdry}}\cup A_{\text{Br}})$ are unable to reach $\mathcal{W}_E(A_{\text{Br}})$, i.e. no causal signals originating from $\chi_{\text{con}}(A_{\text{Bdry}}\cup A_{\text{Br}})$ can influence events in $\mathcal{W}_E(A_{\text{Br}})$. Due to the trivial trajectory of the connected surface in the $xz$-plane, it suffices to ask when the points on the connected surface are spacelike separated from $\mathcal{W}_E(\text{Vir}(A_{\text{Br}}))$. The reason is that in our particular setup of brane and boundary intervals the connected surface always lies entirely the $\mathcal{R}_M$ region\footnote{Since we are dealing with all the surfaces lying in $\mathcal{R}_M$ the sufficient condition we derive will also be necessary analogous to the AdS$_3$ case.} (depicted in Figure \ref{BTZRMLRegions}). 

\begin{figure}[t]
\centering
\includegraphics[width=100mm]{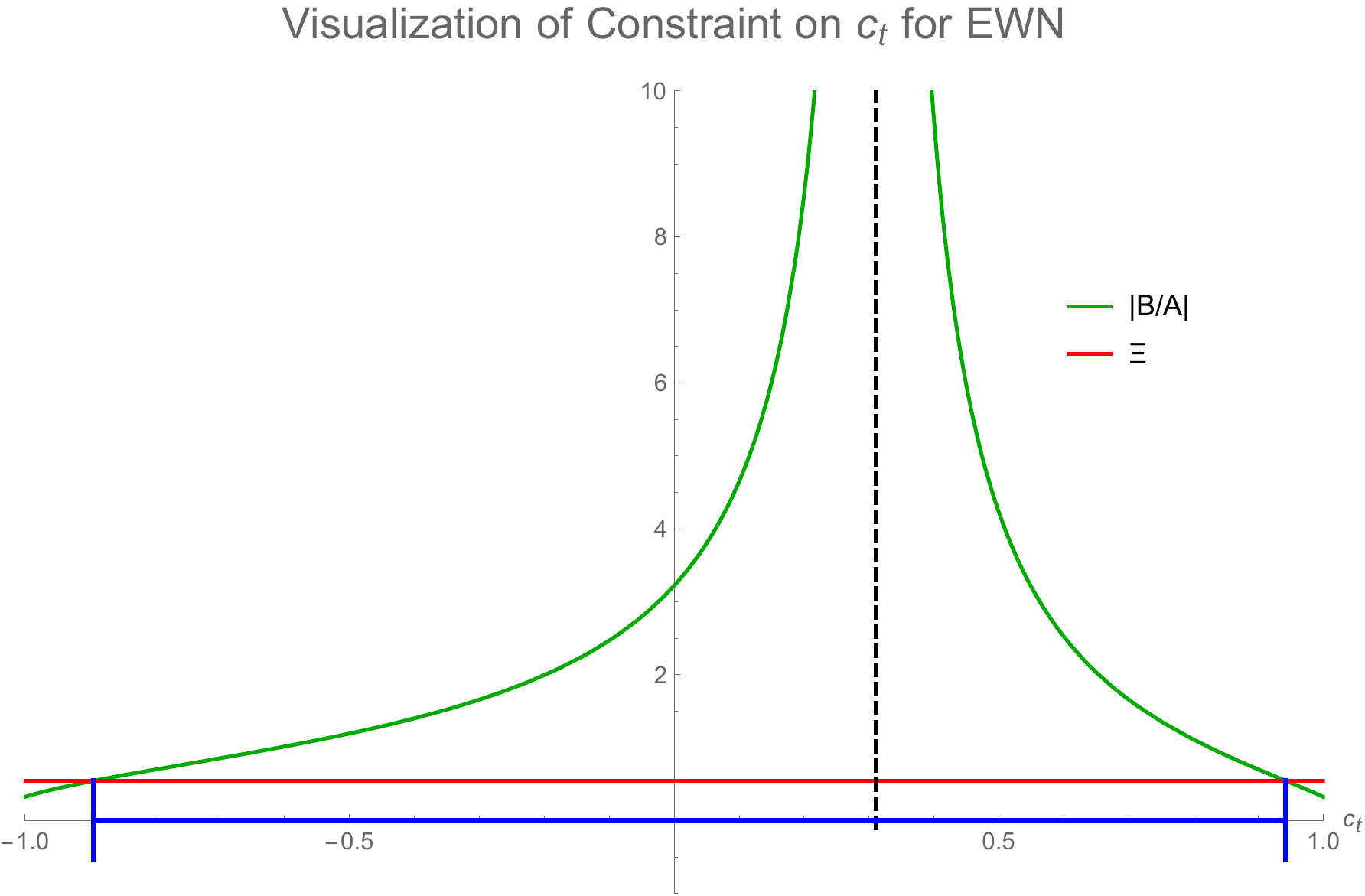}
    \caption{Visualization of the inequality given in Eq. (\ref{ManipulatedSuffConBTZEWN}). The green line is $|B/A|$ (it has a singularity at $c_t=c_0=\frac{\sin\tau_{\text{Br}}}{\sin y_{\text{Br}}}$) and the red line is $\Xi$. Entanglement wedge nesting is respected when the green line is above the red line. This naturally gives an interval of values for $c_t$ in which entanglement wedge nesting is satisfied (depicted by the blue interval on the horizontal $c_t$ axis). The lower and upper bounds of the interval are $c_-$ for the lower bound and $c_+$ for the upper bound whose expressions are generally given by Eq. (\ref{ctIntervalGeneral}). The smallest interval comes from the limit when $a\to\infty$ (i.e. size of the brane interval diverges), in that case $c_{\pm}$ is given by Eq. (\ref{ctIntervalTightest}). For this particular plot we set $y_{\text{Br}}=2\pi/5$, $\tau_{\text{Br}}=3/10$, and $a/z_+=1/2$. This gives $c_0\approx 0.31$ (location of black dashed asymptote), $c_{-}\approx -0.89$ (left bound of blue interval), and $c_+\approx 0.94$ (right bound of blue interval).  \label{EWNBTZIneqVis}}
\end{figure}

The points on the connected surface are spacelike (or possibly null) separated from points in $\mathcal{W}_E(A_{\text{Br}})$ if
\begin{equation}
\label{NiaveCondForSLSepGenPoints}
    \begin{split}
        a'+|t-t_{\text{Br}}|\leq z_+\text{arctanh}\left[\sqrt{1-\frac{1-\frac{z^2}{z_+^2}}{\cosh^2\left(\frac{x}{z_+}\right)}}\right],
    \end{split}
\end{equation}
where $a'$ was defined in Eq. \ref{EntWedgeBTZtoBr}. Points that satisfy this inequality cannot influence events in $\mathcal{W}_E(A_{\text{Br}})$, guaranteeing that $\mathcal{W}_E(A_{\text{Br}})\subseteq \mathcal{W}_E(A_{\text{Bdry}}\cup A_{\text{Br}})$. Substituting the location of the connected RT surface, Eg.\ \eqref{RightExtConnRTSch} into Eq.\ \eqref{NiaveCondForSLSepGenPoints} and rearranging the results as discussed in Appendix \ref{AppendixNaiveConBTZEWN} we obtain
\begin{align}
    \label{ManipulatedSuffConBTZEWN}
    \left|\frac{B}{A}\right|=\left|\frac{\cos\tau_*\cosh\left(\frac{t_{\text{Br}}}{z_+}\right)-c_t\sinh\left(\frac{t_{\text{Br}}}{z_+}\right)}{\cos\tau_*\sinh\left(\frac{t_{\text{Br}}}{z_+}\right)-c_t\cosh\left(\frac{t_{\text{Br}}}{z_+}\right)}\right|\geq \sqrt{1-\frac{1-\frac{z_{Br}^2}{z_+^2}}{\cosh^2\left(\frac{a}{z_+}\right)}} \equiv \Xi(a,z_{Br}).
\end{align}
Whenever this inequality is satisfied, EWN holds. It is possible to derive a specific constraint on the parameter $c_t$ which characterizes the connected surface. If we fix the parameters $y_{\text{Br}}$, $\tau_{\text{Br}}$, and $a$ and allow $c_t$ to vary (i.e. we fix the location and size of the brane interval and move the boundary interval up and down in time) then we can visualize the constraint with the help of the listed facts about the left hand side of \eqref{ManipulatedSuffConBTZEWN} discussed in Appendix \ref{ConnectedRTSurfaceDeivationAppendix}, c.f.\ Figure \ref{EWNBTZIneqVis}.

As we can see the right-hand side of the inequality in Eq. (\ref{ManipulatedSuffConBTZEWN}) is constant. The left-hand side explicitly depends on $c_t$, has a vertical asymptote at $c_t=c_0$ and is non-zero and well defined everywhere else in the regime $c_t\in(-1,1)$. Therefore, the inequality is only satisfied for some values $c_t\in[c_-,c_+]$ with $-1<c_-<c_0<c_+<1$. The specific values of $c_{\pm}$ can be obtained by solving the equations 
\begin{equation}
\begin{split}
    &\frac{B}{A}\bigg\vert_{c_t=c_{\pm}}=\frac{X_\pm\cos^2 y_{\text{Br}}\sin\tau_{\text{Br}}-\sqrt{1-X_\pm^2}\sin^2y_{\text{Br}}\cos\tau_{\text{Br}}}{\left(X_\pm\cos\tau_{\text{Br}}-\sqrt{1-X_\pm^2}\sin\tau_{\text{Br}}\right)\sin y_{\text{Br}}\cos\tau_{\text{Br}}}=\mp\Xi(a,z_{Br})\\
    &X_{\pm}=c_\pm\sin y_{\text{Br}},\\
\end{split}
\end{equation}
and solutions are,\footnote{Details of getting solution given in Appendix \ref{AppendixNaiveConBTZEWN}.}
\begin{equation}
    \label{ctIntervalGeneral}
    \begin{split}
        &c_-=-\left[1+\frac{\left[\cos(2\tau_{\text{Br}})-\cos(2y_{\text{Br}})\right]\left[\Xi^2\cos^2\tau_{\text{Br}}-\cos^2y_{\text{Br}}\right]}{2\cos^2\tau_{\text{Br}}\left(\sin y_{\text{Br}}-\Xi\sin\tau_{\text{Br}}\right)^2}\right]^{-\frac{1}{2}}\\
        &c_+=\left[1+\frac{\left[\cos(2\tau_{\text{Br}})-\cos(2y_{\text{Br}})\right]\left[\Xi^2\cos^2\tau_{\text{Br}}-\cos^2y_{\text{Br}}\right]}{2\cos^2\tau_{\text{Br}}\left(\sin y_{\text{Br}}+\Xi\sin\tau_{\text{Br}}\right)^2}\right]^{-\frac{1}{2}}.\\
    \end{split}
\end{equation}
The tightest interval occurs when we take the limit of the expressions above as $\Xi\to 1$ (i.e. when $a\to\infty$) to give,\footnote{In Appendix \ref{BTZConnCondition} we show that the connected phase will dominate in this regime. Intuitively, the thermal extremal surface area goes to infinity while the connected surface area remains finite as $a\to\infty$.}
\begin{equation}
    \label{ctIntervalTightest}
    \begin{split}
       c_-=-\frac{1}{\sqrt{1+\left(\frac{\sin\tau_{\text{Br}}+\sin y_{\text{Br}}}{\cos\tau_{\text{Br}}}\right)^2}}, \qquad 
       c_+=\frac{1}{\sqrt{1+\left(\frac{\sin\tau_{\text{Br}}-\sin y_{\text{Br}}}{\cos\tau_{\text{Br}}}\right)^2}}.\\
    \end{split}
\end{equation}

\begin{comment}
\subsection{Constraints from EWN}
\label{DiscussionOfBTZResultSec}
\end{comment}

Now that we have derived a condition for when EWN is respected in the connected phase we can continue and describe a nice geometric interpretation of what the tightest constraints given in Eq. (\ref{ctIntervalTightest}) mean. First lets recall that a priori we just required that $|c_t|<1$ so the connected RT surfaces are spacelike. Now suppose we have a connected curve which starts on the left boundary at some Kruskal time $|\tau_{\text{Bdry}}|<\pi/2$. For $c_t=0$ the connected surface in Kruskal coordinates has a trivial horizontal profile and will end on the brane in the right exterior at the same Kruskal time as it started with at the boundary. However for more general values of $c_t$ less than unity one can imagine that there are certain values of $c_t$ where the surface would go through the past or future singularity before it hits the brane. 
\begin{figure}[h!]
\centering
\includegraphics[width=130mm]{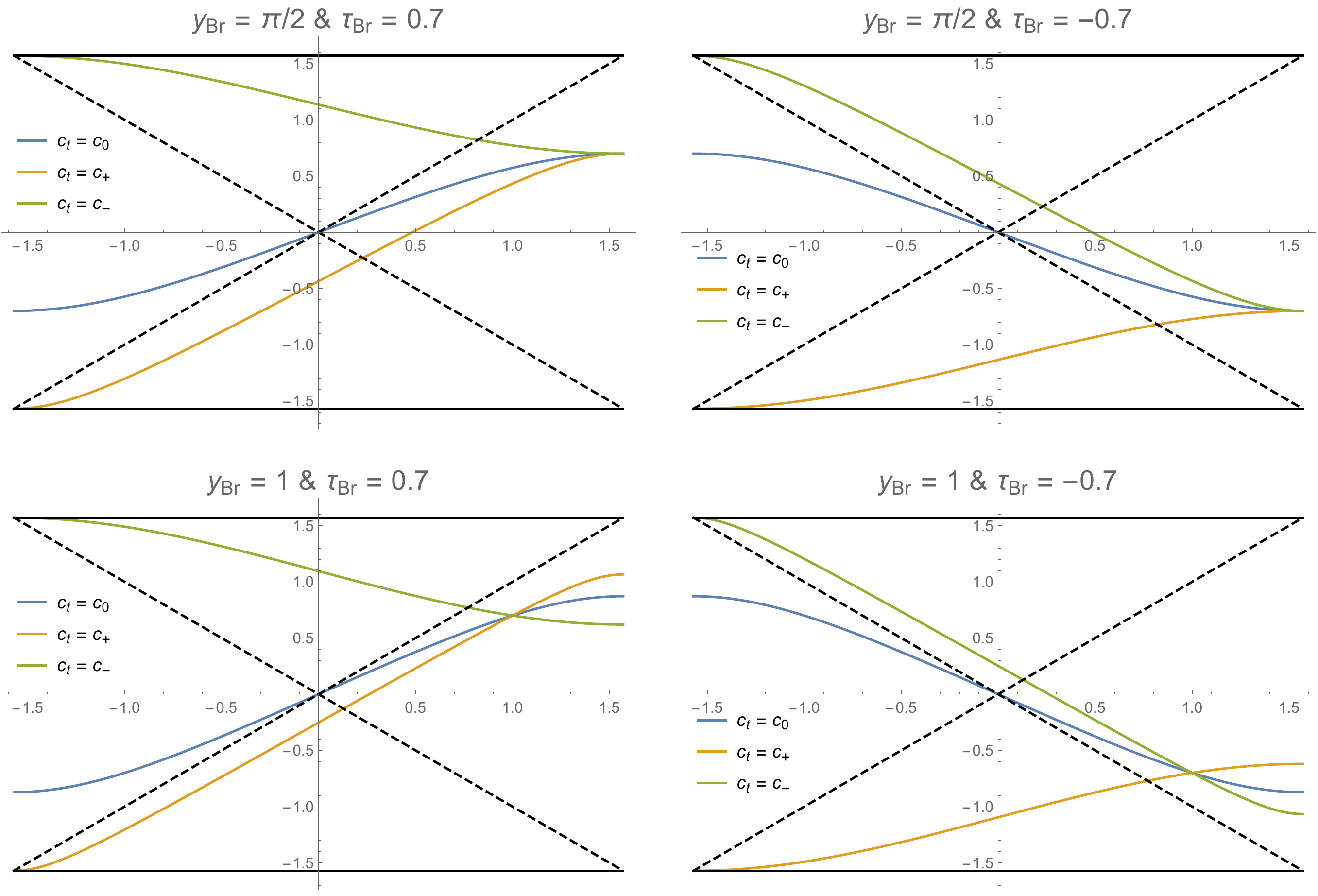}
    \caption{This figure shows connected surfaces which correspond to $c_{\pm}$ given in Eq. (\ref{ctIntervalTightest}) as well as $c_0$ which goes through the bifurcate horizon for different values of $\tau_{\text{Br}}$ and $y_{\text{Br}}$. \label{CriticalValConSurPlot}}
\end{figure}
It turns out the condition that EWN hold for intervals on the brane in the right exterior exactly excludes the configurations that would allow the connected surface to go through the singularity. In particular, if we fix the location of a constant time interval in the right exterior on the brane (i.e. fix $\tau_{\text{Br}}$ and $y_{\text{Br}}$) then the condition that EWN holds translates to
\begin{equation}
    \label{jtboundsinter}
    c_-=-\frac{1}{\sqrt{1+\left(\frac{\sin\tau_{\text{Br}}+\sin y_{\text{Br}}}{\cos\tau_{\text{Br}}}\right)^2}} \leq c_t\leq \frac{1}{\sqrt{1+\left(\frac{\sin\tau_{\text{Br}}-\sin y_{\text{Br}}}{\cos\tau_{\text{Br}}}\right)^2}}=c_+.
\end{equation}
In Figure \ref{CriticalValConSurPlot} we plot the connected surface in Kruskal coordinates in the $\tau y$-plane for $c=c_+$, $c=c_-$, and $c=c_0$ and see that the condition prevents the connected surface from going through the future and past singularities.

This is consistent with expectations from \cite{Lee:2022efh} where the authors argued that connected RT surfaces that pass the ``singularity'' of a planar BTZ back hole cross, so that the orientation of the entanglement wedge flips. In practice, this means that the two connected components become timelike separated. Just like in Section \ref{ChapterDerivingConstraintForEWN}, we can conclude that a violation of EWN comes from disconnected parts of the RT surface being able to causally influence each other.

\section{Concluding Remarks}
\label{ConclusionDiss}

In this work, we have used basic examples to study the idea of associating an entanglement wedge for subregions located on cutoff surfaces and analyzed conditions for entanglement wedge nesting (EWN) to hold. More specifically, we focused on models where the cutoff surface takes the form of an end-of-the-world (ETW) brane in three dimensions, although our general prescription is independent of this. The prescription used to construct the entanglement wedge was to consider boundary and cutoff subregions, $A$ and $B$, and naively apply the RT formula. Throughout the paper we chose $A$ and $B$ to be spacelike separated with respect to the bulk metric. Within this prescription, care has to be taken near the cutoff surface.

We demonstrated that even though the subregions $A$ and $B$ are spacelike separated, EWN can still fail. This is in contrast to standard AdS/CFT holography (which we recover from our results by sending the cutoff to the asymptotic boundary, i.e. $\theta_0\to\pi/2$ in the AdS example and $y_{\text{Br}}\to\pi/2$ in the BTZ example) where spacelike separation of $A$ and $B$ implies EWN. Instead, to ensure EWN holds one needs stronger conditions, which we derived for two exemplary setups in Sections \ref{ChapterDerivingConstraintForEWN} and \ref{EWNBTZChapterOfPaper}. The failure of EWN was caused in both cases by an unexpected timelike separation of RT surfaces. We would like to interpret this feature, where RT surfaces are not spacelike separated, even though the associated regions on the boundary or cutoff are, as a bulk manifestation of the non-local physics in the theory on the cutoff surface. It indicates that non-locality can make spacelike separated regions on the cutoff non-independent. We also related our construction to the restricted maximin procedure of \cite{Grado-White:2020wlb}. We pointed out that while restricted maximin automatically satisfies EWN for spacelike separated subregions, it can give rise to RT surfaces with non-trivial profiles in time for constant time subregions in static geometries. We expect restricted maximin and our prescription agree when all RT surfaces are spacelike separated (or equivalently when EWN is respected).

Our understanding of holography at a finite cutoff is poor and we hope that a careful study of properties of entanglement wedges in the presence of a cutoff can shed light on the properties of holographic duals to finite subregions. It would in particular be interesting to generalize our results to more complicated geometries and interval configurations, also in higher dimensions. Optimistically, this could provide a set of rules which determine under which conditions cutoff and boundary subregions can be treated as truly independent. It is interesting to note that violations of the conditions we obtained for EWN throughout this paper have always started in arbitrarily small bulk neighborhoods of the entangling surfaces. Perhaps this will always be the case if a non-trivial violation of EWN occurs, such that it could be sufficient to focus on those regions which would significantly simplify the analyses. Even if this is not generally the case one might at least formulate some concrete necessary conditions for EWN to hold again, this may provide insight as to how one can naturally identify independent subregions in cutoff holography. It may also be interesting to explore is how much our conclusions are affected through the particular model of cutoff surface we choose. Here we chose surfaces that are often employed in doubly holography setups and one might wonder if things may change for more general choices of cutoff surfaces.\footnote{It was pointed out in \cite{Omiya:2021olc} that the ETW branes considered in double holography fails to satisfy the Gao-Wald theorem \cite{Gao:2000ga}. We believe that our results can be regarded as a consequence of such violations of the Gao-Wald theorem. Based on this, we might speculate that more general cutoff surfaces also feature such violations of the Gao-Wald theorem and therefore our conclusions may carry over in a similar manner in more general contexts.}  

Another suggestive feature which warrants further investigation can be seen in Figure \ref{WedgeIntBrane}. This figure shows how the intersection of the entanglement wedge $\mathcal W_E(B)$ with the cutoff brane grows as we include further away regions $\mathcal W_E(B) \to \mathcal W_E(A \cup B)$ (in this case on the asymptotic boundary). It is natural to interpret the intersection of the entanglement wedge with the cutoff surface as a domain of dependence, i.e., the spacetime subregion within which we can predict the state uniquely given a state on $B$ (or $A \cup B$). In a standard local quantum field theory, given a subsystem $B$, we can predict the state within the naive domain of dependence of $B$. Having access to a spacelike separated subregion $A$ does not affect this. For the theory on the cutoff, given $B$ we do not have access to the full naive domain of dependence. However, including far away regions increases the domain in which we can predict the state. Notably, as discussed, e.g., in \cite{Neuenfeld:2023svs}, regions close to the boundary of the naive domain of dependence of $B$ correspond to UV information about the state on $B$. This seems to indicate that for holographic cutoff theories IR information is stored locally, while UV information is stored non-locally in stark contrast to naive expectations. It would be very interesting to find a model which realizes such a behavior explicitly.

Finally, another interesting avenue of research goes back to our discussion near the end of Section \ref{BackgroundMotiSec} of boundary conditions on the brane. In particular, when we impose Neumann boundary conditions on the brane we should think of the constraints that we obtain from EWN as constraints on brane world gravity. The rough procedure would be to understand how the location of the brane subregion, $B$, depends on various gravitational couplings in the brane-world theory of gravity.\footnote{We fix the location of $A$ on a conformal boundary and location of $B$ is determined dynamically by understanding where the connected RT surface ends on the brane which should depend on the gravitational couplings on the brane.} By precisely understanding this relation one could then straightforwardly translate the constraints we found here to constraints on gravity couplings. We leave such an investigation for later work.    

\acknowledgments
I thank Niayesh Afshordi, Dominik Neuenfeld, Donald Marolf, Xi Dong, Vincent Chen, Samir Mathur, Bin Guo, Hao Geng, and Takato Mori for helpful discussions and comments as well as Dominik Neuenfeld for initial collaboration. 

The early stages of this work have also appeared in Krishan Saraswat's Ph.D. thesis \cite{Saraswat:2023knb}, which was supported by the University of Waterloo, Natural Sciences and Engineering Research Council of Canada (NSERC), and the Perimeter Institute for Theoretical Physics. Research at Perimeter Institute is supported in part by the Government of Canada through the Department of Innovation, Science and Economic Development Canada and by the Province of Ontario through the Ministry of Colleges and Universities. 

From August 2023 - 2024 this work was supported by Clifford Johnson, the University of Southern California, and the US Department of Energy for support under grant DE-SC 0011687.

\appendix
\addtocontents{toc}{\protect\setcounter{tocdepth}{1}}

\section{Supplemental Calculations for Section 2}
\subsection{General Solution for Extremal Curves in Poincare AdS\texorpdfstring{$_3$}{3}}
\label{AppDerivingGenSolRTSurfAdS3}
In this appendix we will find the general solution to the equations of motion that are obtain from the Lagrangian
\begin{equation}
    \mathcal{L}=\int dx L\frac{\sqrt{1+\dot{z}^2-\dot{t}^2}}{z},
\end{equation}
which we read off from the function given in Eq. (\ref{PoincareAdS3RTFunctional}).
Since there is no explicit dependence on $x$ there is a conserved Hamiltonian corresponding to $x$-translations which can be written as
\begin{equation}
    H_x=-\frac{L}{z\sqrt{1+\dot{z}^2-\dot{t}^2}}.
    \label{xHam}
\end{equation}
Furthermore, since there is no explicit time dependence we get the following equation for $\dot{t}$,
\begin{equation}
    c_t=-\frac{L\dot{t}}{z\sqrt{1+\dot{z}^2-\dot{t}^2}}.
    \label{tEOM}
\end{equation}
Combining \ref{xHam} and \ref{tEOM} we will obtain the following equations for $t(x)$ and $z(x)$,
\begin{equation}
    \begin{split}
        &\dot{t}=k, \qquad
        \dot{z}^2=\frac{L^2}{H_x^2z^2}+k^2-1,\\
        \label{diffEqzandt}
    \end{split}
\end{equation}
where $k, H_x$ are constants. The general solution to the equations above is
\begin{equation}
    \begin{split}
        &t(x)=kx+c_1, \qquad
        z(x)=\sqrt{\frac{L^2}{H_x^2(1-k^2)}-(1-k^2)(x+c_2)^2},\\
    \end{split}
\end{equation}
which is what we wrote in Eq. (\ref{GenSolPoincareAdS3RtSurf}). In the special case where $k=0$ the RT surface remains on a constant time slice and traces out a half circle in the bulk given by
\begin{equation}
\label{GeneralSolkeq0}
   \begin{split}
   &t(x)=c_1, \qquad
z(x)=\sqrt{\frac{L^2}{H_x^2}-\left(x+c_2\right)^2}.\\
  \end{split} 
\end{equation}
It is straightforward to see that this is a half circle in the $z-x$ plane. It has a center at $(t=c_1,x=-c_2,z=0)$ with a radius $\frac{L}{|H_x|}$.

\subsection{Spacelike Separated Points From Line Segments AdS\texorpdfstring{$_3$}{3}}
\label{SpacelikePointsFromCurve}
Consider a spacelike line segment in AdS$_3$ Poincare space parameterized in terms of $\lambda$
\begin{equation}
\begin{split}
    &t=t(\lambda), \qquad x=x(\lambda), \qquad z=z(\lambda),\\
    &-\left(\frac{dt}{d\lambda}\right)^2+\left(\frac{dx}{d\lambda}\right)^2+\left(\frac{dz}{d\lambda}\right)^2>0, \qquad \lambda\in[\lambda_L,\lambda_R].\\
\end{split}
\end{equation}
To understand the set of points that are causally disconnected from the line segment we can consider an arbitrary point on the line segment $(t(\lambda_0),x(\lambda_0),z(\lambda_0))$, where $\lambda_0\in(\lambda_L,\lambda_R)$. We consider the lightcone centered at the point
\begin{equation}
    (t-t(\lambda_0))^2=(x-x(\lambda_0))^2+(z-z(\lambda_0))^2.
\end{equation}
Next consider another point on the curve that is arbitrarily close to $x_0$, with coordinates $(t(\lambda_0+\delta \lambda),x(\lambda_0+\delta \lambda),z(\lambda_0+\delta \lambda))$. The lightcone at that point is
\begin{equation}
    (t-t(\lambda_0+\delta \lambda))^2=(x-x(\lambda_0-\delta \lambda))^2+(z-z(\lambda_0+\delta \lambda))^2.
\end{equation}
Since $\delta \lambda$ is small we expand to obtain
\begin{equation}
\begin{split}
    &(t-t(\lambda_0))^2=(x-x(\lambda_0))^2+(z-z(\lambda_0))^2\\
    &-2(-(t-t(\lambda_0))t'(\lambda_0)+(x-x(\lambda_0))x'(\lambda_0)+(z-z(\lambda_0))z'(\lambda_0))\delta \lambda\\
    &-\left(-x'(\lambda_0)^2+t'(\lambda_0)^2-z'(\lambda_0)^2-(t-t(\lambda_0))t''(\lambda_0)+(z-z(\lambda_0))z''(\lambda_0)\right)\delta \lambda^2+\cdot\cdot\cdot.\\
\end{split}
\end{equation}
We can find intersections of the light cones by solving
\begin{equation}
\begin{split}
 &0=-2(-(t-t(\lambda_0))t'(\lambda_0)+(x-x(\lambda_0))x'(\lambda_0)+(z-z(\lambda_0))z'(\lambda_0))\\
    &-\left(-x'(\lambda_0)^2+t'(\lambda_0)^2-z'(\lambda_0)^2-(t-t(\lambda_0))t''(\lambda_0)+(z-z(\lambda_0))z''(\lambda_0)\right)\delta \lambda+\cdot\cdot\cdot.\\  
\end{split}
\end{equation}
Setting the leading term to zero describes a plane of intersection after a small perturbation away from $\lambda_0$ is made,
\begin{equation}
    -(t-t(\lambda_0))t'(\lambda_0)+(x-x(\lambda_0))x'(\lambda_0)+(z-z(\lambda_0))z'(\lambda_0)=0.
\end{equation}
Note that the point $(t(\lambda_0),x(\lambda_0),z(\lambda_0))$ is contained in the plane. Furthermore, we can express any tangent vector to the plane as follows,
\begin{equation}
    n^\mu=(n^t,n^x,n^z)=\left((t-t(\lambda_0))x'(\lambda_0),(t-t(\lambda_0))t'(\lambda_0)-(z-z(\lambda_0))z'(\lambda_0),(z-z(\lambda_0))x'(\lambda_0)\right),
\end{equation}
with an appropriate choice of $t,z$. Next consider the tangent vector to the curve at $\lambda=\lambda_0$ given by
\begin{equation}
    \mathcal{T}^\mu=(\mathcal{T}^t,\mathcal{T}^x,\mathcal{T}^z)=(t'(\lambda_0),x'(\lambda_0),z'(\lambda_0)).
\end{equation}
Using these expressions we can explicitly show that $g_{\mu\nu}\mathcal{T}^\mu n^{\mu}=0$, where $g_{\mu\nu}=z^{-2}\eta_{\mu\nu}$. This demonstrates that the plane of intersection of cones near each other is along a plane that is orthogonal to the tangent vector along the curve with respect to the AdS$_3$ inner product.

This fact is important to note because we can apply procedure for any point $\lambda_0\in(\lambda_L,\lambda_R)$. It follows that we can generate the null surface which separates the spacelike and timelike regions in ``most'' of the bulk by shooting orthogonal geodesics from the curve. This prescription, however, is not the full story in fact we must also deal with the endpoints of the line segments. This is not difficult from our previous analysis we simply analyzed light cones to each point and the endpoints we will do a similar analysis. We start by noting the null surface we described by shooting null geodesics orthogonally is valid all the way up to $\lambda=\lambda_R,\lambda_L$. Recall that the set of orthogonal null geodesics will lie on the following orthogonal planes to the curve endpoints
\begin{equation}
    \begin{split}
        -(t-t(\lambda_{L,R}))t'(\lambda_{L,R})+(x-x(\lambda_{L,R}))x'(\lambda_{L,R})+(z-z(\lambda_{L,R}))z'(\lambda_{L,R})=0.
    \end{split}
\end{equation}
These two planes will divide the spacetime into three regions:
\begin{itemize}
    \item $\mathcal{R}_L$ are the set of points that are to the ``left'' of the orthogonal plane located at $(t,x,z)=(t(\lambda_L),x(\lambda_L),z(\lambda_L))$. 
    \item $\mathcal{R}_R$ are the set of points that are to the ``right'' of the orthogonal plane located at $(t,x,z)=(t(\lambda_R),x(\lambda_R),z(\lambda_R))$
    \item $\mathcal{R}_M$ are the set of points $(t,x,z)$ that are enclosed by the planes and contain the line segment and is precisely where our procedure of finding spacelike points by emitting orthogonal null geodesics to the curve will be valid.   
\end{itemize}
To find the set of spacelike separated points in regions $\mathcal{R}_{L,R}$ it suffices to consider the light cones centered at the endpoints. This completes our formal discussion of how one can construct the null surfaces that allow one to determine what points are spacelike separated from a line segment.

To conclude this Appendix, lets apply this formalism to the RT curves in AdS$_3$ with $0 \leq k<1$. Recall that we can write the extremal RT surface anchored to an interval with endpoints $(t_L,x_L)$ and $(t_R,x_R)$ as follows
\begin{equation}
\begin{split}
    &t(x)=k\left(x-\frac{x_R-x_L}{2}\right)+\frac{t_R+t_L}{2}\\
    &z(x)=\sqrt{1-k^2}\sqrt{\left(\frac{x_R-x_L}{2}\right)^2-\left(x-\frac{x_R+x_L}{2}\right)^2}\\
    &k=\frac{t_R-t_L}{x_R-x_L}<1.\\
\end{split}
\end{equation}
The normal planes to the tangent at the conformal boundary at $x=x_L,x_R$ are given by
\begin{equation}
\begin{split}
    &0=-(t-t_L)t'(x_L)+(x-x_L)+zz'(x_L)\\
    &0=-(t-t_R)t'(x_R)+(x-x_R)+zz'(x_R).\\
\end{split}
\end{equation}
Generally one will find that this particular parametrization in terms of $x$ of the RT surface at the endpoints leads to a divergences for $z'(x_{L,R})$ terms. This is not a problem with the formalism but rather a problem of the way we parameterized the surface. Lets instead parameterize the surface in terms of $z$. We can write $x(z)$ as
\begin{equation}
    x(z)=\frac{x_R+x_L}{2}\pm\sqrt{\left(\frac{x_R-x_L}{2}\right)^2-\frac{z^2}{1-k^2}},
\end{equation}
where the $\pm$ defines the two branches of the RT surface when parameterize in terms of $z$. We can plug this into $t(x)$ to get
\begin{equation}
    t(z)=k\left(x_L\pm\sqrt{\left(\frac{x_R-x_L}{2}\right)^2-\frac{z^2}{1-k^2}}\right)+\frac{t_R-t_L}{2}.
\end{equation}
In this parametrization the equation of the plane normal to the tangent is given by
\begin{equation}
    0=-(t-t_{L,R})\frac{dt}{dz}\bigg\vert_{z=0}+(x-x_{L,R})\frac{dx}{dz}\bigg\vert_{z=0}+z.
\end{equation}
Once can explicitly check the derivatives are zero so the plane is given by
\begin{equation}
    z(t,x)=0.
\end{equation}
This shows that the RT curve ends on the conformal boundary orthogonally. This also implies that we can construct the null surface to which separates spacelike and timelike is generated by the geodesics shot orthogonal from the RT surface. In other words the entire AdS$_3$ spacetime is $\mathcal{R}_M$ so there is no need to worry about the light cones at the endpoints.

\subsection{Extremal Curves with \texorpdfstring{$|k|\leq 1$}{|k|<=1} as Intersection of Light Cones}
\label{ExtremalRTCurveLC}
Lets go back to the line element of Poincare AdS$_3$ it is given by:
\begin{equation}
    ds^2=\frac{-dt^2+dx^2+dz^2}{z^2}.
\end{equation}
We can see that under a Lorentz transformation of the form
\begin{equation}
    \begin{split}
        &t'=\gamma(t+\beta x), \qquad
        x'=\gamma(x+\beta t),\qquad
        z'=z,\\
    \end{split}
\end{equation}
gives
\begin{equation}
    ds'^2=\frac{-dt'^2+dx'^2+dz'^2}{z'^2}.
\end{equation}
So extremal surfaces after Lorentz transforms remain extremal surfaces anchored to a boosted boundary interval. This allows us to simply take the extremal surface anchored to constant time slice boundary interval and boost the whole setup to get the extremal surface for a tilted interval.
In particular, we could start with a constant time slice interval with endpoints $(t_0,x_L)$ and $(t_0,x_R)$. The extremal surface is given by the following parametric equation in $x$
\begin{equation}
    \begin{split}
    \label{ConstTimeSlicExtSSuf}
        &t(x)=t_0, \qquad z(x)=\sqrt{\left(\frac{x_R-x_L}{2}\right)^2+\left(x-\frac{x_R+x_L}{2}\right)^2}.\\
    \end{split}
\end{equation}
One can check that the expression in Eq. (\ref{ConstTimeSlicExtSSuf}) is actually the intersection of the following two light cone surfaces
\begin{equation}
    \begin{split}
        &t-t_0=-\frac{x_R-x_L}{2}+\sqrt{z^2+\left(x-\frac{x_R+x_L}{2}\right)^2} \quad t<t_0\\
        & t-t_0=\frac{x_R-x_L}{2}-\sqrt{z^2+\left(x-\frac{x_R+x_L}{2}\right)^2} \quad t>t_0.\\
    \end{split}
\end{equation}
Next consider the case of a tilted interval on the boundary through a Lorentz boost with endpoints at $(t_L',x_L')$ and $(t_R',x_R')$ and $|\beta|=\frac{|t_R'-t_L'|}{|x_R'-x_L'|}<1$. Since we were able to write the constant time slice extremal surface in terms of the intersection of two lightcones we should be able to write the new extremal surface in terms of the intersection of the lightcones with Lorentz transformed tips. In particular in the original coordinates we had a lightcone tip to the future of the extremal surface at $(t_+,x_+,z_+)=\left(t_0+\frac{x_R-x_L}{2},\frac{x_R+x_L}{2},0\right)$ and to the past of the extremal surface at $(t_-,x_-,z_-)=\left(t_0-\frac{x_R-x_L}{2},\frac{x_R+x_L}{2},0\right)$. The future tip is Lorentz transformed to
\begin{equation}
\begin{split}
    &(t_+',x_+',z_+')=\left(\frac{t_R'+t_L'}{2}+\frac{x_R'-x_L'}{2},\frac{x_R'+x_L'}{2}+\beta\frac{x_R'-x_L'}{2},0\right).\\
\end{split}
\end{equation}
The past tip is transformed to
\begin{equation}
\begin{split}
    &(t_-',x_-',z_-')=\left(\frac{t_R'+t_L'}{2}-\frac{x_R'-x_L'}{2},\frac{x_R'+x_L'}{2}-\beta\frac{x_R'-x_L'}{2},0\right).\\
\end{split}
\end{equation}
Now we consider intersection of the two lightcones,
\begin{equation}
    \begin{split}
        &t'-t'_+=-\sqrt{z'^2+\left(x'-x'_+\right)^2} \quad t'<t'_+\\
        & t'-t'_-=\sqrt{z'^2+\left(x'-x'_-\right)^2} \quad t'>t'_-.\\
    \end{split}
\end{equation}
To find the intersection we solve,
\begin{equation}
    (t'-t'_+)^2-(x'-x'_+)^2=(t'-t'_-)^2-(x'-x'_-)^2.
\end{equation}
With some simple algebra we find the intersection to be along the curve
\begin{equation}
\begin{split}
    &t'_{int}(x')=\frac{x'_+-x'_-}{t'_+-t'_-}\left(x'-\frac{x'_++x'_-}{2}\right)+\frac{t'_++t'_-}{2}=\beta\left(x'-\frac{x_R'+x_L'}{2}\right)+\frac{t_R'+t_L'}{2}\\
    &z'_{int}(x')=\sqrt{\left[\beta\left(x'-\frac{x_+'+x_-'}{2}\right)+\frac{t'_+-t_-'}{2}\right]^2-(x'-x_-')^2}\\
    &=\frac{1}{\gamma}\sqrt{\left(\frac{x_R'-x_L'}{2}\right)^2-\left(x'-\frac{x_R'+x_L'}{2}\right)^2},\\
\end{split}
\end{equation}
where $\beta=\frac{t_R'-t_L'}{x_R'-x_L'}$ and $\gamma=\frac{1}{\sqrt{1-\beta^2}}$. This indeed is the extremal surface we derived in Eq. (\ref{GenSolPoincareAdS3RtSurf}) with the following identification of parameters
\begin{equation}
\begin{split}
    &\beta=\frac{t_R'-t_L'}{x_R'-x_L'}=k, \qquad\frac{x'_R-x_L'}{2}=\frac{L}{|H_x|(1-k^2)}\\
    &c_1=\frac{t_R'+t_L'-k(x_R'+x_L')}{2}, \qquad c_2=-\frac{x_R'-x_L'}{2}.\\
\end{split}
\end{equation}

\subsection{Expression for Null Evolution of $\chi_{i}$ for $|k_{i}|<1$}
\label{NullEvoEquationschij}
Before writing the null evolution of $\chi_i$ it is useful to discuss some basic facts about identifying spacelike separated regions to extremal curves in AdS$_3$ with $|k|<1$.

Since every extremal curve with $0\leq |k|<1$ can be expressed as an intersection of certain bulk lightcones it is easy to find the set of spacelike separated regions to curves. Recall in Appendix \ref{ExtremalRTCurveLC} we were able to show that the expression for the extremal surface written in Eq. (\ref{GenSolPoincareAdS3RtSurf}) can be expressed as the intersection of the following lightcones
\begin{equation}
\label{LigconeForIntersecExtrSurf}
    \begin{split}
        &t-t_+=-\sqrt{z^2+\left(x-x_+\right)^2}, \qquad t<t_+,\\
        & t-t_-=\sqrt{z^2+\left(x-x_-\right)^2}, \qquad t>t_-,\\
    \end{split}
\end{equation}
where $t_{\pm}=c_1-kc_2\pm\frac{L}{|H_x|(1-k^2)}$ and $x_{\pm}=-c_2\pm \frac{kL}{|H_x|(1-k^2)}$. We can express the intersection curve in terms of the coordinates of the cone tips as
\begin{equation}
    \begin{split}
        &t(x)=k\left(x-\frac{x_++x_-}{2}\right)+\frac{t_++t_-}{2},\\
        &z(x)=\sqrt{1-k^2}\sqrt{\left(\frac{t_+-t_-}{2}\right)^2-\left(x-\frac{x_++x_-}{2}\right)^2},\\
    \end{split}
\end{equation}
where $k=\frac{x_+-x_-}{t_+-t_-}$ and $x\in\left[\frac{x_++x_-}{2}-\frac{t_+-t_-}{2},\frac{x_++x_-}{2}+\frac{t_+-t_-}{2}\right]$. This allows us to express the set of spacelike separated points from the extremal curve towards the AdS$_3$ boundary as\footnote{In Appendix \ref{SpacelikePointsFromCurve} we show that to understand the points that are spacelike separated to a line segment we can consider null surfaces generated by emitting null geodesics orthogonally from the RT curve.}  
\begin{equation}
    \begin{split}
    \label{SLTowardsBdry}
        &|t-t_+|>\sqrt{z^2+(x-x_+)^2}, \quad t(x)\leq t \leq t_+,\\
        &|t-t_-|>\sqrt{z^2+(x-x_-)^2}, \quad t_-\leq t \leq t(x),\\
    \end{split}
\end{equation}
while the set of spacelike separated points away from the AdS$_3$ boundary are given by
\begin{equation}
    \begin{split}
    \label{SLAwayBdry}
        &|t-t_+|<\sqrt{z^2+(x-x_+)^2}, \quad t\leq t(x)\\
        &|t-t_-|<\sqrt{z^2+(x-x_-)^2}, \quad t \geq t(x).\\
    \end{split}
\end{equation}
The lightcones given in Eq. (\ref{LigconeForIntersecExtrSurf}) are generated by the family of null geodesics that are emitted orthogonally from the curve. These facts are useful when writing down expressions for the null sheets emitted from $\chi_i$ in $\mathcal{R}_{M,i}$.

Here we write the piecewise defined expression for the null sheets/congruence emitted from $\chi_{i}$ in the regime where $|k_i|<1$, where $i=\{1,2\}$. 

\begin{equation}
\label{DefOfFullNullEvolutionchi1}
\begin{split}
    &z^{(\pm)}_{\text{Null},i}(x,t)=\begin{cases} 
      z^{(\pm)}_{\text{Null},\mathcal{R}_{R,i}}(x,t)\qquad \text{if}& \frac{\partial_x z_{\sigma_i}}{|\partial_x z_{\sigma_i}|} (z_{\sigma_i}(x,t)-z^{(\pm)}_{\text{Null},\mathcal{R}_{R,i}}(x,t))\geq 0  \\
       z^{(\pm)}_{\text{Null},\mathcal{R}_{M,i}}(x,t) \qquad \text{if}& \frac{\partial_x z_{\sigma_i}}{|\partial_x z_{\sigma_i}|} (z_{\sigma_i}(x,t)-z^{(\pm)}_{\text{Null},\mathcal{R}_{M,i}}(x,t))<0 \\ 
   \end{cases}\\
&z^{(\pm)}_{\text{Null},\mathcal{R}_{R,i}}(x,t)=b_i\cos\theta_0\pm\sqrt{(t-t_b)^2-(x-b_i\sin\theta_0)^2}\\
&z^{(\pm)}_{\text{Null},\mathcal{R}_{M,i}}(x,t)=\Theta[\pm(t_i(x)-t)]\sqrt{(t-t_{+,i})^2-(x-x_{+,i})^2}+\Theta[\pm(t-t_i(x))]\sqrt{(t-t_{-,i})^2-(x-x_{-,i})^2}\\
&z_{\sigma_i}(x,t)=\frac{2\Delta z_i}{\Delta t^2-\Delta x_i^2+\Delta z_i^2}\left(-x\Delta x_i+t\Delta t\right)+\frac{\Delta z_i\left[b_i^2-a_i^2+t_a^2-t_b^2\right]}{\Delta t^2-\Delta x_i^2+\Delta z_i^2}\\
    &t_i(x)=t_a+\frac{\Delta t}{\Delta x_i}(x+a_i),\\
\end{split}
\end{equation}
where $\Delta t=t_b-t_a$, $\Delta x_i=a_i+b_i\sin\theta_0$, and $\Delta z_i=b_i\cos\theta_0$. The $\pm$ signs appearing in Eq. (\ref{DefOfFullNullEvolutionchi1}) determine if the null points are have larger $z$ than $\chi_i$ (given by ``$+$'' sign) or a smaller $z$ than $\chi_i$ (given by ``$-$'' sign) , in the $xz$-plane for each fixed time slice. What we plot in the Figure \ref{chi2SLTimeSlice} is $z_{\text{Null},i}(x)=\left[z^{(+)}_{\text{Null},i}(x,t=T)\cup z^{(-)}_{\text{Null},i}(x,t=T)\right]$ at various fixed time slices $t=T$.\footnote{NOTE: The Heaviside step function, $\Theta$, used in Eq. (\ref{DefOfFullNullEvolutionchi1}) has a value of $1/2$ when the argument is zero (i.e. $\Theta(0)=\frac{1}{2}$)}

\section{Supplemental Calculations for Section 3}
\subsection{Analysis of Conditions for EWN}
\label{AppendixNiaveConAdS3}

In this Appendix we will analyze the conditions given in Eq. (\ref{SuffEWNA}), Eq. (\ref{RLConstraint}), and Eq. (\ref{NaiveSLSeparationPoincareCond})  which will give us sufficient conditions for entanglement wedge nesting to hold. 

We begin by considering the condition in Eq. (\ref{SuffEWNA}), satisfying this condition ensures that no violations of EWN can occur for $\mathcal{W}_E(A)$. For $\chi_i$ the condition reads
\begin{equation}
    \left(\frac{a_2-a_1}{2}+|t_{i}(x)-t_a|\right)^2\leq z_{i}(x)^2+\left(x+\frac{a_2+a_1}{2}\right)^2,
\end{equation}
where $z_{i}(x)$ and $t_{i}(x)$ are given in Eq. (\ref{Z1RT}). Along $\chi_i$ we have the following relation between $x$ and $|t_i(x)-t_a|$,
\begin{equation}
\begin{split}
    &x(\delta t_{i,a})=-a_i+\frac{|t_i-t_a|}{|k_i|}=-a_i+\frac{|\delta t_{i,a}|}{|k_i|}\\
    &\delta t_{i,a}=t_i-t_a.\\
\end{split}
\end{equation}
Substituting $x=x(\delta t_{i,a})$ into the bound gives
\begin{equation}
    \begin{split}
&\left(\frac{a_2-a_1}{2}+|\delta t_{i,a}|\right)^2-z_{i}\left(-a_i+\frac{|\delta t_{i,a}|}{|k_i|}\right)^2-\left(-a_i+\frac{|\delta t_{i,a}|}{|k_i|}+\frac{a_2+a_1}{2}\right)^2\\
&=\frac{|\delta t_{i,a}|(a_i+b_i\sin\theta_0)}{|k_i|}\left[\left(|k_i|+\frac{a_2-a_1}{2(a_i+b_i\sin\theta_0)}\right)^2-\frac{\left(\frac{a_1+a_2}{2}\right)^2+b_i^2+2\left(\frac{a_2+a_1}{2}\right)b_i\sin\theta_0}{(a_i+b_i\sin\theta_0)^2}\right]\leq 0.\\
    \end{split}
\end{equation}
So to satisfy the inequality throughout $\mathcal{M}_{\text{phys}}$ we must require
\begin{equation}
\begin{split}
    &|k_i|\leq -\frac{a_2-a_1}{2(a_i+b_i\sin\theta_0)}+\frac{\sqrt{\left(\frac{a_1+a_2}{2}\right)^2+b_i^2+2\left(\frac{a_2+a_1}{2}\right)b_i\sin\theta_0}}{a_i+b_i\sin\theta_0}\\
    &\Leftrightarrow |\Delta t|\leq -\frac{a_2-a_1}{2}+\sqrt{\left(\frac{a_1+a_2}{2}\right)^2+b_i^2+2\left(\frac{a_2+a_1}{2}\right)b_i\sin\theta_0}.\\
    &=-\frac{a_2-a_1}{2}+\sqrt{b_i\cos^2\theta_0+\left(b_i\sin\theta_0+\frac{a_2+a_1}{2}\right)^2},\\
\end{split}
\end{equation}
which is the result in Eq. (\ref{MostStringentEWNACondd}) with $i=1$.

Next we consider the conditions given in Eq. (\ref{NaiveSLSeparationPoincareCond}). For $\chi_i$ we need to consider,

\begin{equation}
\begin{split}
    &\left[|t_i(x)-t_b|+R_b\right]^2-\left[x-\frac{b_1+b_2}{2\sin\theta_0}\right]^2-z_i(x)^2< 0\\
    &R_b=\frac{\sqrt{b_1^2+b_2^2+2b_1b_2\cos(2\theta_0)}}{2\sin\theta_0}.\\
\end{split}
\end{equation}
On the $\chi_i$ curve there is a relation between $|t_i(x)-t_b|$ and $x$ given by,
\begin{equation}
\begin{split}
    &x(\delta t_i)=b_i\sin\theta_0-\frac{|t_i-t_b|}{|k_i|}=b_i\sin\theta_0-\frac{|\delta t_i|}{|k_i|}\\
    &\delta t_i=t_i-t_b.\\
\end{split}
\end{equation}
We plug this into the inequality and simplify to obtain
\begin{equation}
    \begin{split}
        & \left[|\delta t_i|+R_b\right]^2-\left[b_i\sin\theta_0-\frac{|\delta t_i|}{|k_i|}-\frac{b_1+b_2}{2\sin\theta_0}\right]^2-z_i\left(b_i\sin\theta_0-\frac{|\delta t_i|}{|k_i|}\right)^2\\
        &=\frac{|\delta t_i|(a_i+b_i\sin\theta_0)}{|k_i|}\left[\left[|k_i|+\frac{R_b}{a_i+b_i\sin\theta_0}\right]^2-\frac{\left(2a_i+\frac{b_1+b_2}{\sin\theta_0}\right)^2}{4(a_i+b_i\sin\theta_0)^2}\right]< 0.\\
    \end{split}
\end{equation}
So to satisfy the inequality throughout $\mathcal{M}_{\text{phys}}$ we require that
\begin{equation}
\begin{split}
    &|k_i|< -\frac{R_b}{a_i+b_i\sin\theta_0}+\frac{2a_i+\frac{b_1+b_2}{\sin\theta_0}}{2(a_i+b_i\sin\theta_0)}\\
   &\Leftrightarrow |\Delta t|< a_i+\frac{b_1+b_2-\sqrt{b_1^2+b_2^2+2b_1b_2\cos(2\theta_0)}}{2\sin\theta_0}. \\ 
\end{split}
\end{equation}
Which reproduces the bound we had in Eq. (\ref{SuffConEWNNaivedelt}).

Finally, we will consider the condition in Eq. (\ref{RLConstraint}) with $i=j$. For $\chi_i$, we shall consider,
\begin{equation}
    (t_i(x)-t_b)^2-(x-b_i\sin\theta_0)^2-(z_i(x)-b_i\cos\theta_0)^2<0.
\end{equation}
We claim that the inequality above is satisfied as long as $|\Delta t|<\Delta t_c=\sqrt{\Delta x_1^2+\Delta z_1^2}$. To demonstrate this we begin by noting that along $\chi_i$ we can express $x$ as $x=b_i\sin\theta_0-\frac{|\delta t_i|}{|\Delta t|}\Delta x_i$, where $\delta t_i(x)=t_i(x)-t_b$ and we also write $b_i\cos\theta_0=\Delta z_i$. Then we have:
\begin{equation}
    0>\delta t_i(x)^2-\frac{\delta t_i(x)^2}{k_i^2}-(z_i(x)-\Delta z_i)^2=\delta t_i(x)^2\left(\frac{\Delta t^2-\Delta x_i^2}{\Delta t^2}\right)-\Delta z_i^2-z_i(x)^2+2\Delta z_iz_i(x).
\end{equation}
We can also express $z_i(x)$ in terms of $\Delta x_i$, $\Delta z_i$, $\Delta t$, and $\delta t_i(x)$. Using Eq. (\ref{Z1RT}) we can write,
\begin{equation}
\label{ziInDelta}
    z_i(x)^2=\left(1-\frac{|\delta t_i(x)|}{|\Delta t|}\right)\left[\frac{|\delta t_i(x)|}{|\Delta t|}(-\Delta t^2+\Delta x_i^2)+\Delta z_i^2\right].
\end{equation}
Putting everything together we arrive at the following identity:
\begin{equation}
\begin{split}
    &(t_i(x)-t_b)^2-(x-b_i\sin\theta_0)^2-(z_i(x)-b_i\cos\theta_0)^2\\
    &=-\left[\frac{|\delta t_i|(-\Delta t^2+\Delta x_i^2+\Delta z_i^2)+2\Delta z_i^2(|\Delta t|-|\delta t_i|)}{\Delta t}\right]+2\Delta z_iz_i(x)\\
\end{split}
\end{equation}
All we need to do now is demonstrate that the expression in the second line is negative. To do this, we note that the terms in the square brackets are non-negative and $\Delta z_iz_i(x)\geq 0$. Therefore it suffices to show that,
\begin{equation}
    -\left[\frac{|\delta t_i|(-\Delta t^2+\Delta x_i^2+\Delta z_i^2)+2\Delta z_i^2(|\Delta t|-|\delta t_i|)}{\Delta t}\right]^2+4\Delta z_i^2z_i(x)^2<0.
\end{equation}
Expanding the left-hand side of the inequality above and rearranging the terms gives,
\begin{equation}
\begin{split}
    &-\frac{\delta t_i^2(-\Delta t^2+\Delta x_i^2+\Delta z_i^2)}{\Delta t^2}\\
    &+4\Delta z_i^2\left[z_i(x)^2-\left(1-\frac{|\delta t_i(x)|}{|\Delta t|}\right)\left(\frac{|\delta t_i(x)|}{|\Delta t|}(-\Delta t^2+\Delta x_i^2)+\Delta z_i^2\right)\right]\\
\end{split}
\end{equation}
The set of terms in the second line are zero due to Eq. (\ref{ziInDelta}). We arrive at the following conclusion, 

\begin{equation}
\begin{split}
     &-\left[\frac{|\delta t_i|(-\Delta t^2+\Delta x_i^2+\Delta z_i^2)+2\Delta z_i^2(|\Delta t|-|\delta t_i|)}{\Delta t}\right]^2+4\Delta z_i^2z_i(x)^2\\
    &=-\delta t_i^2\left[-\Delta t^2+\Delta x_i^2+\Delta z_i^2\right]^2<0\\
\end{split}
\end{equation}
This proves that anytime $|\Delta t|<\Delta t_c$ the constraints in Eq. (\ref{RLConstraint}) with $i=j$ are satisfied.

\subsection{Conditions when $\chi_1\cap \mathcal{R}_M=\varnothing$ and $\chi_2\cap \mathcal{R}_R=\varnothing$}
\label{chi1RMCondAppendix}
In this appendix we will be understanding when $\chi_1\cap \mathcal{R}_M=\varnothing$ and $\chi_2\cap \mathcal{R}_R=\varnothing$. We will formulate the conditions by first fixing the spatial location of $A$ and $B$ (i.e. we fix $a_1,a_2,b_1,b_2,\theta_0$) and vary $\Delta t$ or equivalently $k_{1,2}$ and study what regions $\chi_{1,2}$ land in.

Lets begin by writing down the condition that needs to be satisfied to prove that for a certain placement of $A$ and $B$ we have that $\chi_1\cap \mathcal{R}_M=\varnothing$. To begin we must recall that the plane the separates the regions $\mathcal{R}_M$ and $\mathcal{R}_L$ is given by the expression written in Eq . (\ref{LeftBraneLine}). Since the profile along $t$ of the plane is trivial it suffices to simply refer to the plane as a ray in $xz$-plane and in this appendix this ray will be given by
\begin{equation}
\label{z1toLeftCondition}
    z_{\sigma_L}(x)=-\frac{b_1\sin(2\theta_0)}{b_2+b_1\cos(2\theta_0)}x+\frac{b_1(b_1+b_2)\cos\theta_0}{b_2+b_1\cos(2\theta_0)}.
\end{equation}
The region to the immediate right of this ray is $\mathcal{R}_M$, therefore the condition that $\chi_1\cap \mathcal{R}_M=\varnothing$ is equivalent to saying that $z_1(x)$ is to the left of the ray $z_{\sigma_L}(x)$. This can be formally written out as the following condition  
\begin{equation}
\begin{split}
     &z_{\sigma_L}(x)-z_1(x)\geq 0 \\
     &x\in[-a_1,b_1\sin\theta_0],\\
\end{split}
\end{equation}
where $z_1(x)$ is given by Eq. (\ref{Z1RT}). The strategy here will be to fix the parameters $a_1,a_2,b_1,b_2,\theta_0$ and vary $\Delta t$ (i.e. we vary $k_1$ and study if the inequality given in Eq. (\ref{z1toLeftCondition}) is respected). As a first step we will analyze the inequality to linear order at the point where the brane intersects with $\chi_1$. In this case we obtain
\begin{equation}
\begin{split}
    &0\leq z_{\sigma_L}(x)-z_1(x)\\
    &=\frac{a_1+b_1\sin\theta_0}{2b_1\cos\theta_0}\left[k_1^2-\left(1-\frac{(b_1\cos\theta_0)^2}{(a_1+b_1\sin\theta_0)^2}\left(1+\frac{4\sin\theta_0(a_1+b_1\sin\theta_0)}{b_2+b_1\cos(2\theta_0)}\right)\right)\right](b_1\sin\theta_0-x)\\
    &+\mathcal{O}((x-b_1\sin\theta_0)^2).\\
\end{split}
\end{equation}
This places a constraint on $|k_1|$ of the form
\begin{equation}
\begin{split}
    &|k_1|\geq k_{1,*}\\
    &k_{1,*}=\sqrt{1-\frac{(b_1\cos\theta_0)^2}{(a_1+b_1\sin\theta_0)^2}\left(1+\frac{4\sin\theta_0(a_1+b_1\sin\theta_0)}{b_2+b_1\cos(2\theta_0)}\right)}.\\
\end{split}
\end{equation}
So whenever this holds we will know that at least in a neighborhood of the point where the brane and $\chi_1$ intersect $\chi_1\cap \mathcal{R}_M=\varnothing$.\footnote{The reader may be concerned that the expression for $k_{1,*}$ is not always going to be real. This is indeed true so the reader should really interpret the inequality is being true when $k_{1,*}\in\mathbb{R}$. In the case when it is not real we can still show that $\chi_1\cap \mathcal{R}_M=\varnothing$.} 

Now we will prove that this is not only true in a neighborhood of the point but for all $x\in[-a_1,b_1\sin\theta_0]$. Start by noting that
\begin{equation}
    \frac{\partial}{\partial|k_1|}z_1(x)=-\frac{|k_1|(a_1+x)(b_1\sin\theta_0-x)}{z_1(x)}\leq 0.
\end{equation}
This implies that if $|k_1|<|\tilde{k}_1|$ then we have that,
\begin{equation}
\label{largerkIssmallerz}
    z_1(x)\vert_{|k_1|}\geq z_1(x)\vert_{|\tilde{k}_1|}. 
\end{equation}
Now consider what happens when we set $|k_1|=k_{1,*}$. We find that
\begin{equation}
\begin{split}
    &z_{\sigma_L}(x)^2-z_1(x)^2\vert_{|k_1|=k_{1,*}}=\left[\frac{b_1\cos\theta_0(b_1+b_2+2a_1\sin\theta_0)(x-b_1\sin\theta_0)}{(b_2+b_1\cos(2\theta_0))(a_1+b_1\sin\theta_0)}\right]^2\geq 0\\
    &\Rightarrow z_{\sigma_L}(x)-z_1(x)\vert_{|k_1|=k_{1,*}}\geq 0.\\
\end{split}
\end{equation}
This shows that when $|k_1|=k_{1,*}$ we have $\chi_1 \cap\mathcal{R}_M=\varnothing$ not just in small neighborhoods of $x=b_1\sin\theta_0$ but for all $x\in[-a,b_1\sin\theta_0]$. Now consider any $|k_1|>k_{1,*}$. Due to Eq. (\ref{largerkIssmallerz}) we can write
\begin{equation}
\begin{split}
    &z_1(x)\vert_{|k_1|>k_{1,*}}\leq z_1(x)\vert_{|k_1|=k_{1,*}}\\
    &\Rightarrow z_1(x)\vert_{|k_1|>k_{1,*}}-z_{\sigma_L}(x)\leq z_1(x)\vert_{|k_1|=k_{1,*}}-z_{\sigma_L}(x)\leq 0\\
    &\Rightarrow z_{\sigma_L}(x)-z_1(x)\vert_{|k_1|>k_{1,*}} \geq 0.\\
\end{split}
\end{equation}
This proves the statement that whenever $|k_1|\geq k_{1,*}$ then $\chi_1\cap \mathcal{R}_M=\varnothing$.

Now consider the case when $k_{1,*}$ becomes imaginary in this case we have that
\begin{equation}
\label{k1sbecomesIm}
    k_{1,*}^2=1-\frac{(b_1\cos\theta_0)^2}{(a_1+b_1\sin\theta_0)^2}\left(1+\frac{4\sin\theta_0(a_1+b_1\sin\theta_0)}{b_2+b_1\cos(2\theta_0)}\right)<0.
\end{equation}
We claim that when $k_{1,*}^2<0$ then $\chi_1\cap \mathcal{R}_{M}=\varnothing$ for any value of $|k_1|$. To prove this it suffices to show that
\begin{equation}
\label{sufficentk=0Cond}
    z_{\sigma_L}(x)-z_1(x)|_{k_1=0}\geq 0.
\end{equation}
The reason for this is because of the result we wrote in Eq. (\ref{largerkIssmallerz}), it shows that $-z_1(x)|_{k_1=0}\geq -z_1(x)\Rightarrow z_{\sigma_L}(x)-z_1(x)|_{k_1=0}\geq z_{\sigma_L}(x)-z_1(x)$. Therefore, satisfying the bounds written in Eq. (\ref{sufficentk=0Cond}) for $k_1=0$ is sufficient to have a bound for general values of $k_1$. A simple computation reveals that
\begin{equation}
\begin{split}
    &z_{\sigma_L}(x)^2-z_1(x)^2\vert_{k_1=0}\\
    &=\left[-1+\frac{(b_1\cos\theta_0)^2}{(a_1+b_1\sin\theta_0)^2}\left(1+\frac{4\sin\theta_0(a_1+b_1\sin\theta_0)}{b_2+b_1\cos(2\theta_0)}\right)\right](a_1+b_1\sin\theta_0)(b_1\sin\theta_0-x)\\
    &+\left[1+\left(\frac{b_1\sin(2\theta_0)}{b_2+b_1\cos(2\theta_0)}\right)^2\right](b_1\sin\theta_0-x)^2.\\
\end{split}
\end{equation}
The quadratic term in $b_1\sin\theta_0-x$ is already positive and the linear term will also be positive whenever Eq. (\ref{k1sbecomesIm}) is satisfied which is exactly when $k_{1,*}$ becomes imaginary. This proves our claim that whenever $k_{1,*}^2<0$ we know that $\chi_1\cap\mathcal{R}_M=\varnothing$.

Now lets turn to the task of deriving conditions for having $\chi_2\cap \mathcal{R}_R=\varnothing$. In this case the plane that separates $\mathcal{R}_M$ and $\mathcal{R}_R$ is given by the expression in Eq. (\ref{RightBraneLine}). Just as before, the plane will have a trivial profile in $t$ and we will just refer to it as a ray in the $xz$-plane. Explicitly, the ray will be given by
\begin{equation}
    z_{\sigma_R}(x)=-\frac{b_2\sin(2\theta_0)}{b_1+b_2\cos(2\theta_0)}x+\frac{b_2(b_1+b_2)\cos\theta_0}{b_1+b_2\cos(2\theta_0)}.
\end{equation}
To the right of this ray we will have the region $\mathcal{R}_R$ so the condition to have $\chi_2\cap\mathcal{R}_R=\varnothing$ is equivalent to saying that $\chi_2$ is to the left of the ray. This can be expressed as\footnote{Notice that we actually have two different cases which did not appear when considering when $\chi_1\cap\mathcal{R}_M=\varnothing$. This was because the slope of the ray $z_{\sigma_L}(x)$ was always negative. However for $z_{\sigma_R}(x)$ the sign of the slope of the ray will depend on the sign of the quantity $(b_1+b_2\cos(2\theta_0))$, hence we need to multiply the inequality in Eq. (\ref{z2toLeftCondition}) by the pre-factor of $\frac{b_1+b_2\cos(2\theta_0)}{|b_1+b_2\cos(2\theta_0)|}$ to account for the two cases.}
\begin{equation}
    \begin{split}
    \label{z2toLeftCondition}
        &\frac{b_1+b_2\cos(2\theta_0)}{|b_1+b_2\cos(2\theta_0)|}\left[z_{\sigma_R}(x)-z_2(x)\right]\geq 0\\
        &x\in[-a_2,b_2\sin\theta_0],\\
    \end{split}
\end{equation}
where $z_2(x)$ is given in Eq. (\ref{Z1RT} with $i=2$). 

Lets begin by considering the case when $b_1+b_2\cos(2\theta_0)>0$ (in this case the slope of the ray is negative). To linear order around $x=b_2\sin\theta_0$ we require
\begin{equation}
\begin{split}
    &0\leq z_{\sigma_R}(x)-z_2(x)\\
    &=\frac{a_2+b_2\sin\theta_0}{2b_2\cos\theta_0}\left[k_2^2-\left(1-\frac{(b_2\cos\theta_0)^2}{(a_2+b_2\sin\theta_0)^2}\left(1+\frac{4\sin\theta_0(a_2+b_2\sin\theta_0)}{b_1+b_2\cos(2\theta_0)}\right)\right)\right](b_2\sin\theta_0-x)\\
    &+\mathcal{O}((b_2\sin\theta_0-x)^2).\\
\end{split}
\end{equation}
Using the linear order term we obtain the following condition on $|k_2|$
\begin{equation}
\begin{split}
    &|k_2|\geq k_{2,*}\\
    &k_{2,*}=\sqrt{1-\frac{(b_2\cos\theta_0)^2}{(a_2+b_2\sin\theta_0)^2}\left(1+\frac{4\sin\theta_0(a_2+b_2\sin\theta_0)}{b_1+b_2\cos(2\theta_0)}\right)}.\\
\end{split}
\end{equation}
This demonstrates that when $|k_2|\geq k_{2,*}$ (and also assuming that $k_{2,*}\in\mathbb{R}$) then in sufficiently small neighborhoods of the point where the brane intersects $\chi_2$ we have that $\chi_2\cap \mathcal{R}_R=\varnothing$. Using similar reasoning as we did in the study of when $\chi_1\cap\mathcal{R}_M=\varnothing$ the condition that $|k_2|\geq k_{2,*}$ implies that $\chi_2\cap \mathcal{R}_R=\varnothing$ throughout the whole domain given by $x\in[-a_2,b_2\sin\theta_0]$. It is also straightforward to show that $\chi_2\cap \mathcal{R}_R=\varnothing$ anytime $k_{2,*}^2<0$.\footnote{In particular we just go through analogous steps from Eqs.(\ref{largerkIssmallerz}-\ref{sufficentk=0Cond}) with $z_{\sigma_L}(x)\to z_{\sigma_R}(x)$, $z_{1}(x)\to z_{2}(x)$, and $k_{1,*}\to k_{2,*}$. The expressions will take on similar forms and everything follows in a similar way as before.}

This now leaves us with the case where $b_1+b_2\cos(2\theta_0)<0$ (i.e. the ray separating $\mathcal{R}_M$ and $\mathcal{R}_R$ will have positive slope). In this case we can easily argue that $\chi_2\cap\mathcal{R}_R=\varnothing$. Start by noting that the region $\mathcal{M}_{\text{phys}}$ is described by $z\geq\Theta(x)\cot\theta_0 x$. Consider the following quantity,
\begin{equation}
    \Theta(x)\cot\theta_0 x-z_{\sigma_R}(x)= \begin{cases} 
      -\frac{(b_1+b_2)(b_2\sin\theta_0-x)\cot\theta_0}{b_1+b_2\cos(2\theta_0)} & x \geq 0 \\
      -z_{\sigma_R}(x) & x<0. \\ 
   \end{cases}
\end{equation}
Lets analyze the quantity above in the domain $x\in[-a_2,b_2\sin\theta_0]=[-a_2,0)\cup [0,b_2\sin\theta_0]$. Since $b_1+b_2\cos(2\theta_0)<0$ and $b_2\sin\theta_0-x>0$ it follows that when $x\in [0,b_2\sin\theta_0]$ we know $ \Theta(x)\cot\theta_0 x-z_{\sigma_R}(x)\geq 0$. Now lets consider the domain $x\in[-a_2,0)$. Here it suffices to consider $z_{\sigma_R}(x=0)=\frac{b_2(b_1+b_2)\cos\theta_0}{b_1+b_2\cos(2\theta_0)}$. Using this we can conclude that $\Theta(x)\cot\theta_0 x-z_{\sigma_R}(x)\geq 0$ in the region $x\in[-a_2,0)$. Combining everything we conclude that
\begin{equation}
    \begin{split}
    \label{BranesigmaRRelation}
        &\Theta(x)\cot\theta_0 x-z_{\sigma_R}(x)\geq 0\\
        &x\in[-a_2,b_2\sin\theta_0].\\
    \end{split}
\end{equation}
Using the result above along with the fact that $z_2(x)\geq \Theta(x)\cot\theta_0 x$ for all $x\in[-a_2,b_2\sin\theta_0]$ we can show
\begin{equation}
\begin{split}
    &z_2(x)-z_{\sigma_R}(x)\geq \Theta(x)\cot\theta_0 x-z_{\sigma_R}(x)\geq 0\\
    &x\in[-a_2,b_2\sin\theta_0].\\
\end{split}
\end{equation}
This proves that whenever $b_1+b_2\cos(2\theta_0)<0$ then $\chi_2 \cap\mathcal{R}_R=\varnothing$.

Finally, if $b_1+b_2\cos(2\theta_0)=0$ then the ray separating $\mathcal{R}_R$ and $\mathcal{R}_M$ is a vertical line in the $xz$-plane (i.e. described by a line with $x=b_2\sin\theta_0$) in this case it is trivial to see that $\chi_2 \cap\mathcal{R}_R=\varnothing$.  

Lets now summarize the main results of this appendix:
\begin{itemize}
    \item $\chi_1\cap \mathcal{R}_M= \varnothing$ if and only if:
    \begin{itemize}
        \item $|k_1|\geq \text{Re}\left[k_{1,*}\right]$.\footnote{Here we are combining the cases when $k_{1,*}$ is real or imaginary by simply putting bounds using the real part of $k_{1,*}$.} 
    \end{itemize}
    \item $\chi_2\cap\mathcal{R}_R=\varnothing$ if and only if:
    \begin{itemize}
        \item $|k_2|\geq \text{Re}\left[k_{2,*}\right]$ when $b_1+b_2\cos(2\theta_0)>0$.
        \item  Or if $b_1+b_2\cos(2\theta_0)\leq 0$.
    \end{itemize}
\end{itemize}
These results will be important towards finding the relevant necessary and sufficient constraints on $|\Delta t|$ such that EWN is respected in the connected phase in Section \ref{PoincareAdS3SuffCondSec}.

\subsection{Condition for Dominance of Connected Phase}
\label{ConnectedConditionAppend}
Below we will derive an inequality that determines when the connected phase will dominate. We start by writing expressions for $\chi_{\text{dis}}(A)$, $\chi_{\text{dis}}(B)$, and $\chi_{\text{con}}(A\cup B)$.

The regulated area for $\chi_{\text{dis}}(A)$ is
\begin{equation}
\begin{split}
    &\mathcal{A}^{reg}_{\chi_{\text{dis}}(A)}=\int_{x_L(\epsilon)}^{x_R(\epsilon)}\frac{L}{z}\sqrt{1+\left(\frac{dz}{dx}\right)^2}dx,\\
\end{split}
\end{equation}
where $x_{R}(\epsilon)=\frac{-a_1-a_2+\sqrt{(a_2-a_1)^2-4\epsilon^2}}{2}$, $x_{L}(\epsilon)=\frac{-a_1-a_2-\sqrt{(a_2-a_1)^2-4\epsilon^2}}{2}$, and $z=z(x)$ is given by Eq. (\ref{chidisAexp}). Performing the integral we obtain
\begin{equation}
\begin{split}
    &\mathcal{A}^{reg}_{\chi_{\text{dis}}(A)}=\frac{L}{2}\ln\left[\frac{x+a_2}{x+a_1}\right]\bigg\vert_{x_L(\epsilon)}^{x_R(\epsilon)}=2L\left[\ln\left(\frac{\Delta x_A}{\epsilon}\right)+\mathcal{O}(\epsilon^2)\right].\\
\end{split}
\end{equation}
The regulated area for $\chi_{\text{con}}(A\cup B)$ is,
\begin{equation}
    \mathcal{A}_{\chi_{\text{con}}(A\cup B)}=\mathcal{A}_{\chi_{1}}+\mathcal{A}_{\chi_{2}},
\end{equation}
where $\mathcal{A}_{\chi_{i}}^{reg}$ is given by
\begin{equation}
    \begin{split}
        &\mathcal{A}^{reg}_{\chi_{i}}=\int_{x^{(i)}_{L}(\epsilon)}^{b_{i}\sin\theta_0}\frac{L}{z_{i}}\sqrt{-k_{i}^2+1+\left(\frac{dz_{i}}{dx}\right)^2}dx,\\
    \end{split}
\end{equation}
and $x_L^{(i)}(\epsilon)=-a_i+\frac{\Delta x_i}{-\Delta t^2+\Delta x_i^2+\Delta z_i^2}\epsilon^2+\mathcal{O}(\epsilon^4)$, $\Delta t=t_b-t_a$, $\Delta x_{i}=a_{i}+b_{i}\sin\theta_0$, $z_{i}=b_{i}\cos\theta_0$, and $z_i$ is given by Eq. (\ref{Z1RT}).  Doing the integral and also a series expansion in $\epsilon$ we obtain
\begin{equation}
    \mathcal{A}^{reg}_{\chi_{i}}=L\left[\ln\left(\frac{-\Delta t^2+\Delta x^2_{i}+\Delta z^2_{i}}{\epsilon\Delta z_{i}}\right)+\mathcal{O}(\epsilon^2)\right],
\end{equation}
which gives the following the regulated length of $\chi_{\text{con}}(A\cup B)$,
\begin{equation}
    \mathcal{A}^{reg}_{\chi_{\text{con}}(A\cup B)}=L\left[\ln\left(\frac{\left(-\Delta t^2+\Delta x^2_{1}+\Delta z^2_{1}\right)\left(-\Delta t^2+\Delta x^2_{2}+\Delta z^2_{2}\right)}{\epsilon^2\Delta z_1\Delta z_2}\right)+\mathcal{O}(\epsilon^2)\right].
\end{equation}

The area for $\chi_{\text{dis}}(B)$ is given by
\begin{equation}
    \mathcal{A}_{\chi_{\text{dis}}(B)}=\int_{b_1\sin\theta_0}^{b_2\sin\theta_0}\frac{L}{z}\sqrt{1+\left(\frac{dz}{dx}\right)^2}dx,
\end{equation}
where we use $z=z(x)$ from Eq. (\ref{chidisBexp}) to obtain,
\begin{equation}
\begin{split}
    &\mathcal{A}_{\chi_{\text{dis}}(B)}=-L \text{arctanh}\left[\frac{-2x+\frac{b_1+b_2}{\sin\theta_0}}{\sqrt{-4b_1b_2+\left(\frac{b_1+b_2}{\sin\theta_0}\right)^2}}\right]\bigg\vert_{b_1\sin\theta_0}^{b_2\sin\theta_0}\\
    &=\frac{L}{2}\ln\left[\frac{\left(b_1+b_2\cos(2\theta_0)-\sqrt{b_1^2+b_2^2+2b_1b_2\cos(2\theta_0)}\right)\left(b_2+b_1\cos(2\theta_0)+\sqrt{b_1^2+b_2^2+2b_1b_2\cos(2\theta_0)}\right)}{\left(b_2+b_1\cos(2\theta_0)-\sqrt{b_1^2+b_2^2+2b_1b_2\cos(2\theta_0)}\right)\left(b_1+b_2\cos(2\theta_0)+\sqrt{b_1^2+b_2^2+2b_1b_2\cos(2\theta_0)}\right)}\right]\\
    &=\frac{L}{2}\ln\left[\frac{\left(b_1+b_2\cos(2\theta_0)-2R_b\sin\theta_0\right)\left(b_2+b_1\cos(2\theta_0)+2R_b\sin\theta_0\right)}{\left(b_2+b_1\cos(2\theta_0)-2R_b\sin\theta_0\right)\left(b_1+b_2\cos(2\theta_0)+2R_b\sin\theta_0\right)}\right].\\
\end{split}
\end{equation}
Note that here we did not use any regulator. This is because the distances between points on the brane away from the origin when $\theta_0\neq \pi/2$ are finite. 

Using the expressions above we can write the condition in which the connected phase dominates as
\begin{equation}
    \lim_{\epsilon\to 0}\left[\mathcal{A}^{reg}_{\chi_{\text{dis}}(A)}+\mathcal{A}_{\chi_{\text{dis}}(B)}-\mathcal{A}^{reg}_{\chi_{\text{con}}(A\cup B)}\right]> 0,
\end{equation}
which can be rewritten as follows
\begin{equation}
\label{GeneralizedConPhaseIneq}
    \frac{\left(-\Delta t^2+\Delta x_1^2+\Delta z_1^2\right)\left(-\Delta t^2+\Delta x_2^2+\Delta z_2^2\right)}{\Delta x_A^2\Delta z_1\Delta z_2}<e^{\mathcal{A}_{\chi_{\text{dis}}(B)}/L}.
\end{equation}
We can rearrange the inequality given in Eq. (\ref{GeneralizedConPhaseIneq}) and complete the square in $\Delta t^2$ to obtain
\begin{equation}
    \begin{split}
        &\left[\Delta t^2-\frac{\Delta x_1^2+\Delta x_2^2+\Delta z_1^2+\Delta z_2^2}{2}\right]^2\\
        &-\left(\frac{\Delta x_1^2-\Delta x_2^2+\Delta z_1^2-\Delta z_2^2}{2}\right)^2-\Delta x_A^2\Delta z_1\Delta z_2e^{\mathcal{A}_{\chi_{\text{dis}}(B)}/L}<0.\\
    \end{split}
\end{equation}
Using the facts that all the $\Delta$ quantities are non-negative and also $-\Delta t^2+\Delta x_{i}^2+\Delta z_{i}^2>0$ we can isolate for $\Delta t^2$ and arrive at
\begin{equation}
\begin{split}
     &\Delta t^2>\Delta t^2_{\text{con}}\\
     &\Delta t^2_{\text{con}}=\frac{\Delta x_1^2+\Delta x_2^2+\Delta z_1^2+\Delta z_2^2}{2}-\sqrt{\left(\frac{\Delta x_1^2-\Delta x_2^2+\Delta z_1^2-\Delta z_2^2}{2}\right)^2+\Delta x_A^2\Delta z_1\Delta z_2e^{\mathcal{A}_{\chi_{\text{dis}}(B)}/L}},\\
\end{split}
\label{DeltatForConnectedPhase}
\end{equation}
which is precisely what we mention in Eq. (\ref{ConnectedConditionDominance}) with $\Delta x_A=2a_-$.

In Section \ref{DominanceOfConnectedPhaseSection}, we discuss when the connected phase becomes dominant. In particular, we argued that the conditions we wrote for EWN would be non-trivial only if $\Delta t_{EWN}-\Delta t_{\text{con}}>0$. This is because satisfying such a condition would imply that as $|\Delta t|\to\Delta t_{EWN}$ the connected phase would dominate and our condition in Eq. (\ref{FinalEWNCondition}) will need to be enforced in the connected phase. In particular, since $\Delta t_{EWN}=\min\{\Delta t_{A,EWN},\Delta t_{2,EWN}\}$, we should analyze $\Delta t_{2,EWN}-\Delta t_{\text{con}}$ and $\Delta t_{A,EWN}-\Delta t_{\text{con}}$ and determine their sign for various parameters. Since there are many parameters involved, we adopt the following approach in our analysis. First, we fix $b_{1,2}$ and $a_{\pm}$. Then we plot $\Delta t_{A,EWN}-\Delta t_{\text{con}}$ or $\Delta t_{2,EWN}-\Delta t_{\text{con}}$ as a function of $\theta_0\in(0,\pi/2)$. This will give a single curve. Whenever the curve is above the $x$-axis the connected phase will dominate, and when it is on or below the $x$-axis the disconnected phase will dominate. 

\begin{figure}[ht!]
\centering
\includegraphics[width=140mm]{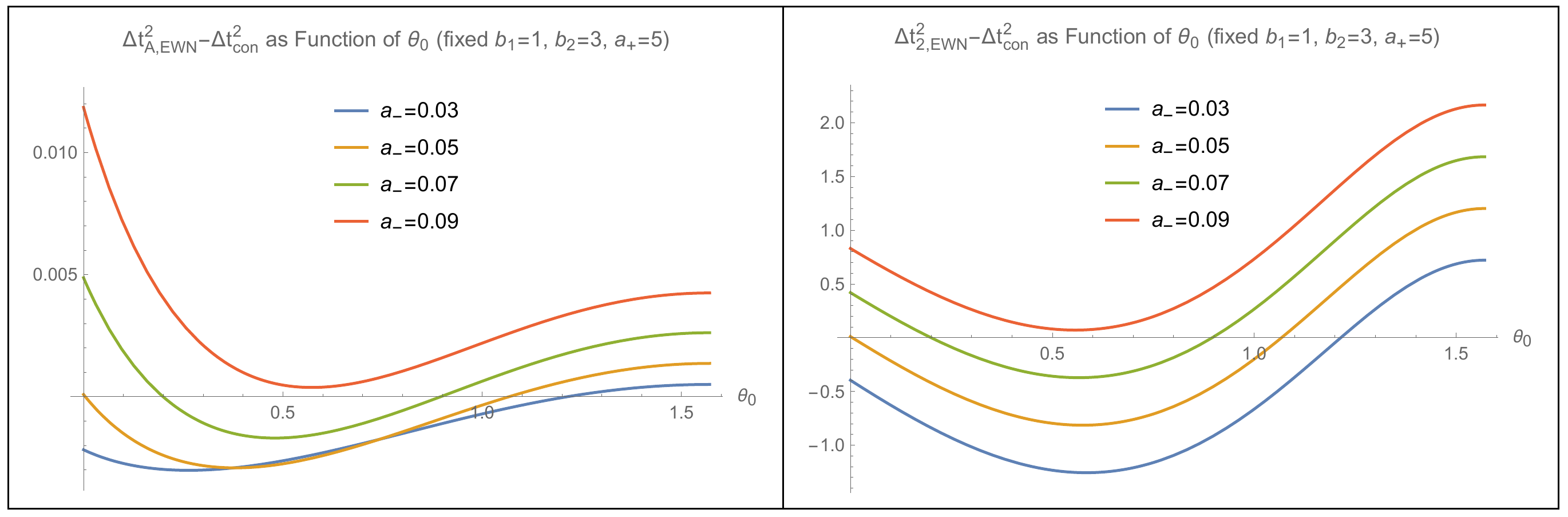}
\caption{Plot of $\Delta t_{A,EWN}^2-\Delta t_{\text{con}}^2$ (left frame) and $\Delta t_{2,EWN}^2-\Delta t_{\text{con}}^2$ (right frame) as function of $\theta_0$ for various choices of $a_-$ with $b_1=1, b_2=3, a_+=5$. \label{connectedGenricval}}
\end{figure}

For example, in Figure \ref{connectedGenricval} we plot both $\Delta t_{A,EWN}^2-\Delta t_{\text{con}}^2$ (left frame) and $\Delta t_{A,EWN}^2-\Delta t_{\text{con}}^2$ (right frame) as functions of $\theta_0$ for fixed $b_{1,2}$ and $a_+$, and each colored line in the graph corresponds to a particular choice of $a_-$.\footnote{The reader might be wondering why we take the difference of squares rather than difference of the quantities themselves. It is just to reduce the number of square roots in our expression which can cause reduce numerical error when we get close to machine precision. We know the quantities themselves are always non-negative so can do this without affecting our conclusions.}  At first glance, we can see for certain choices of parameters the lines can be above or below the $x$-axis. We can also see that when comparing between the left and right frames, lines of the same color (i.e. the lines that are plotted for same parameters $b_{1,2}$ and $a_{\pm}$) are above or below the $x$-axis for the exact same interval(s) in $\theta_0$. This can be seen visually in the plots, and verified through more precise numerical calculations. This is not just true for the particular sample we show but also more generally. This in fact is likely not a coincidence; it is a consequence of another fact that we numerically verified for many different choices of parameters. Which is, the roots of $\Delta t_{A,EWN}-\Delta t_{\text{con}}$ and $\Delta t_{2,EWN}-\Delta t_{\text{con}}$ are precisely given by solutions to $2a_-=\ell_*$, where we introduced $\ell_*$ in Eq. (\ref{DefinitionOfellstar}). In fact, recalling that by definition $\ell_*=2a_-\Leftrightarrow\Delta t_{A,EWN}=\Delta t_{2,EWN}$ makes it very clear why the roots match for the same colored lines in Figure \ref{connectedGenricval}. 

\begin{figure}[ht!]
\centering
\includegraphics[width=140mm]{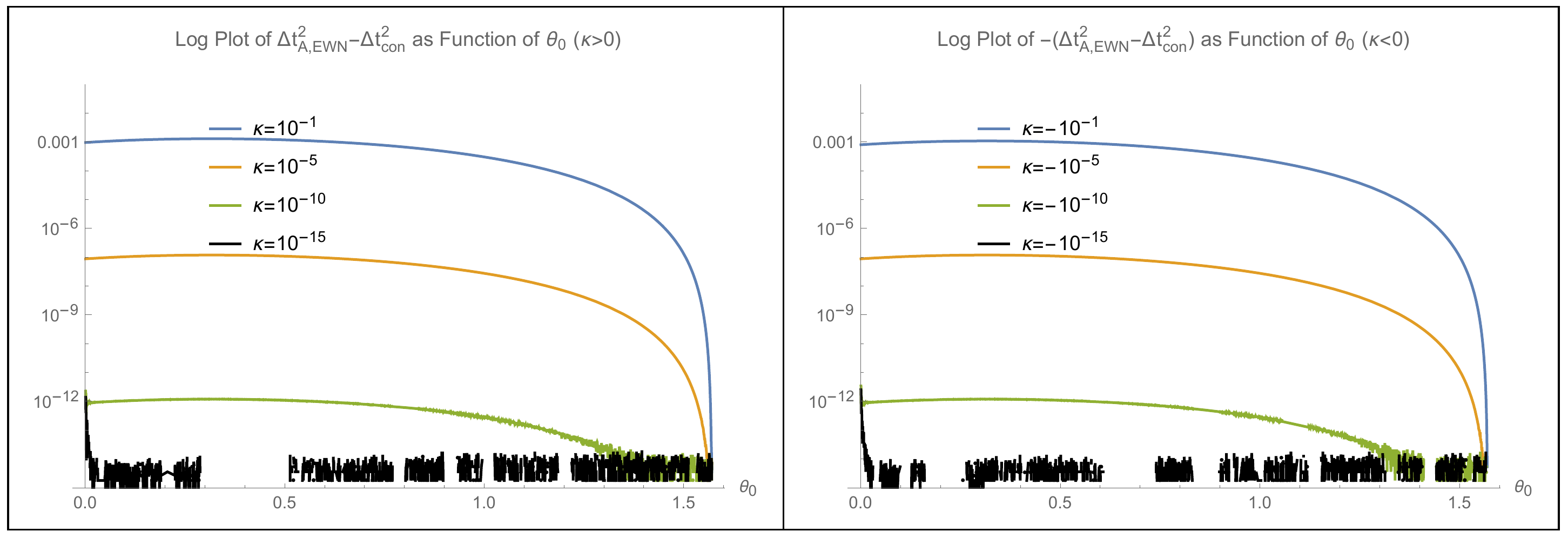}
\caption{Log plot of $\Delta t_{A,EWN}^2-\Delta t_{\text{con}}^2$ for various $\kappa>0$ (left frame) and $-(\Delta t_{A,EWN}^2-\Delta t_{\text{con}}^2)$ for various $\kappa<0$ (right frame) as function of $\theta_0$ with $b_1=1, b_2=3, a_+=5$.  \label{conConditionNearBounds}}
\end{figure}

What all these numerical results suggest is that anytime $2a_->\ell_*$ we have $\Delta t_{A,EWN}-\Delta t_{\text{con}}>0\Rightarrow \Delta t_{2,EWN}-\Delta t_{\text{con}}>0$\footnote{The implications occurs because we defined $\ell_*$ such that when $2a_->\ell_*\Leftrightarrow\Delta t_{A,EWN}<\Delta t_{2,EWN}$. This implies that if we know $\Delta t_{A,EWN}-\Delta t_{\text{con}}>0$ then we also know $\Delta t_{2,EWN}-\Delta t_{\text{con}}>0$.} and anytime $2a_-<\ell_*$ we have $\Delta t_{A,EWN}-\Delta t_{\text{con}}<0\Rightarrow \Delta t_{2,EWN}-\Delta t_{\text{con}}<0$.\footnote{The implications occurs because we defined $\ell_*$ such that when $2a_-<\ell_*\Leftrightarrow\Delta t_{2,EWN}<\Delta t_{A,EWN}$. This implies that if we know $\Delta t_{A,EWN}-\Delta t_{\text{con}}<0$, then we also know $\Delta t_{2,EWN}-\Delta t_{\text{con}}<0$.} We can verify this further through a more refined analysis by defining $a_-=\frac{\ell_*}{2}(1+\kappa)$. When we fix $\kappa$ to $\kappa<0$ we will always have $2a_-<\ell_*$ and when $\kappa>0$ we will always be in the case where $2a_->\ell_*$. In Figure \ref{conConditionNearBounds} we make a Log plot $\Delta t_{A,EWN}^2-\Delta t_{\text{con}}^2$ (for $\kappa>0$, in left frame of Figure \ref{conConditionNearBounds}) and a Log plot $-(\Delta t_{A,EWN}^2-\Delta t_{\text{con}}^2)$ (for $\kappa<0$, in right frame of Figure \ref{conConditionNearBounds}) as a function of $\theta_0$. 
In the left frame we can see that when $\kappa>0$ the value of $\Delta t_{A,EWN}^2-\Delta t_{\text{con}}^2>0$ and in the right frame we see when $\kappa<0$ the value of $-(\Delta t_{A,EWN}^2-\Delta t_{\text{con}}^2)>0\Leftrightarrow \Delta t_{A,EWN}^2-\Delta t_{\text{con}}^2<0$. Furthermore, as $|\kappa|\to 0$, the value of $\Delta t_{A,EWN}^2-\Delta t_{\text{con}}^2\to 0$ as expected (one can check that similar results appear for more general choices of parameters as well, the wild oscillations we see in the curves as $\kappa\to 0$ in the plots are artifacts of approaching machine precision). 

Together, all these numerical results allow us to fill in the first two rows of Table \ref{tab:my_label}, which tell us if the connected or disconnected phase dominates as $|\Delta t|\to \Delta t_{2,EWN},\Delta t_{A,EWN}$.        

\subsection{Numerical Analysis of Spacelike Separation of $\chi_{1,2}$}
\label{AppendixProvingCasesWhenNotCauchy}
\begin{figure}[ht!]
\centering
\includegraphics[width=100mm]{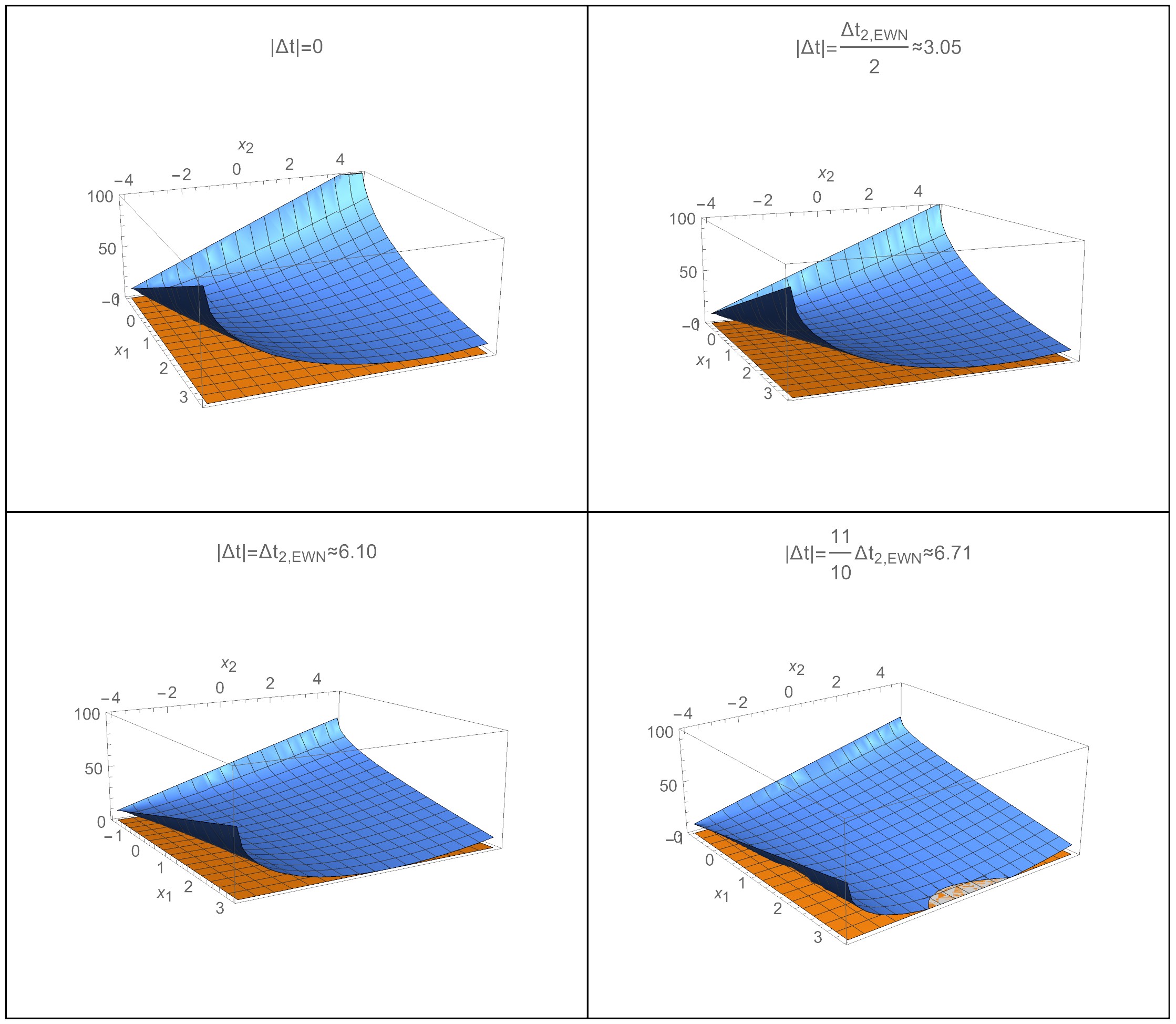}
\caption{Same plot of $S$ (blue surface) and zero-plane (orange plane) for sample parameters $b_1=7,b_2=10,a_1=1,a_2=4,\theta_0=1/2$ for various values of $|\Delta t|$ shown in each frame.  \label{SLSheetPlotschi12}}
\end{figure}

\begin{figure}[ht!]
\centering
\includegraphics[width=100mm]{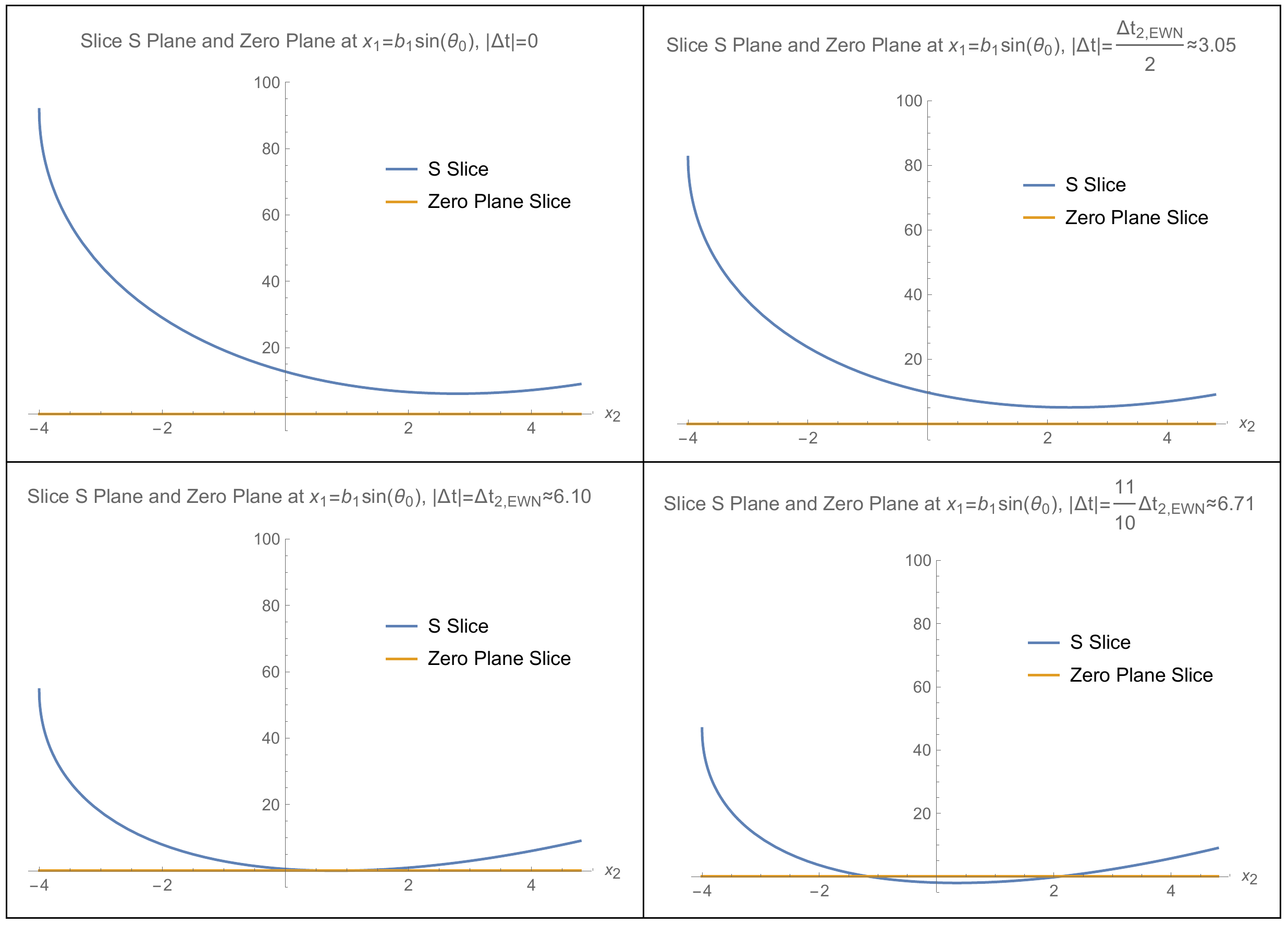}
\caption{Slice of surfaces depicted in Figure \ref{SLSheetPlotschi12} along $x_1=b_1\sin\theta_0$ plane. \label{SLSheetSlicePlots}}
\end{figure}

This Appendix discusses the details of our numerical analysis which further support the conjecture in Section \ref{SigmaABCauchyExploreSection} that $|\Delta t|<\Delta t_{2,EWN}\Leftrightarrow\chi_1$ and $\chi_2$ are spacelike separated. To explore the conjecture numerically we define a function of two variables
\begin{equation}
    S(x_1,x_2)=-(t_1(x_1)-t_2(x_2))^2+(x_1-x_2)^2+(z_1(x_1)-z_2(x_2))^2,
\end{equation}
over the range of parameters $x_i\in[-a_i,b_i\sin\theta_0]$ where $z_i(x),t_i(x)$ are defined in Eq. (\ref{Z1RT}). We know that $\chi_1$ will be spacelike separated from $\chi_2$ iff $S(x_1,x_2)>0$. We analyze this numerically by first fixing the values of $a_{1,2},b_{1,2},\theta_0$ and $\delta t$, where $|\Delta t|=\Delta t_{2,EWN}+\delta t$ and plotting the surface $S(x_1,x_2)$. Then we can determine if $\chi_1$ is spacelike to $\chi_2$ by simply checking if the surface $S$ is above the zero-plane. What we find in our numerical investigation is that in general whenever $\delta t<0$ then $S$ remains above the zero plane and whenever $\delta t>0$ a portion of $S$ dips below the zero plane. For example, in Figure \ref{SLSheetPlotschi12} we plot $S$ as a blue surface and the zero plane as the orange surface with $b_1=7,b_2=10,a_1=1,a_2=4,\theta_0=1/2$ we have four frames for this choice of parameters with $|\Delta t|=0$ (top left frame), $|\Delta t|=\frac{1}{2}\Delta t_{2,EWN}$ (top right frame), $|\Delta t|=\Delta t_{2,EWN}$ (bottom left frame), $|\Delta t|=\frac{11}{10}\Delta t_{2,EWN}$ (bottom right frame). 
We can clearly see that when $|\Delta t|<\Delta t_{2,EWN}$ the blue surface stays above the zero plane, indicating that $\chi_1$ and $\chi_2$ are spacelike. We can also see that the shape of the blue plane is such that it gets closest to the zero plane along the $x_1=b_1\sin\theta_0$ slice. In Figure \ref{SLSheetSlicePlots} we plot this particular slice in four frames with the same set of parameters as the frames in Figure \ref{SLSheetPlotschi12}. 
We can clearly see that when $|\Delta t|=\Delta t_{2,EWN}$ the blue plane will marginally touch the zero plane. In fact, the two bottom frames in Figure \ref{SLSheetSlicePlots} are exactly the results we proved in Section \ref{SigmaABCauchyExploreSection} the two top frames of Figures \ref{SLSheetPlotschi12} and \ref{SLSheetSlicePlots} corroborate our claim that, indeed, $|\Delta t|<\Delta t_{2,EWN}\Leftrightarrow \chi_1$ and $\chi_2$ are spacelike separated. For more generic parameters, one can check that the results are similar.

\section{Supplemental Calculations for Section 4}
\subsection{Derivation of Thermal RT Surface}
\label{ThermalRTSurfaceDeivationAppendix}
In this subsection we will derive the equations of motion to determine the thermal RT surface. We define the Lagrangian associated to the functional given in Eq.  ($\ref{ThermalRTFunctional}$) as
\begin{equation}
    \mathcal{L}(t,\dot{t},z,\dot{z};x)=\frac{L}{z}\sqrt{-\tilde{f}(z)\dot{t}^2+\frac{\dot{z}^2}{\tilde{f}(z)}+1}.
\end{equation}
Since the Lagrangian has no explicit $x$ dependence so we can define a conserved Hamiltonian associated with $x$-translations given by
\begin{equation}
    \mathcal{H}_x=\dot{z}\frac{\partial\mathcal{L}}{\partial \dot{z}}+\dot{t}\frac{\partial\mathcal{L}}{\partial \dot{t}}-\mathcal{L}=-\frac{L^2}{z^2\mathcal{L}}.
\end{equation}
We have the following equation of motion for $t$
\begin{equation}
    \frac{\partial\mathcal{L}}{\partial \dot{t}}=c_t,
\end{equation}
where $c_t$ is constant. Plugging this into the Lagrangian we obtain the following expression for $\dot{t}$
\begin{equation}
    \dot{t}=-c_t\frac{\mathcal{L}z^2}{L^2\left(1-\frac{z^2}{z_+^2}\right)}=\frac{c_t}{\mathcal{H}_x\left(1-\frac{z^2}{z_+^2}\right)}.
\end{equation}
Since $z<z_+$ we know that sign of the derivative will not change along the RT curve. Since we want to anchor the thermal RT surface to a constant time slice interval we must require that $c_t=0$. This in turn means $\dot{t}=0 \Rightarrow t(x)=t_{\text{Bdry}}$. If we plug this back into the equation for $\mathcal{H}_x$ and rearrange for $\dot{z}$ we get
\begin{equation}
    \dot{z}^2=\frac{\left(z_+^2-z^2\right)\left(L^2-\mathcal{H}_x^2z^2\right)}{\mathcal{H}_x^2z_+^2z^2}.
\end{equation}
We can see that $\dot{z}=0$ when $z=z_{tp}=L/|\mathcal{H}_x|$, this is the turning point for the RT curve where the derivative $\dot{z}$ goes to zero. We want an RT surface that starts at $x=-a$ extends into the bulk and then returns to the boundary at $x=a$. Rearranging the expression $\dot{z}$ we obtain the following integral equation for the ``left portion'' of the RT curve where $\dot{z}\geq 0$ 
\begin{equation}
    \int_{-a}^xdx=\int_0^{z(x)}\frac{z_+z}{\sqrt{(z_+^2-z^2)(z_{tp}^2-z^2)}}dz.
\end{equation}
We can explicitly integrate each side of the expression and with some algebraic/trigonometric manipulations we arrive at
\begin{equation}
    z_{left}(x)=\sqrt{z_+^2-\left[z_+\cosh\left(\frac{x+a}{z_+}\right)-z_{tp}\sinh\left(\frac{x+a}{z_+}\right)\right]^2}.
\end{equation}
By requiring that $z_{tp}=z(x_{tp})$ we obtain the following relation between $x_{tp}$ and $z_{tp}$
\begin{equation}
    z_{tp}=z_+\tanh\left(\frac{a+x_{tp}}{z_+}\right).
\end{equation}
Plugging this back into the expression for $z_{left}(x)$ gives
\begin{equation}
    z_{left}(x)=z_+\sqrt{1-\frac{\cosh^2\left(\frac{x-x_{tp}}{z_+}\right)}{\cosh^2\left(\frac{a+x_{tp}}{z_+}\right)}}.
\end{equation}
For the ``right portion'' of the RT curve we have to perform the following integrals
\begin{equation}
    \int_{x}^adx=-\int_{z(x)}^0 \frac{z_+z}{\sqrt{(z_+^2-z^2)(z_{tp}^2-z^2)}}dz=\int_{0}^{z(x)} \frac{z_+z}{\sqrt{(z_+^2-z^2)(z_{tp}^2-z^2)}}dz.
\end{equation}
Based on our previous work on the left portion we can show the right portion of the RT curve will take the form
\begin{equation}
    z_{right}(x)=\sqrt{z_+^2-\left[z_+\cosh\left(\frac{a-x}{z_+}\right)-z_{tp}\sinh\left(\frac{a-x}{z_+}\right)\right]^2}.
\end{equation}
Once again, the requirement that $z_{tp}=z(x_{tp})$ gives
\begin{equation}
    z_{tp}=z_+\tanh\left(\frac{a-x_{tp}}{z_+}\right).
\end{equation}
Using this we write
\begin{equation}
    z_{right}(x)=z_+\sqrt{1-\frac{\cosh^2\left(\frac{x-x_{tp}}{z_+}\right)}{\cosh^2\left(\frac{a-x_{tp}}{z_+}\right)}}.
\end{equation}
The condition that the two portions of the RT curves should be smoothly glued at $x=x_{tp}$ and $z=z_{tp}$ (i.e. $z_{left}(x_{tp})=z_{right}(x_{tp})$) gives the following equation
\begin{equation}
    \sqrt{1-\frac{\cosh^2\left(\frac{x-x_{tp}}{z_+}\right)}{\cosh^2\left(\frac{a-x_{tp}}{z_+}\right)}}=\sqrt{1-\frac{\cosh^2\left(\frac{x-x_{tp}}{z_+}\right)}{\cosh^2\left(\frac{a+x_{tp}}{z_+}\right)}}.
\end{equation}
We can see that this requires the setting of $x_{tp}=0$ as expected. So our final result for the thermal RT surface in Schwarzschild coordinates anchored to the endpoints of a constant time slice interval on the boundary centered around $x=0$ of size $2a$ is
\begin{equation}
    \begin{split}
        &t(x)=t_{\text{Bdry}}\\
        &z(x)=z_+\sqrt{1-\frac{\cosh^2\left(\frac{x}{z_+}\right)}{\cosh^2\left(\frac{a}{z_+}\right)}}.\\
    \end{split}
\end{equation}
This gives the result we wrote in Eq. (\ref{ThermalRTSchBdryAnchor}). \\

\noindent \textbf{Schwarzschild to Kruskal Coordinates:}

\noindent Now that we have derived the thermal RT surface in Schwarzschild coordinates we can use the expression given for the thermal RT surface along with the transformations given in Eq. (\ref{SchwToKruskalCoordTrans}). Lets begin by computing the thermal surface in the left exterior which is anchored to an interval on the boundary at time $t=t_L=t_{\text{Bdry}}$. We have the following expressions for $u$ and $v$
\begin{equation}
    \begin{split}
        &v(x)=\frac{\tau(x)-y(x)}{2}=\arctan\left[e^{-\frac{t_{\text{Bdry}}}{z_+}}\sqrt{\frac{z_+-z_L(x)}{z_++z_L(x)}}\right]\\
        &u(x)=\frac{\tau(x)+y(x)}{2}=\arctan\left[-e^{\frac{t_{\text{Bdry}}}{z_+}}\sqrt{\frac{z_+-z_L(x)}{z_++z_L(x)}}\right]\\
        &z_L(x)=z_+\sqrt{1-\frac{\cosh^2\left(\frac{x}{z_+}\right)}{\cosh^2\left(\frac{a}{z_+}\right)}}.\\
    \end{split}
\end{equation}
With these expressions we can write
\begin{equation}
    \begin{split}
        &\sin\left(\tau(x)\right)=\sin\left[v(x)+u(x)\right]=-\sqrt{\frac{(z_L^2-z_+^2)\sinh^2\left(\frac{t_{\text{Bdry}}}{z_+}\right)}{z_L^2\sinh^2\left(\frac{t_{\text{Bdry}}}{z_+}\right)-z_+^2\cosh^2\left(\frac{t_{\text{Bdry}}}{z_+}\right)}}\\
        &=-\frac{\cosh\left(\frac{x}{z_+}\right)\sinh\left(\frac{t_{\text{Bdry}}}{z_+}\right)}{\sqrt{\cosh^2\left(\frac{a}{z_+}\right)+\cosh^2\left(\frac{x}{z_+}\right)\sinh^2\left(\frac{t_{\text{Bdry}}}{z_+}\right)}}\\
        &\sin\left(y(x)\right)=\sin\left[u(x)-v(x)\right]=-\frac{\cosh\left(\frac{t_{\text{Bdry}}}{z_+}\right)\cosh\left(\frac{x}{z_+}\right)}{\sqrt{\cosh^2\left(\frac{a}{z_+}\right)+\cosh^2\left(\frac{x}{z_+}\right)\sinh^2\left(\frac{t_{\text{Bdry}}}{z_+}\right)}}.\\
    \end{split}
\end{equation}
This gives Eq. (\ref{ThermalRTKruskalLeftExt}).

Next we proceed with writing the thermal RT surface anchored to a constant time slice interval $t=t_{\text{Br}}$ on an ETW brane in the right exterior. We have the following expressions for $u$ and $v$ in the right exteriors 
\begin{equation}
    \begin{split}
        &v(x)=\frac{\tau(x)-y(x)}{2}=\arctan\left[-e^{-\frac{t_{\text{Br}}}{z_+}}\sqrt{\frac{z_+-z_R(x)}{z_++z_R(x)}}\right]\\
        &u(x)=\frac{\tau(x)+y(x)}{2}=\arctan\left[e^{\frac{t_{\text{Br}}}{z_+}}\sqrt{\frac{z_+-z_R(x)}{z_++z_R(x)}}\right]\\
        &z_R(x)=z_+\sqrt{1-\frac{\cosh^2\left(\frac{x}{z_+}\right)}{\cosh^2\left(\frac{a}{z_+}\right)}\left(1-\frac{z_{Br}^2}{z_+^2}\right)}.\\
    \end{split}
\end{equation}
With similar manipulations as before we find that a thermal RT surface in the right exterior anchored to a constant time slice interval ($t=t_{\text{Br}}$) centered around $x=0$ of length $2a$ is given by
\begin{equation}
\begin{split}
    &\sin\left(\tau(x)\right)=\sin[v(x)+u(x)]=\frac{\sqrt{1-\frac{z_{Br}^2}{z_+^2}}\cosh\left(\frac{x}{z_+}\right)\sinh\left(\frac{t_{\text{Br}}}{z_+}\right)}{\sqrt{\cosh^2\left(\frac{a}{z_+}\right)+\left(1-\frac{z_{Br}^2}{z_+^2}\right)\cosh^2\left(\frac{x}{z_+}\right)\sinh^2\left(\frac{t_{\text{Br}}}{z_+}\right)}}\\
    &\sin(y(x))=\sin[u(x)-v(x)]=\frac{\sqrt{1-\frac{z_{Br}^2}{z_+^2}}\cosh\left(\frac{x}{z_+}\right)\cosh\left(\frac{t_{\text{Br}}}{z_+}\right)}{\sqrt{\cosh^2\left(\frac{a}{z_+}\right)+\left(1-\frac{z_{Br}^2}{z_+^2}\right)\cosh^2\left(\frac{x}{z_+}\right)\sinh^2\left(\frac{t_{\text{Br}}}{z_+}\right)}}.\\
\end{split}
\end{equation}
This gives the expressions in Eq. (\ref{ThermalRTRightExtBr}). We can check that the following identities hold,
\begin{equation}
\begin{split}
    & \frac{\cos^2\left(y(\pm a)\right)}{\cos^2\left(\tau(\pm a)\right)}=\frac{z_{Br}^2}{z_+^2}\\
    &\frac{\sin(y(x))+\sin(\tau(x))}{\sin(y(x))-\sin(\tau(x))}=e^{\frac{2t_{\text{Br}}}{z_+}},\\
\end{split}
\end{equation}
which is exactly what we would expect from the coordinate transformations given in Eq. (\ref{KruskaltoSchCoordTrans}).

\subsection{Derivation of Connected RT Surface}
\label{ConnectedRTSurfaceDeivationAppendix}
\noindent \textbf{Derivation in Kruskal Coordinates:}

\noindent We will go over the derivation of finding the connected RT surface that connects the brane and boundary intervals that are placed in the same position in the ``$x$-direction'' (i.e. brane and boundary intervals have endpoints that are located at the same $x$-coordinate location). The connected RT surface will consist of two line segments that go through the black hole horizon and connect the endpoints of the intervals on the brane and boundary which may exist on different time slices. To obtain such line segments we must extremize the functional given in Eq. (\ref{ConnRTKruskFunc}). The Lagrangian associated to the functional is
\begin{equation}
    \mathcal{L}=\frac{\sqrt{L^2\left(-\dot{\tau}^2+1\right)+\frac{r_+^2}{L^2}\cos^2\tau\dot{x}^2}}{\cos y}.
\end{equation}
Since there is no explicit $x$-dependence we have
\begin{equation}
    \frac{\partial\mathcal{L}}{\partial \dot{x}}=c_x,
\end{equation}
where $c_x$ is constant. Plugging in the actual Lagrangian gives the following expression for $\dot{x}$
\begin{equation}
    \dot{x}=c_x\frac{L^2}{r_+^2}\frac{\cos^2y}{\cos^2\tau}\mathcal{L}.
\end{equation}
We can see that within the bulk geometry a spacelike RT surface will have a derivative that is monotonic for $c_x\neq 0$. However, we also know that we must require the RT curve to start and end at the same $x$-coordinate. This is only possible if $c_x=0$. Therefore, we restrict ourselves to the set of RT surfaces with $\dot{x}=0$. This means we can deal with a much simpler Lagrangian
\begin{equation}
    \mathcal{L}=L\frac{\sqrt{1-\dot{\tau}^2}}{\cos y},
\end{equation}
which is what we wrote in Eq. (\ref{ConnRTKruskSimpleVerFunc}). Since we no longer have any explicit $\tau$ dependence on the Lagrangian we have another conserved quantity namely,
\begin{equation}
    c_\tau L=\frac{\partial \mathcal{L}}{\partial \dot{\tau}}.
\end{equation}
The extra factor $L$ is just a convenient normalization. Plugging in the appropriate Lagrangian gives
\begin{equation}
    \dot{\tau}=\pm\frac{1}{\sqrt{1+\frac{1}{c_\tau^2\cos^2y}}}.
\end{equation}
The sign of the above solution is determined by $\text{sign}(\tau_{\text{Br}}-\tau_{\text{Bdry}})$. In particular, when $\operatorname{sign}{\tau_{\text{Br}}-\tau_{\text{Bdry}}}>0$ we have
\begin{equation}
    \int_{\tau_{\text{Bdry}}}^{\tau(y)} d\tau=\int_{-\pi/2}^y\frac{dy'}{\sqrt{1+\frac{1}{c_\tau^2\cos^2y'}}}.
\end{equation}
The integrals can be done explicitly to obtain
\begin{equation}
    \tau(y)-\tau_{\text{Bdry}}=\arctan\left[\frac{c_\tau \sin y'}{\sqrt{1+c_\tau^2\cos^2y'}}\right]\bigg\vert_{-\pi/2}^y.
\end{equation}
Using the relation between $\arcsin$ and $\arctan$ allows us to write
\begin{equation}
    \tau(y)-\tau_{\text{Bdry}}=\arcsin\left[\frac{c_\tau}{\sqrt{1+c_\tau^2}}\sin y\right]+\arcsin\left[\frac{c_\tau}{\sqrt{1+c_\tau^2}}\right].
\end{equation}
This gives us the expression for the connected RT surface in Eq. (\ref{ConnRTSurfKruskCord}). Note that in general that $\text{sign}(c_\tau)=\text{sign}(\tau_{\text{Br}}-\tau_{\text{Bdry}})$. We can see for a fixed $\tau_{\text{Bdry}}$ the value of $c_\tau$ fixes where in time the connected surface connects to the brane.\\ 

\noindent \textbf{Connected Surface in Schwarzschild Coordinates:}

\noindent Now that we have obtained an expression for the connected RT curve in Kruskal coordinates we can express it in the right exterior in Schwarzschild coordinates with the help of the coordinate transformations given in Eq. (\ref{KruskaltoSchCoordTrans}). Since we are interested analyzing entanglement wedge nesting in the right exterior in Schwarzschild coordinates. We will find it convenient to use the following expression for the connected surface in Kruskal coordinates 
\begin{equation}
    \begin{split}
        &\tau(y)=\tau_{\text{Br}}+\arcsin(c_t\sin y)-\arcsin(c_t\sin y_{\text{Br}})=\tau_*+\arcsin(c_t\sin y)\\
        &c_t=\frac{c_\tau}{\sqrt{1+c_\tau^2}}.\\
    \end{split}
\end{equation}
The connected RT surface ends on a brane located at $y=y_{\text{Br}}$ where $0\leq y_{\text{Br}}\leq \pi/2$ and at $\tau=\tau_{\text{Br}}$ where $|\tau_{\text{Br}}|\leq y_{\text{Br}}$. Recall the following relation between Kruskal and Schwarzschild coordinates
\begin{equation}
    \begin{split}
        &  t=\frac{L^2}{2r_+}\ln\left[\frac{\sin y+\sin\tau}{\sin y-\sin\tau}\right], \qquad r=r_+\frac{\cos\tau}{\cos y}.\\
    \end{split}
\end{equation}
We can invert the relations above to obtain
\begin{equation}
\label{relationinver}
    \begin{split}
        &\sin y= \pm\frac{1}{\sqrt{1+\frac{1}{\left(\frac{r^2}{r_+^2}-1\right)\cosh^2\left(\frac{r_+t}{L^2}\right)}}}\\
        &\sin \tau=\pm \frac{\tanh\left(\frac{r_+t}{L^2}\right)}{\sqrt{1+\frac{1}{\left(\frac{r^2}{r_+^2}-1\right)\cosh^2\left(\frac{r_+t}{L^2}\right)}}}\Rightarrow \cos\tau=\frac{\sqrt{\frac{r^2}{r^2-r_+^2}}}{\cosh\left(\frac{r_+t}{L^2}\right)\sqrt{1+\frac{1}{\left(\frac{r^2}{r_+^2}-1\right)\cosh^2\left(\frac{r_+t}{L^2}\right)}}},\\
    \end{split}
\end{equation}
where the ``$+$'' is for the right exterior, the ``$-$'' is for the left exterior. We focus on the right exterior and take the ``$+$'' sign. From the equation of the connected surface we have
\begin{equation}
\label{consurfEQBr}
    \begin{split}
        &c_t\sin y=\sin(\tau-\tau_*)=\cos\tau_*\sin\tau-\sin\tau_*\cos\tau.\\
    \end{split}
\end{equation}
\begin{comment}
It is straightforward to see that $\text{sign}(c_t)=\text{sign}\sin(\tau-\tau_*)$. Recall that $\tau_*=\tau_{\text{Br}}-\arcsin(c_t\sin y_{\text{Br}})$. Since $|\tau_{\text{Br}}|<y_{\text{Br}}$ we have the following bounds:
\begin{equation}
    \begin{split}
        &-\pi\leq \tau_{\text{Br}}-y_{\text{Br}}<\tau_*<\tau_{\text{Br}}+y_{\text{Br}}\leq\pi\\
        &|\tau-\tau_*|< y_{\text{Br}}\\
    \end{split}
\end{equation}
Due to the first inequality we can we know $sign(c_t)=sign(\tau-\tau_*)$. 
\end{comment}
Now we plug in the expressions we had for $\sin\tau$ and $\sin y$ in Eq. (\ref{relationinver}) into Eq. (\ref{consurfEQBr}) to obtain
\begin{equation}
    \begin{split}
        &c_t=\pm \cos\tau_*\tanh\left(\frac{r_+t}{L^2}\right)-\sin\tau_*\frac{\sqrt{\frac{r^2}{r^2-r_+^2}}}{\cosh\left(\frac{r_+t}{L^2}\right)}\\
        &\Rightarrow c_t\cosh\left(\frac{r_+t}{L^2}\right)=\pm\cos\tau_*\sinh\left(\frac{r_+t}{L^2}\right)-\sin\tau_*\sqrt{\frac{r^2}{r^2-r_+^2}}.\\
    \end{split}
\end{equation}
Rearranging the expression above and recalling that $z=L^2/r$, we obtain the following expression for the connected RT surface in the right exterior in Schwarzschild coordinates
\begin{equation}
    \begin{split}
        &\left(\frac{z}{z_+}\right)^2=1-\frac{\sin^2\tau_*}{\left[\cos\tau_*\sinh\left(\frac{r_+t}{L^2}\right)-c_t\cosh\left(\frac{r_+t}{L^2}\right)\right]^2}\\
        &=1-\frac{\sin^2\tau_*}{\left[A\cosh\left(\frac{r_+\Delta t}{L^2}\right)+B\sinh\left(\frac{r_+\Delta t}{L^2}\right)\right]^2}\\
        &A=\cos\tau_*\sinh\left(\frac{r_+t_{\text{Br}}}{L^2}\right)-c_t\cosh\left(\frac{r_+t_{\text{Br}}}{L^2}\right)\\
        &B=\cos\tau_*\cosh\left(\frac{r_+t_{\text{Br}}}{L^2}\right)-c_t\sinh\left(\frac{r_+t_{\text{Br}}}{L^2}\right)\\
        &\Delta t=t-t_{\text{Br}}.\\
    \end{split}
\end{equation}
An important point to note in the expression above is the issue of the sign of $\Delta t$. In particular, depending on the precise value of $c_t$ the surface would extend away from the brane toward positive $t$ or negative $t$. For example, if $c_t$ was fixed such that the connected surface moves toward larger $t>t_{\text{Br}}$ then we certainly cannot have $\Delta t<0$ (at least in a arbitrarily small neighborhood of the brane). So the central question is how to understand if we should be in the case where $\Delta t>0$ or $\Delta t<0$ for a given $c_t$. At first glance this may appear to be a rather complicated issue due to the expressions we wrote. However, there is a useful ``trick'' we can use to circumvent this problem.  

To begin we do an expansion around the $\Delta t=0$ (near the brane) to get
\begin{equation}
    \begin{split}
        \left(\frac{z}{z_+}\right)^2=\left[1-\frac{\sin^2\tau_*}{A^2}\right]+\frac{2r_+\sin^2\tau_*}{A^2L^2}\frac{B}{A}\Delta t.
    \end{split}
\end{equation}
One can check that the zeroth order term will evaluate to
\begin{equation}
    \begin{split}
        &1-\frac{\sin^2\tau_*}{A^2}=\frac{\cos^2y_{\text{Br}}}{\cos^2\tau_{\text{Br}}}=\frac{z_{\text{Br}}^2}{z_+^2}.\\
    \end{split}
\end{equation}
This makes sense since $\Delta t=0$ should place us on the brane. Now lets focus on the linear order term. The linear order term will dictate if the brane is moving above or below the $t=t_{\text{Br}}$. In particular we want to $z$ to increase so the surface moves outward away from the brane this means the linear term should have an overall positive sign. This in turn means that we must require that $\text{sign}(B/A)=\text{sign}(\Delta t)$. With the identity we showed for the zeroth order term we are able to write
\begin{equation}
    \begin{split}
        &\left(\frac{z}{z_+}\right)^2=1-\frac{1-\frac{z_{Br}^2}{z_{+}^2}}{\left[\cosh\left(\frac{r_+\Delta t}{L^2}\right)+\frac{B}{A}\sinh\left(\frac{r_+\Delta t}{L^2}\right)\right]^2}.\\
    \end{split}
\end{equation}
Now we have a precise characterization of how the value of $c_t$ affects if $\Delta t>0$ or if $\Delta t<0$. In particular, if $B/A>0\Rightarrow \Delta t>0$ and if $B/A<0\Rightarrow \Delta t<0$. We can ``automate/circumvent'' these cases by strictly dealing with the following expression involving absolute values   
\begin{equation}
    \begin{split}
        &\left(\frac{z}{z_+}\right)^2=1-\frac{1-\frac{z_{Br}^2}{z_{+}^2}}{\left[\cosh\left(\frac{r_+|\Delta t|}{L^2}\right)+\left|\frac{B}{A}\right|\sinh\left(\frac{r_+|\Delta t|}{L^2}\right)\right]^2}.\\
    \end{split}
\end{equation}
This precisely gives us the expression for the connected surface in Schwarzschild coordinates in the right exterior given in Eq. (\ref{RightExtConnRTSch}).

For the sake of completeness, we can do a similar analysis and write down the connected surface in the left exterior as well. It is given by
\begin{equation}
\begin{split}
\label{LeftExteriorConnSch}
    &\left(\frac{z}{z_+}\right)^2=1-\frac{1}{\left[\cosh\left(\frac{r_+|\tilde{\Delta t|}}{L^2}\right)+\left|\frac{\tilde{B}}{\tilde{A}}\right|\sinh\left(\frac{r_+|\tilde{\Delta t|}}{L^2}\right)\right]^2}\\
    &\tilde{A}=c_t\cosh\left(\frac{r_+t_{\text{Bdry}}}{L^2}\right)-\cos\tilde{\tau}_*\sinh\left(\frac{r_+t_{\text{Bdry}}}{L^2}\right)\\
    &\tilde{B}=c_t\sinh\left(\frac{r_+t_{\text{Bdry}}}{L^2}\right)-\cos\tilde{\tau}_*\cosh\left(\frac{r_+t_{\text{Bdry}}}{L^2}\right)\\
    &\tilde{\Delta t}=t-t_{\text{Bdry}}, \qquad \tilde{\tau}_*=\tau_{\text{Bdry}}+\text{arcsin} (c_t).\\
\end{split}
\end{equation}\\

\noindent \textbf{Useful Properties of ``$B/A$'':}

\noindent It is worthwhile to take the time to more carefully understand the properties of the ratio $B/A$. We can explicitly write it as follows in terms of $c_t$, $y_{\text{Br}}$, and $\tau_{\text{Br}}$
\begin{equation}
    \frac{B}{A}=\frac{c_t\cos^2 y_{\text{Br}}\tan\tau_{\text{Br}}-\sin y_{\text{Br}}\sqrt{1-c_t^2\sin^2 y_{\text{Br}}}}{\cos\tau_{\text{Br}}\left(c_t\sin y_{\text{Br}}-\tan\tau_{\text{Br}}\sqrt{1-c_t^2\sin^2y_{\text{Br}}}\right)}.
\end{equation}
Since we want to strictly consider intervals on the brane in the right exterior we must require that $|\tau_{\text{Br}}|<y_{\text{Br}}$ we also note that $|c_t|<1$. With this we conclude that: $ \lim_{c_t\to-1} (B/A)\geq 0$, $ \lim_{c_t\to1} (B/A)\leq 0$, $B/A$ is singular at $c_t=c_0=\frac{\sin \tau_{\text{Br}}}{\sin y_{\text{Br}}}$, and finally that $\frac{d}{dc_t}\left(\frac{B}{A}\right)$ is non-negative (i.e. monotonically increasing as $c_t$ increases) when $c_t\neq c_0$. 
Using these facts we can easily deduce that $B/A\geq 0$ when $-1<c_t<c_0$ and $B/A\leq 0$ when $c_0<c_t<1$. So the sign of $B/A$ swaps at $c_t=c_0$. One can easily check that the point of sign swapping occurs exactly when the connected extremal surface passes through the bifurcate horizon at $\tau=y=0$. These facts will become important when we discuss conditions for entanglement wedge nesting.

\subsection{Null Geodesics and Lightcones for Planar BTZ Background}
\label{NullGeodesAndLCBTZ}

In this appendix we will derive expressions the null geodesics propagating in the exterior spacetime of a planar BTZ black hole in Schwarzschild coordinates. The equations of motion for the geodesics will be obtained from the following Lagrangian (where we parameterize the geodesics trajectories using $z=L^2/r$)
\begin{equation}
\begin{split}
    &\mathcal{L}=\frac{L^2}{z^2}\left[-\tilde{f}(z)\dot{t}^2+\frac{1}{\tilde{f}(z)}+\dot{x}^2\right], \qquad \text{with} \qquad \tilde{f}(z)=1-\frac{z^2}{z_+^2}.\\
\end{split}
\end{equation}
Using the fact that the Lagrangian has no explicit $x$ or $t$ dependence we have the following conserved quantities
\begin{equation}
\begin{split}
    &-\frac{\tilde{f}(z)}{z^2}\dot{t}=a_t, \qquad \frac{\dot{x}}{z^2}=a_x.\\
\end{split}
\end{equation}
Dividing one equation by the other and squaring gives
\begin{equation}
    \left(\frac{\tilde{f}(z)\dot{t}}{\dot{x}}\right)^2=\alpha^2,
\end{equation}
where $\alpha$ is some constant. Finally we can use the condition for having a null geodesic (i.e. $\mathcal{L}=0$) to obtain the following first order equations that can be easily integrated,
\begin{equation}
\label{NullGeoApexEmittedBTZ}
\begin{split}
    &\frac{dx}{dz}=\pm\frac{1}{\sqrt{\alpha^2-\tilde{f}(z)}}\Rightarrow x-x_0=\pm z_+\text{arctanh}\left[\frac{z}{z_+\sqrt{\alpha^2-1+\frac{z^2}{z_+^2}}}\right]\Bigg\vert_{z_0}^z\\
    &\frac{dt}{dz}=\frac{\pm \alpha}{\tilde{f}(z)\sqrt{\alpha^2-\tilde{f}(z)}}\Rightarrow t-t_0=\pm z_+\text{arctanh}\left[\frac{z\alpha}{z_+\sqrt{\alpha^2-1+\frac{z^2}{z_+^2}}}\right]\Bigg\vert_{z_0}^z.\\
\end{split}
\end{equation}
In the special case where we choose a point $(t_0,x_0,z_0=0)$ on the boundary we can use the expressions above to obtain a fairly simple expression for a boundary lightcone
\begin{equation}
\label{boundaryStartLCBTZSch}
    \begin{split}
        &\left(\frac{t(x,z)-t_0}{z_+}\right)^2=\text{arctanh}^2\left[\frac{\sqrt{z^2+(z_+^2-z^2)\tanh^2\left(\frac{x-x_0}{z_+}\right)}}{z_+}\right]\\
         &=\text{arctanh}^2\left[\sqrt{1-\frac{1-\frac{z^2}{z_+^2}}{\cosh^2\left(\frac{x-x_0}{z_+}\right)}}\right].\\
    \end{split}
\end{equation}
For more general cases the expression will be much more complicated and we will not explicitly write it here (nor will we require it in our discussions).

\subsection{Entanglement Wedges For Constant Time Intervals in BTZ Background}
\label{AppendixEntangleWedgesConstTimeIntervalsBTZ}

An important fact about the thermal RT curves given in Eq. (\ref{ThermalRTSchBdryAnchor}), which are anchored to constant time slice intervals on the boundary, is that they can be understood in terms of the intersection of certain lightcones whose apexes live on certain points on a conformal boundary (recall these lightcones are described by the equation given in Eq. (\ref{boundaryStartLCBTZSch})). In particular, for the thermal RT surface given in Eq. (\ref{ThermalRTSchBdryAnchor}) we can define two lightcones. One lightcone is generated by past directed null geodesics originating from the point $(t_0=t_{\text{Bdry}}+a,x_0=0,z=0)$ given by,
\begin{equation}
    t_{p.d.}(x,z)=t_{\text{Bdry}}+a-z_+\text{arctanh}\left[\sqrt{1-\frac{1-\frac{z^2}{z_+^2}}{\cosh^2\left(\frac{x}{z_+}\right)}}\right].
\end{equation}
The other lightcone consists of future oriented null geodesics originating from the point $(t_0=t_{\text{Bdry}}-a,x_0=0,z=0)$ given by,
\begin{equation}
    t_{f.d.}(x,z)=t_{\text{Bdry}}-a+z_+\text{arctanh}\left[\sqrt{1-\frac{1-\frac{z^2}{z_+^2}}{\cosh^2\left(\frac{x}{z_+}\right)}}\right].
\end{equation}
The surfaces will intersect along a curve on the constant time slice $t=t_{\text{Bdry}}$ we can find this curve by setting
\begin{equation}
    t_{p.d.}(x,z)=t_{f.d.}(x,z)\Rightarrow z=z_+\sqrt{1-\frac{\cosh^2\left(\frac{x}{z_+}\right)}{\cosh^2\left(\frac{a}{z_+}\right)}}.
\end{equation}
This is precisely the RT curve anchored to a constant time boundary interval. The region enclosed by $t_{p.d.}$ and $t_{f.d.}$ is called the causal wedge we can express the null boundary of the causal wedge in the following compact notation
\begin{equation}
    a-|t-t_{\text{Bdry}}|=z_+\text{arctanh}\left[\sqrt{1-\frac{1-\frac{z^2}{z_+^2}}{\cosh^2\left(\frac{x}{z_+}\right)}}\right],
\end{equation}
where $a-|t-t_{\text{Bdry}}|\geq 0$, which is what we wrote in Eq. (\ref{NullBdryOfEntangleWedgeBTZ}). Points inside the causal wedge satisfy
\begin{equation}
\label{randomlabel}
    \begin{split}
        &a-|t-t_{\text{Bdry}}|>z_+\text{arctanh}\left[\sqrt{1-\frac{1-\frac{z^2}{z_+^2}}{\cosh^2\left(\frac{x}{z_+}\right)}}\right].\\
    \end{split}
\end{equation}
Since the boundary of the causal wedge contains the set of points in the RT curve we can conclude a well-known fact that in the BTZ black hole background the causal and entanglement wedges are the same. We give an example of what it may look like in Figure \ref{EntangWedgeBdryBTZ}. 

\begin{figure}
\centering
\includegraphics[width=120mm]{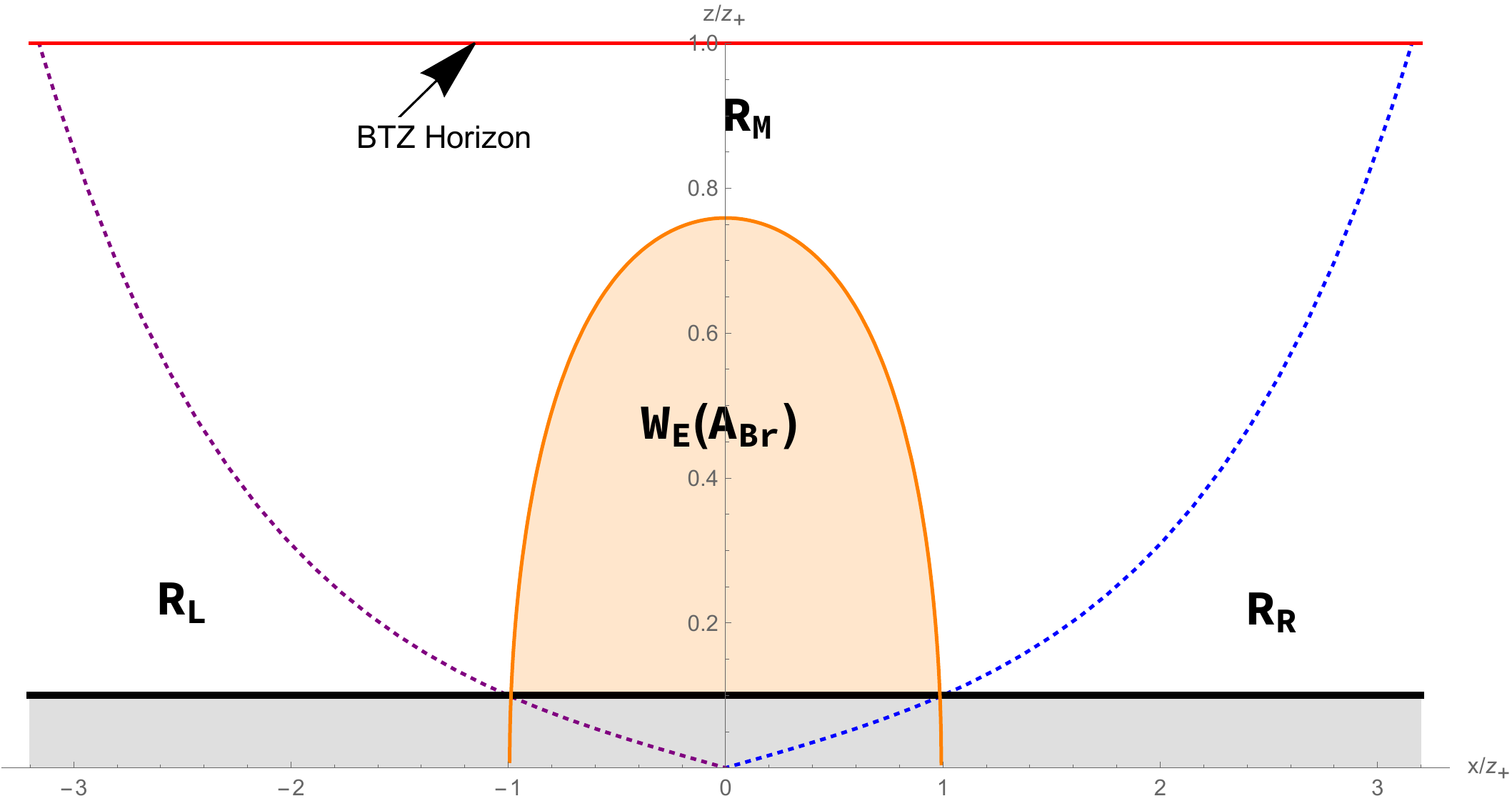}
\caption{We depict how the planes defined in Eq. (\ref{SigmaRLBTZ}) split the BTZ exterior geometry into $\mathcal{R}_{L,M,R}$. In particular, $z_{\sigma_R,BTZ}$ ($z_{\sigma_L,BTZ}$) are depicted by the dotted blue (purple) curves. The shaded orange shaded region is $\mathcal{W}_E(A_{\text{Br}})\subset \mathcal{R}_M$. BTZ horizon is solid red line, shaded gray region in behind the brane and thick black line is the brane. }
\label{BTZRMLRegions}
\end{figure}

For a constant time slice on the Brane in the BTZ background we take a similar approach to as we did for Poincare AdS$_3$. For the interval on the brane $A_{\text{Br}}$ we continue it past the brane and allow it to end on the boundary this defines a new constant time slice interval, the ``virtual interval'', denoted $\text{Vir}(A_{\text{Br}})$ which is defined as 
\begin{equation}
    \begin{split}
        &\text{Vir}(A_{\text{Br}})=\{(t,x,z)|t=t_{\text{Br}},x\in[-a',a'],z=0\}\\
        &a'=z_+\text{arccosh}\left[\frac{\cosh\left(\frac{a}{z_+}\right)}{\sqrt{1-\frac{z_{Br}^2}{z_+^2}}}\right].\\
    \end{split}
\end{equation}
Using this ``virtual interval'' we can construct its entanglement wedge in the full BTZ black hole background. Now we claim that the entanglement wedge associated to $A_{\text{Br}}$ can be identified as $\mathcal{W}_E(A_{\text{Br}})=\mathcal{W}_E(\text{Vir}(A_{\text{Br}}))\cap\mathcal{M}_{\text{phys}}$. First we need to introduce the BTZ analogues of planes $\sigma_{L,R}$ which will enable us to identify $\mathcal{R}_{L,M,R}$ which we discussed in the AdS$_3$ examples in Section \ref{ConstructWEBSec}. In the BTZ case the key is to appeal to the lightcones describing $\mathcal{W}_E(\text{Vir}(A_{\text{Br}}))$ whose apexes are located at $(t=t_{\text{Br}}\pm a',x=0,z=0)$. The set of geodesics emitted from these apexes generates the null sheets that enclose $\mathcal{W}_E(\text{Vir}(A_{\text{Br}}))$. In particular, referring to our results in Appendix \ref{NullGeodesAndLCBTZ} in Eq. (\ref{NullGeoApexEmittedBTZ}) the set of null geodesics emitted from the apexes have the following profile in the $xz$-plane
\begin{equation}
    x(z)=\pm z_+\text{arctanh}\left[\frac{z}{z_+\sqrt{\alpha^2-1+\frac{z^2}{z_+^2}}}\right].
\end{equation}
The $+$ ($-$) branch are the set of null geodesics emitted toward $x>0$ ($x<0$). These rays form a foliation of the spacetime in the $xz$-plane parametrized by $\alpha$. We can trivially extend these rays along the $t$ direction and define a foliation of the exterior BTZ geometry. The advantage of this foliation is that each individual null geodesic emitted from the apexes is restricted to a single leaf. The values for $\alpha$ for the null geodesics that intersect the entangling surface on the brane is given by
\begin{equation}
\label{alphacrit}
    \alpha^2=\alpha_c^2=1+\frac{\frac{z_{\text{Br}}^2}{z_+^2}}{\sinh^2\left(\frac{a}{z_+}\right)}.
\end{equation}
This defines two planes in the BTZ background described by
\begin{equation}
\label{SigmaRLBTZ}
\begin{split}
    &z_{\sigma_R,BTZ}(t,x)=z_{Br}\frac{\sinh\left(\frac{x}{z_+}\right)}{\sinh\left(\frac{a}{z_+}\right)}\Theta(x)\\
    &z_{\sigma_L,BTZ}(t,x)=-z_{Br}\frac{\sinh\left(\frac{x}{z_+}\right)}{\sinh\left(\frac{a}{z_+}\right)}\Theta(-x).\\
\end{split}
\end{equation}
As the subscripts suggest these are the BTZ analogues of the equations for $\sigma_R,\sigma_L$ we discussed in AdS$_3$. Now it is straightforward to split the bulk into $\mathcal{R}_{L,M,R}$ and we depict this in Figure \ref{BTZRMLRegions} which (roughly speaking) is the BTZ analogue of Figure \ref{IllustratingSuffCondFig} obtained in the AdS$_3$ analysis. 
It should be stressed that Figure \ref{BTZRMLRegions} should be properly understood as a the $t=t_{\text{Br}}$ slice of the setup. When we go to a different slice $t=t_{\text{Br}}$ the dotted lines which separate $R_{L,M,R}$ remain fixed, however, the solid black line representing the brane will shift away from $z=z_{Br}$ to a different position given by Eq. (\ref{BraneTraj}). It will be more convenient to express the trajectory of the brane in $z$ rather than $r$ coordinates. Using the transformation $z=L^2/r$ the trajectory of the brane reads
\begin{equation}
    z(t)=z_+\frac{\cos y_{\text{Br}}}{\sqrt{1-\sin^2y_{\text{Br}}\tanh^2\left(\frac{t}{z_+}\right)}}.
\end{equation}
We will find it useful to express $y_{\text{Br}}$ in terms of $z_{Br}$ and $t_{Br}$. Using the fact that $z(t_{Br})=z_{Br}$ we will find that
\begin{equation}
    \sin^2y_{\text{Br}}=\frac{1-\frac{z_{Br}^2}{z_+^2}}{1+\left(1-\frac{z_{Br}^2}{z_+^2}\right)\tanh^2\left(\frac{t}{z_+}\right)}.
\end{equation}
Using this, we obtain the final form for the trajectory of the brane
\begin{equation}
\label{FinalBraneTraj}
    Z_{\star}(t)=\sqrt{\frac{2Z_{Br}^2\cosh^2T}{1+Z_{Br}^2\cosh(2T)+(1-Z_{Br}^2)\cosh(2T_{Br})}},
\end{equation}
where to reduce clutter we defined the horizon scaled variables $Z=z/z_+$, $Z_{Br}=z_{Br}/z_+$, $T=t/z_+$, and $T_{Br}=t_{Br}/z_+$. Taking the derivative of the brane trajectory with respect to $T$ one can easily see that it vanishes at $T=0$ and is positive(negative) for $T>0$($T<0$). Consider some choice of $T_{Br}\geq0$ the interval $A_{\text{Br}}$ and $\bar{\chi}_{\text{ther}}(A_{\text{Br}})$ lie on the $T=T_{Br}$ time slice. We claim that any portion of the null congruence originating from $\bar{\chi}_{\text{ther}}(A_{\text{Br}})$ behind the brane heading towards the boundary at $z=0$ will remain behind the brane at all times. By proving this we will be able to conclude that only null congruences from $\chi_{\text{ther}}(A_{\text{Br}})$ are needed to fully enclose a co-dimension one region (i.e. no need to amend with lightcones near the interval endpoints) and that $\mathcal{W}_E(A_{\text{Br}})=\mathcal{W}_E(\text{Vir}(A_{\text{Br}}))\cap \mathcal{M}_{\text{phys}}$.   

Begin by writing down the trajectory of null geodesics that originate from $\bar{\chi}_{\text{ther}}(A_{\text{Br}})$ and travel towards to boundary. These are given by the results in Eq. (\ref{NullGeoApexEmittedBTZ}), using these, we can write the relevant trajectories in the $tz$-plane as
\begin{equation}
\begin{split}
    &Z_{\text{ray}}(T)=\sqrt{\frac{(\alpha^2-1)\sinh^2\left(|T-T_{Br}|-A'\right)}{1+(\alpha^2-1)\cosh^2\left(|T-T_{Br}|-A'\right)}}\\
    &A'=\text{arccosh}\left[\frac{\cosh(A)}{1-Z_{Br}^2}\right], \qquad A=\frac{a}{z_+}.\\
    \end{split}
\end{equation}
It is straightforward to see that when $|T-T_{Br}|=A'$ all the geodesics will intersect at the boundary at the apexes of the cones we described earlier and when $T=T_{Br}$ the null geodesics will lie on $\bar{\chi}_{\text{ther}}(A_{\text{Br}})$. So we are interested in the time window $0\leq|T-T_{Br}|\leq A'$. Each value of $\alpha\geq 1$ corresponds to a particular ray in the congruence emitted toward the boundary from $\bar{\chi}_{\text{ther}}(A_{\text{Br}})$. The rays which begin from behind the brane at $T=T_{Br}$ are precisely the ones with $1\leq |\alpha|\leq\alpha_{c}$ where $\alpha_c$ is given in Eq. (\ref{alphacrit}) (note that these rays will live in $\mathcal{R}_{L,R}$  and when $\alpha=\pm\alpha_c$ the rays follow trajectories in the $xz$-plane corresponding to the dotted lines in Figure \ref{BTZRMLRegions}). We claim the family of rays we identified in the discussion above will always remain behind the brane. To prove this we compute
\begin{equation}
\label{RayNestingInA}
    \begin{split}
        &Z_{\text{ray}}^2\vert_{A=A_2}-Z_{\text{ray}}^2\vert_{A=A_1}\\
        &=\frac{(\alpha^2-1)\sinh^2\left(|T-T_{Br}|-A_2'\right)}{1+(\alpha^2-1)\cosh^2\left(|T-T_{Br}|-A_2'\right)}-\frac{(\alpha^2-1)\sinh^2\left(|T-T_{Br}|-A_1'\right)}{1+(\alpha^2-1)\cosh^2\left(|T-T_{Br}|-A_1'\right)}\\
        &=\frac{\alpha^2(\alpha^2-1)\sinh(A'_2-A'_1)\sinh(A'_1+A'_2-2|T-T_{Br}|)}{[1+(\alpha^2-1)\cosh^2\left(|T-T_{Br}|-A_2'\right)][1+(\alpha^2-1)\cosh^2\left(|T-T_{Br}|-A_1'\right)]}\\
        &A'_{1,2}=\text{arccosh}\left[\frac{\cosh(A_{1,2})}{1-Z_{Br}^2}\right].\\
    \end{split}
\end{equation}
Analyzing the expression above we can clearly see that when $A_2>A_1$ we have that $Z_{\text{ray}}^2\vert_{A=A_2}-Z_{\text{ray}}^2\vert_{A=A_1}>0\Rightarrow Z_{\text{ray}}\vert_{A=A_2}-Z_{\text{ray}}\vert_{A=A_1}>0$.\footnote{Note that we need to make a restriction on $|T-T_{Br}|<A_1'$ since $Z_{\text{ray}}\vert_{A=A_1}$ terminates on the boundary at $|T-T_{Br}|=A_1'$ whereas $Z_{\text{ray}}\vert_{A=A_2}$ terminates later.} This means the trajectory of $Z_{\text{ray}}\vert_{A=A_1}$ is bounded/nested in by $Z_{\text{ray}}\vert_{A=A_2}$. Next we consider the following computation
\begin{equation}
    \begin{split}
        &Z_{\text{ray}}^2\vert_{\alpha=\alpha_2}-Z_{\text{ray}}^2\vert_{\alpha=\alpha_1}\\
        &=\frac{(\alpha_2^2-1)\sinh^2\left(|T-T_{Br}|-A'\right)}{1+(\alpha_2^2-1)\cosh^2\left(|T-T_{Br}|-A'\right)}-\frac{(\alpha_1^2-1)\sinh^2\left(|T-T_{Br}|-A'\right)}{1+(\alpha_1^2-1)\cosh^2\left(|T-T_{Br}|-A'\right)}\\
        &=\frac{(\alpha_2^2-\alpha_1^2)\sinh^2(|T-T_{Br}|-A')}{[1+(\alpha_2^2-1)\cosh^2\left(|T-T_{Br}|-A'\right)][1+(\alpha_1^2-1)\cosh^2\left(|T-T_{Br}|-A'\right)]}.\\
    \end{split}
\end{equation}
We can clearly see that anytime $\alpha_2>\alpha_1$ we have $Z_{\text{ray}}^2\vert_{\alpha=\alpha_2}-Z_{\text{ray}}^2\vert_{\alpha=\alpha_1}>0\Rightarrow Z_{\text{ray}}\vert_{\alpha=\alpha_2}-Z_{\text{ray}}\vert_{\alpha=\alpha_1}>0$. This means the trajectory of $Z_{\text{ray}}^2\vert_{\alpha=\alpha_1}$ is bounded/ nested in the trajectory of $Z_{\text{ray}}^2\vert_{\alpha=\alpha_2}$. In particular, all the rays that begin behind the brane with $\alpha^2<\alpha_c^2$ will be bounded by the rays with $\alpha=\alpha_c$. So we need only show that rays with $\alpha^2=\alpha_c^2$ remains behind the brane when $|T-T_{Br}|>0$. This can be expressed mathematically as satisfying the bound
\begin{equation}
\label{conditionBehindBrane}
    Z_{\star}^2-Z_{\text{ray}}^2|_{\alpha=\alpha_c}\geq 0.
\end{equation}
To prove the statement above we will make use of the fact that $\lim_{A\to\infty}Z_{\text{ray}}^2|_{\alpha=\alpha_c}>Z_{\text{ray}}^2|_{\alpha=\alpha_c}$ which was demonstrated through the calculation in Eq. (\ref{RayNestingInA}). We can explicitly compute\footnote{Note that the computation is made under the implicit assumption that $T_{Br}\geq 0$ with $T<T_{Br}$. This is sufficient because when $T_{Br}>0$ we know that the brane advances towards larger $Z$ when $T>T_{Br}$ so there is no chance of rays starting behind the brane to end up in front. However, when $T<T_{Br}$ the brane can move towards smaller values of $Z$ which can pose a problem so we really only need to check the regime where $T<T_{Br}$. Furthermore, when $T_{Br}<0$ the analysis will be similar in that case we only care about what will happen when $T>T_{Br}$.} 
\begin{equation}
\begin{split}
    &Z_{\star}^2-\lim_{A\to\infty}Z_{\text{ray}}^2|_{\alpha=\alpha_c}\\
    &=Z_{Br}^2\left[\frac{e^{-2T_{Br}}(-1+e^{2|\Delta T|})(1+2e^{2T_{Br}}+e^{2|\Delta T|})(1-Z_{Br}^2)}{[Z_{Br}^2+e^{2|\Delta T|}(1-Z_{Br}^2)][2+2(1-Z_{Br}^2)\cosh(2T_{Br})+2Z_{Br}^2\cosh(2(T_{Br}-|\Delta T|))]}\right].\\
    &\Delta T=T-T_{Br}\\
\end{split}
\end{equation}
By inspection of the expression above we can see that $Z_{\star}^2-\lim_{A\to\infty}Z_{\text{ray}}^2|_{\alpha=\alpha_c}\geq 0$ and saturation occurs only when $T=T_{Br}$. With this we can conclude that the condition in Eq. (\ref{conditionBehindBrane}) is satisfied which proves the claim that rays emitted from $\bar{\chi}_{\text{ther}}(A_{\text{Br}})$ towards the boundary which start behind the brane always remain behind the brane. This allows us to conclude that $\mathcal{W}_E(A_{\text{Br}})=\mathcal{W}_E(\text{Vir}(A_{\text{Br}}))\cap\mathcal{M}_{\text{phys}}$. We can explicitly write an inequality (which is in Eq. (\ref{EntWedgeBTZtoBr})) which characterizes the set of points in $\mathcal{W}_{E}(A_{\text{Br}})$ as
\begin{equation}
    \begin{split}
    &a'-|t-t_{\text{Br}}|>z_+\text{arctanh}\left[\sqrt{1-\frac{1-\frac{z^2}{z_+^2}}{\cosh^2\left(\frac{x}{z_+}\right)}}\right]\\
    &a'=z_+\text{arccosh}\left[\frac{\cosh\left(\frac{a}{z_+}\right)}{\sqrt{1-\frac{z_{Br}^2}{z_+^2}}}\right].\\
    \end{split}
\end{equation}
Where it is implicitly understood that we only keep points that are not cutoff by the ETW brane. In Figure \ref{EntangWedgeBraneBTZ} we given an example of what $\mathcal{W}_E(A_{\text{Br}})$ might look like.

\subsection{Analysis of Condition for EWN}
\label{AppendixNaiveConBTZEWN}
In this appendix we will manipulate the constraint given by Eq. (\ref{NiaveCondForSLSepGenPoints}) to obtain the inequality given in Eq. (\ref{ManipulatedSuffConBTZEWN}).

We begin by defining $\Delta t=t_{\chi}-t_{\text{Br}}$ and then we rearrange the inequality in Eq. (\ref{NiaveCondForSLSepGenPoints}) to obtain
\begin{equation}
\begin{split}
    \frac{z_{\chi}^2}{z_+^2}\geq 1-\frac{\cosh^2\left(\frac{a}{z_+}\right)}{\cosh^2\left(\frac{a'+|\Delta t|}{z_+}\right)}.
\end{split}
\end{equation}
We have the following identity which comes from a trigonometric expansion
\begin{equation}
    \cosh\left(\frac{a'+|\Delta t|}{z_+}\right)=\frac{\cosh\left(\frac{a}{z_+}\right)\cosh\left(\frac{|\Delta t|}{z_+}\right)+\sinh\left(\frac{|\Delta t|}{z_+}\right)\sqrt{\sinh^2\left(\frac{a}{z_+}\right)+\frac{z_{Br}^2}{z_+^2}}}{\sqrt{1-\frac{z_{Br}^2}{z_+^2}}}.
\end{equation}
Using this we obtain
\begin{equation}
\begin{split}
    &\frac{z_{\chi}^2}{z_+^2}\geq 1-\frac{1-\frac{z_{Br}^2}{z_+^2}}{\left[\cosh\left(\frac{|\Delta t|}{z_+}\right)+\sinh\left(\frac{|\Delta t|}{z_+}\right)\sqrt{\tanh^2\left(\frac{a}{z_+}\right)+\frac{z_{Br}^2}{z_+^2\cosh^2\left(\frac{a}{z_+}\right)}}\right]^2}\\
    &\frac{z_{\chi}^2}{z_+^2}\geq 1-\frac{1-\frac{z_{Br}^2}{z_+^2}}{\left[\cosh\left(\frac{|\Delta t|}{z_+}\right)+\sinh\left(\frac{|\Delta t|}{z_+}\right)\sqrt{1-\frac{1-\frac{z_{Br}^2}{z_+^2}}{\cosh^2\left(\frac{a}{z_+}\right)}}\right]^2}.\\
\end{split}
\end{equation}
Next we plug in the explicit expression for $z_{\chi}$ from Eq. (\ref{RightExtConnRTSch}) and with some simple algebraic manipulations we arrive at
\begin{equation}
    \left|\frac{B}{A}\right|\geq \sqrt{1-\frac{1-\frac{z_{Br}^2}{z_+^2}}{\cosh^2\left(\frac{a}{z_+}\right)}}.
\end{equation}
This gives us Eq. (\ref{ManipulatedSuffConBTZEWN}). It is useful to note that when $c_t=0$ we have
\begin{equation}
\begin{split}
    & \left|\frac{B}{A}\right|^2\bigg\vert_{c_t=0}-\left(1-\frac{1-\frac{z_{Br}^2}{z_+^2}}{\cosh^2\left(\frac{a}{z_+}\right)}\right)=\left|\frac{\sin y_{\text{Br}}}{\sin\tau_{\text{Br}}}\right|^2-\left(1-\frac{1-\frac{z_{Br}^2}{z_+^2}}{\cosh^2\left(\frac{a}{z_+}\right)}\right)\\
    &\geq \left|\frac{\sin y_{\text{Br}}}{\sin\tau_{\text{Br}}}\right|^2-1\geq 0, \\
\end{split}
\end{equation}
as long as $|\tau_{\text{Br}}|<y_{\text{Br}}$, which it is. So we can see that EWN is always satisfied in our setup when $c_t=0$. 

We argued in Section \ref{EWNSuffConBTZPlanarSec} that to translate the constraint above to a constraint on $c_t$ simply requires us to find the appropriate solutions to,
\begin{equation}
    \left|\frac{B}{A}\right|=\Xi=\sqrt{1-\frac{1-\frac{z_{Br}^2}{z_+^2}}{\cosh^2\left(\frac{a}{z_+}\right)}}.
\end{equation}
The solutions to this equation will determine the endpoints on an interval in the $c_t$ parameter space over which EWN is satisfied. In particular, the parameter space over which EWN holds will take the form $c_t\in[c_-,c_+]$ where $c_-\leq 0$ and $c_+\geq 0$.

To get $c_-$ we solve
\begin{equation}
    \frac{B}{A}=\Xi,
\end{equation}
more explicitly the equation above reads,
\begin{equation}
\label{intervalendpointsEq}
    \begin{split}
        &\frac{X\cos^2 y_{\text{Br}}\sin\tau_{\text{Br}}-\sqrt{1-X^2}\sin^2y_{\text{Br}}\cos\tau_{\text{Br}}}{X\cos\tau_{\text{Br}}-\sqrt{1-X^2}\sin\tau_{\text{Br}}}=\Xi\sin y_{\text{Br}}\cos\tau_{\text{Br}}\\
        &X=c_t\sin y_{\text{Br}},\\
    \end{split}
\end{equation}
and the relevant solution is given by
\begin{equation}
    c_t=c_-=-\left[1+\frac{\left[\cos(2\tau_{\text{Br}})-\cos(2y_{\text{Br}})\right]\left[\Xi^2\cos^2\tau_{\text{Br}}-\cos^2y_{\text{Br}}\right]}{2\cos^2\tau_{\text{Br}}\left(\sin y_{\text{Br}}-\Xi\sin\tau_{\text{Br}}\right)^2}\right]^{-\frac{1}{2}}.
\end{equation}
For $c_+$ we need to solve
\begin{equation}
    \frac{B}{A}=-\Xi,
\end{equation}
which is the same equation as Eq. (\ref{intervalendpointsEq}) with an added minus sign on the right-hand side. The relevant solution is given by
\begin{equation}
    c_t=c_+=\left[1+\frac{\left[\cos(2\tau_{\text{Br}})-\cos(2y_{\text{Br}})\right]\left[\Xi^2\cos^2\tau_{\text{Br}}-\cos^2y_{\text{Br}}\right]}{2\cos^2\tau_{\text{Br}}\left(\sin y_{\text{Br}}+\Xi\sin\tau_{\text{Br}}\right)^2}\right]^{-\frac{1}{2}}.
\end{equation}
This gives us the bounds given in Eq. (\ref{ctIntervalGeneral}). It is not difficult to see from Figure \ref{EWNBTZIneqVis} that to get the tightest bound we need to move the horizontal line representing the value of $\Xi$ upwards. The highest it will go is when $a\to\infty\Rightarrow \Xi\to 1$ in that case we obtain the following results for $c_\pm$,
\begin{equation}
    \begin{split}
        &c_{\pm}=\pm\frac{1}{\sqrt{1+\left(\frac{\sin\tau_{\text{Br}}\mp\sin y_{\text{Br}}}{\cos\tau_{\text{Br}}}\right)^2}}.\\
    \end{split}
\end{equation}
These are the bounds in Eq. (\ref{ctIntervalTightest}).

\subsection{Condition for Connected Phase}
\label{BTZConnCondition}
In this appendix we will go compute the areas of the connected and disconnected surfaces to determine when a connected phase will exist.

To begin, we consider the areas of the thermal RT surfaces described by Eq. (\ref{ThermalRTSchBdryAnchor}) and Eq. (\ref{ThermalRTAnchoredToBr}). 

For the thermal RT surface anchored to the boundary we need to introduce a regulator/cutoff near the conformal boundary which we will send to zero when we analyze the condition for the connected phase. Specifically we will be interested in computing
\begin{equation}
\begin{split}
    & \mathcal{A}_{therBdry}^{(reg)}=\int_{x_L(\epsilon)}^{x_R(\epsilon)} dx \frac{L}{z(x)}\sqrt{1+\frac{\dot{z}(x)^2}{\tilde{f}(x)}}\\
    & \tilde{f}(x)=\tilde{f}(z(x))=1-\frac{z(x)^2}{z_+^2}\\
    & z(x) = z_+\sqrt{1-\frac{\cosh^2\left(\frac{x}{z_+}\right)}{\cosh^2\left(\frac{a}{z_+}\right)}}\\
    & x_R(\epsilon)=-x_L(\epsilon)=z_+\text{arccosh}\left[\cosh\left(\frac{a}{z_+}\right)\sqrt{1+\frac{\epsilon^2}{z_+^2}}\right].\\
\end{split}
\end{equation}
One can easily compute the integral and show
\begin{equation}
    \mathcal{A}_{therBdry}^{(reg)}=\frac{L}{2}\ln\left[\frac{\sinh^2\left(\frac{x_R(\epsilon)+a}{z_+}\right)}{\sinh^2\left(\frac{a-x_R(\epsilon)}{z_+}\right)}\right]=L\left[\ln\left(\frac{z_+^2}{\epsilon^2}\right)+\ln\left(4\sinh^2\left(\frac{a}{z_+}\right)\right)+\mathcal{O}\left(\frac{\epsilon^2}{z_+^2}\right)\right].
\end{equation}
For the thermal RT surface anchored to the brane there is no need for a regulator we can just do the computation directly. In this case we will be computing
\begin{equation}
\begin{split}
    &\mathcal{A}_{therBr}=\int_{-a}^a dx \frac{L}{z(x)}\sqrt{1+\frac{\dot{z}(x)^2}{\tilde{f}(x)}}\\
    &z(x) = z_+\sqrt{1-\frac{\cosh^2\left(\frac{x}{z_+}\right)}{\cosh^2\left(\frac{a'}{z_+}\right)}}\\
    &a'=z_+\text{arccosh}\left[\frac{\cosh\left(\frac{a}{z_+}\right)}{\sqrt{1-\left(\frac{z_{Br}}{z_+}\right)^2}}\right]\geq a.\\
\end{split}
\end{equation}
Using the results from before we can immediately write
\begin{equation}
    \mathcal{A}_{therBr}=\frac{L}{2}\ln\left[\frac{\sinh^2\left(\frac{a+a'}{z_+}\right)}{\sinh^2\left(\frac{a'-a}{z_+}\right)}\right].
\end{equation}

For the area of the connected surface we will find it convenient to compute the area in Kruskal coordinates. The area integral we will need to compute is given by
\begin{equation}
    \begin{split}
        &\mathcal{A}_{con.}^{(reg)}=2L\int_{y_L(\epsilon)}^{y_{\text{Br}}}\frac{dy}{\cos y}\sqrt{1-\dot{\tau}(y)^2}\\
        &\tau(y)=\tau_{\text{Bdry}}+\arcsin{[c_t\sin y]}+\arcsin{c_t}.\\
    \end{split}
\end{equation}
We can compute the anti-derivative of the integrand and obtain
\begin{equation}
    \mathcal{A}_{con.}^{(reg)}=2L\text{arctanh}\left(\sin y\sqrt{\frac{1-c_t^2}{1-c_t^2\sin^2 y}}\right)\bigg\vert_{y_L(\epsilon)}^{y_{\text{Br}}}.
\end{equation}
We need to expand the result in terms of $\epsilon$. To do this we need to note that the cutoff $z=\epsilon$ is defined in Schwarzschild coordinates. So using the expression for the connected surface given in Eq. (\ref{LeftExteriorConnSch}) we write
\begin{equation}
    \left(\frac{\epsilon}{z_+}\right)^2=1-\frac{1}{\left[\cosh\left(\frac{r_+|\tilde{\Delta t|}}{L^2}\right)+\left|\frac{\tilde{B}}{\tilde{A}}\right|\sinh\left(\frac{r_+|\tilde{\Delta t|}}{L^2}\right)\right]^2}.
\end{equation}
We can then express $\cosh\left(\frac{r_+|\tilde{\Delta t|}}{L^2}\right)$ in terms of $\epsilon$
\begin{equation}
\begin{split}
    & \cosh\left(\frac{r_+|\tilde{\Delta t|}}{L^2}\right)=\frac{-\frac{1}{\sqrt{1-\frac{\epsilon^2}{z_+^2}}}+\left|\frac{\tilde{B}}{\tilde{A}}\right|\sqrt{\frac{\tilde{B}^2}{\tilde{A}^2}+\frac{\epsilon^2}{z_+^2-\epsilon^2}}}{\frac{\tilde{B}^2}{\tilde{A}^2}-1}\\
    &\Rightarrow\frac{r_+|t-t_{\text{Bdry}}|}{L^2}=\frac{\frac{\epsilon^2}{z_+^2}}{2\left|\frac{\tilde{B}}{\tilde{A}}\right|}+\frac{3\frac{\tilde{B}^2}{\tilde{A}^2}-1}{8\left|\frac{\tilde{B}}{\tilde{A}}\right|^3}\frac{\epsilon^4}{z_+^4}+\mathcal{O}(\epsilon^6).\\
\end{split}
\end{equation}
Recall that sign of $\tilde{B}/{A}$ and $\tilde{\Delta t}$ should be the same so we actually can write the expressions without the absolute values and solve for $t$ and obtain (note that $\frac{L^2}{r_+}=z_+$)
\begin{equation}
    \frac{t(\epsilon)}{z_+}=\frac{t_{\text{Bdry}}}{z_+}+\frac{\frac{\epsilon^2}{z_+^2}}{2\left(\frac{\tilde{B}}{\tilde{A}}\right)}+\frac{3\frac{\tilde{B}^2}{\tilde{A}^2}-1}{8\left(\frac{\tilde{B}}{\tilde{A}}\right)^3}\frac{\epsilon^4}{z_+^4}+\mathcal{O}(\epsilon^6).
\end{equation}
Now we use the coordinate transformation in Eq. (\ref{SchwToKruskalCoordTrans}) and obtain
\begin{equation}
\begin{split}
    &y_L(\epsilon)=-\left[\arctan\left(e^{t(\epsilon)/z_+}\sqrt{\frac{z_+-\epsilon}{z_++\epsilon}}\right)+\arctan\left(e^{-t(\epsilon)/z_+}\sqrt{\frac{z_+-\epsilon}{z_++\epsilon}}\right)\right]\\
    &=-\frac{\pi}{2}+\cos(\tau_{\text{Bdry}})\frac{\epsilon}{z_+}+\mathcal{O}(\epsilon)^2.\\
\end{split}
\end{equation}
This allows us to write
\begin{equation}
    \mathcal{A}_{con.}^{(reg)}=L\left[\ln\left(\frac{z_+^2}{\epsilon^2}\right)+\ln\left(\frac{4(1-c_t^2)}{\cos^2\tau_{\text{Bdry}}}\left(\frac{\sqrt{1-c_t^2\sin^2y_{\text{Br}}}+\sin y_{\text{Br}}\sqrt{1-c_t^2}}{\sqrt{1-c_t^2\sin^2y_{\text{Br}}}-\sin y_{\text{Br}}\sqrt{1-c_t^2}}\right)\right)\right]+\mathcal{O}(\epsilon^2).
\end{equation}
Now the condition for the connected phase will be
\begin{equation}
    \lim_{\epsilon\to 0}\left[\mathcal{A}_{therBdry}^{(reg)}+\mathcal{A}_{therBr}-\mathcal{A}_{con.}^{(reg)}\right]>0.
\end{equation}
Plugging in all our expressions gives
\begin{equation}
    \begin{split}
        &\frac{\cos^2\tau_{\text{Bdry}}}{1-c_t^2}\left(\frac{\sqrt{1-c_t^2\sin^2y_{\text{Br}}}-\sin y_{\text{Br}}\sqrt{1-c_t^2}}{\sqrt{1-c_t^2\sin^2y_{\text{Br}}}+\sin y_{\text{Br}}\sqrt{1-c_t^2}}\right)\left(\frac{\sinh\left(\frac{a'+a}{z_+}\right)}{\sinh\left(\frac{a'-a}{z_+}\right)}\right)\sinh^2\left(\frac{a}{z_+}\right)\geq 1\\
        &a'=z_+\text{arccosh}\left[\frac{\cosh\left(\frac{a}{z_+}\right)}{\sqrt{1-\frac{\cos^2 y_{\text{Br}}}{\cos^2\tau_{\text{Br}}}}}\right].\\
    \end{split}
\end{equation}
Noting that $a'\geq a$, by inspection, we can see that as long as $a$ is sufficiently large, the inequality is satisfied, and therefore a connected phase is always guaranteed to exist as long as $a$ is sufficiently large. This makes the constraint we analyzed in the context of EWN relevant in the limit where $a\to\infty$ (i.e. the constraint we wrote in Eq. (\ref{ctIntervalTightest})).

\begin{comment}
Using the expression for $a'$ explicitly along with some trigonometric identities we can write the condition in the form
\begin{equation}
\begin{split}
    &e^{2a/z_+}>1+\frac{2}{\sqrt{\mathcal{R}}}\\
    &\mathcal{R}=\frac{\cos^2\tau_{\text{Bdry}}}{1-c_t^2}\frac{\sqrt{1-c_t^2\sin^2y_{\text{Br}}}-\sin y_{\text{Br}}\sqrt{1-c_t^2}}{\sqrt{1-c_t^2\sin^2y_{\text{Br}}}+\sin y_{\text{Br}}\sqrt{1-c_t^2}}\geq 0\\
    &\tau_{\text{Bdry}}=\tau_{\text{Br}}-\arcsin(c_t\sin y_{\text{Br}})-\arcsin(c_t).\\
\end{split}
\end{equation}
\end{comment}

\bibliography{Ref.bib}
\bibliographystyle{JHEP}

\end{document}